\documentclass[10pt,twocolumn,reprint,amsmath,amssymb,aps,prx,showpacs,longbibliography,superscriptaddress]{revtex4-1}
\usepackage[latin9]{inputenc}
\usepackage{amsmath}
\usepackage{amssymb}
\usepackage{graphicx}

\makeatletter
\usepackage{amsfonts}
\usepackage{graphicx}
\usepackage{graphics}

\makeatother

\begin{document}
\title{Multidimensional Spectroscopy of Time-Dependent Impurities in Ultracold
Fermions}
\author{Jia Wang}
\affiliation{Centre for Quantum Technology Theory, Swinburne University of Technology,
Melbourne 3122, Australia}
\date{\today}
\begin{abstract}
We investigate the system of a heavy impurity immersed in a degenerated
Fermi gas, where the impurity's internal degree of freedom (pseudospin)
is manipulated by a series of radiofrequency (RF) pulses at several
different times. Applying the functional determinant approach, we
carry out an essentially exact calculation of the Ramsey-interference-type
responses to the RF pulses. These responses are universal functions
of the multiple time intervals between the pulses for all time and
can be regarded as multidimensional (MD) spectroscopy of the system
in the time domain. A Fourier transformation of the time intervals
gives the MD spectroscopy in the frequency domain, providing insightful
information on the many-body correlation and relaxation via the cross-peaks,
e.g., the off-diagonal peaks in a two-dimensional spectrum. These
features are inaccessible for the conventional, one-dimensional absorption
spectrum. Our scheme provides a new method to investigate many-body
nonequilibrium physics beyond the linear response regime with the
accessible tools in cold atoms.
\end{abstract}
\maketitle

\section{Introduction}

Spectroscopy, which records the responses of materials to external
electromagnetic fields, has long been and probably will always be
an essential tool to investigate the structures, behaviors, chemical
reactions, and physical processes in materials. Conventional spectroscopy,
such as ordinary nuclear magnetic resonance (NMR) and optical spectroscopy,
usually shows the responses as a function of a single variable, e.g.,
the frequency of the electromagnetic wave, and hence is called one-dimensional
(1D). In contrast, multidimensional (MD) spectroscopy unfolds spectral
information into several dimensions, which improves resolution and
overcomes spectral congestion. In addition, MD spectroscopy carries
rich information on the correlations between resonance peaks and provides
insights into physics that 1D spectroscopy cannot access. One of the
earliest and most widely successful MD spectroscopy is the two-dimensional
(2D) NMR, first proposed by Jean Jeener and later demonstrated by
Richard Ernst and collaborators \citep{ErnstJCP1976,ErnstJCP1979}.
2D NMR can help distinguish overlapping signals in complex molecules
and unveil the couplings between different resonances, which revolutionize,
e.g., molecular dynamics and structural biology \citep{Ernst1987Book,Keeler2011Book}.

As an analog of its NMR counterparts, optical MD coherent spectroscopy
(MDCS) \citep{Mukamel1993JCP,Cundiff2015JApplP} adapts similar technology
for the IR, visible, or UV regions and sheds new light on chemical
kinetics and solid-state physics \citep{Zanni2001COSB,Jonas2003AnnRevPhysChem,Weinacht2009OE,Moran2012JPCL,Karaiskaj2015JCP,XiaoqinLi2015NC,Karaiskaj2016PRL,XiaoqinLi2016NP,XiaoqinLi2017NC}.
In particular, optical 2DCS reveals coherent and incoherent coupling
dynamics between resonances near the energy of neutral and charged
excitons in atomically thin transition metal dichalcogenides (TMD)
\citep{XiaoqinLi2016NL,Tempelaa2019NC}. More recently, people have
believed that these resonances are excitons dressed by Fermi sea electrons,
i.e., quasiparticles named attractive or repulsive exciton-polarons
\citep{Imamoglu2016NP,Imamoglu2020PRX,Dmitry2017PRB,Dmitry2018PRB,Dmitry2021PRB,Muir2022arXiv}.
Polaron, arguably the most celebrated quasiparticle \citep{Landau1933PhysZSoviet,Mahan2000Book},
has also attracted intensive interest in atomic physics experimentally
\citep{Schirotzek2009PRL,Zhang2012PRL,Grimm2012Nature,Kohl2012Nature,Demler2016Science,Hu2016PRL,Jorgensen2016PRL,Scazza2017PRL,Yan2019PRL,Zwierlein2020Science,Sagi2020PRX}
and theoretically \citep{Bruun2014Review,Chevy2006PRA,Lobo2006PRL,Combescot2007PRL,Punk2009PRA,Cui2010PRA,Mathy2011PRL,Schmidt2012PRA,Rath2013PRA,Shashi2014PRA,Li2014PRA,Kroiss2015PRL,Levinsen2015PRL,JiaWang2015PRL,HuHui2016PRA,Goulko2016PRA,HuHui2018PRA,JiaWang2019PRL,PenaArdila2019PRA,Mulkerin2019AnnPhys,Isaule2021PRA,Pessoa2021PRA,Seetharam2021PRL,Kushal2021PRL,HuHui2022PRA}.
A fundamental and quantitative understanding of these nonequilibrium
many-body dynamics between quasiparticles shown in 2D spectroscopy
is fascinating but challenging. While a commonly adapted approach,
the modified optical Bloch equation with phenomenological terms to
include many-body effects \citep{Smirl1996PRB,Buccafusca1999PRB,Cundiff2002PRB},
gives some intuitive interpretations, first-principal calculations
of 2D spectroscopy are rare. Despite some progress \citep{Zachary2001PRB,Ducastelle2018PRB,Veniard2018PRB},
a quantitative but perturbative study of the complete 2D spectroscopy
has only been carried out recently using the nonlinear (four-wave
mixing) Golden Rule \citep{Tempelaa2019NC} and more recently with
functional determinant approach \citep{Reichman2022arXiv}, with parameters
that can only be approximately obtained in a complex solid-state system.

By contrast, we perform an in-principal \textit{exact} calculation
in a much simpler but realistic system: a heavy impurity immersed
in a degenerate Fermi gas. In such a system, a single parameter, scattering
length, can fully describe the interaction between the impurity and
the isolated and non-interacting Fermi gas at ultracold temperature
and be accurately tuned by Feshbach resonance \citep{Chin2010RMP}.
This system is closely related to the Fermi polaron problem, whose
1D spectroscopy shows singularities \citep{Demler2012PRX,Schmidt2018Review}
that are remnants of polaron resonances destroyed by the well-known
Anderson's ``orthogonality catastrophe'' (OC) \citep{Anderson1967PRL}.
Our recent studies rigorously proved that these singularities could
reduce back to polaron resonances if a mechanism exists to prevent
OC, such as a superfluid pairing gap \citep{JiaWang2022PRL,JiaWang2022PRA}.
However, as far as we know, the correlations and coherent dynamics
between the Fermi singularities or polaron resonances in ultracold
gases have never been investigated; our work here is the first numerically
exact calculation of the MD spectroscopy of a polaron-like system
in ultracold gases. Here, we apply the functional determinant approach
(FDA) \citep{Leonid1996JMathPhys,Klich2003Book,Schonhammer2007PRB,Ivanov2013JMathPhys},
a non-perturbative method that rigorously includes all high-order
correlations and beyond mean-field many-body effects. Since exact
solutions of many-body systems are rare, our results can give new
insight, deepen our understanding of Fermi-edge singularity and polaron
physics, and be regarded as a benchmark to access the accuracy of
other approximation calculations of MD spectroscopy.

\begin{figure*}
\includegraphics[width=0.75\linewidth]{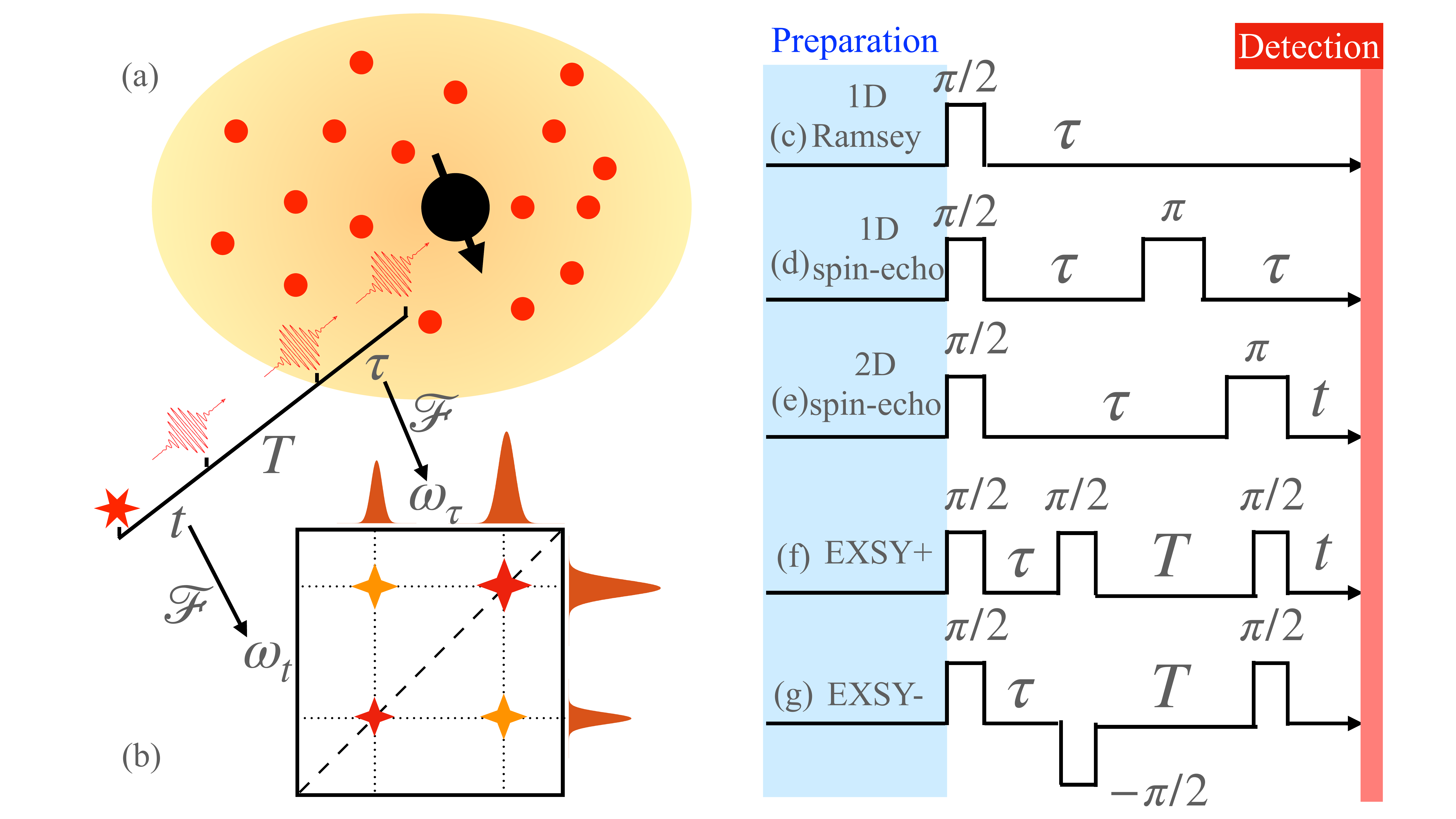}\caption{(a) A sketch of system setup: A localized impurity (the big black
ball with an arrow indicating pseudospin states) is immersed in a
sea of host fermions (red dots) and is manipulated by a series of
RF pulses with time intervals between pulses and detection (indicated
by the red six-point star). (b) is an example of a 2D spectrum, where
the absorption ($\omega_{\tau}$) and emission ($\omega_{t}$) frequencies
are obtained from the Fourier transformation for $\tau$ and $t$,
respectively. The two red crosses on the diagonal (dashed line) mirror
the two singularities in the 1D spectrum shown above and to the right
of the 2D spectrum. The two orange crosses on the off-diagonal are
called cross-peaks, revealing the correlations between the two singularities.
(c) shows a pulse sequence for the 1D Ramsey scheme, (d) for the 1D
spin-echo scheme, and (e) for the 2D spin-echo scheme. (f) and (g)
are EXSY pulse schemes with different pulse phases, which we call
EXSY$+$ and EXSY$-$, respectively. \label{fig:Sketch}}
\end{figure*}

We also propose a realistic experimental scheme to measure the MD
spectroscopy via a generalization of Ramsey spectroscopy. Ramsey spectroscopy
is another technique similar to the NMR, which manipulates internal
(e.g., pseudospin instead of spin) degrees of freedom and observes
the interference determined by the surrounding many-body environment.
Ramsey spectroscopy has found many vital applications in investigating
many-body physics: characterizing quantum correlations \citep{Bloch2004PRL_Ramsey,Boris2004PRA,Demler2010PRL_Ramsey,Demler2013PRL_Ramsey},
measuring topological invariance \citep{Bloch2013NP,Demler2013PRL_Topo},
accessing many-body interactions and beyond mean-field effects \citep{Zwierlein2003PRL,Ketterle2003Science,Rey2009PRL,Yu2010PRL,Ye2013Science,Ye2013Nature},
and studying impurity dynamics closely related to polaron physics
\citep{Goold2011PRA,Demler2012PRX,Schmidt2018Review,Goold2020PRL}.
Ramsey spectroscopy has become a well-established experimental technique
in ultracold atomic gases \citep{Demler2016Science}, thanks to the
unprecedented controllability and rich toolbox atomic physics provides
\citep{Bloch2008RMP,Chin2010RMP}. However, to the best of our knowledge,
all previous studies of Ramsey spectroscopies are 1D. Our work generalizes
the Ramsey spectroscopy to multidimensional, opening the door to exploring
high-order many-body correlations and beyond mean-field dynamics and
providing a perfect meeting point for theoretical and experimental
efforts to examine complex nonequilibrium responses that MD spectroscopy
reveals.

The rest of this paper is organized as follows. In the following section,
we establish our general formalism and show how to apply the exact
FDA approach to calculate MD spectroscopy. Section III is devoted
to presenting our numerical results. Finally, we conclude our paper
by discussing the physics and proposing future extensions in section
IV.

\section{Formalism}

The basic setup of our system is shown in Fig. \ref{fig:Sketch} (a).
We place a localized fermionic or bosonic impurity (the big black
ball) with two internal pseudospins (hyperfine) states $|\uparrow\rangle$
and $|\downarrow\rangle$ (illustrated by the black arrow) in the
background of a single-component ultracold Fermi gas (the red dots).
The localization of impurity can be either achieved by confinement
of a deep optical lattice or treated as an approximation to an impurity
atom with heavy mass. At ultralow temperature, the background Fermi
gas is considered non-interacting. We also assume the fermionic background
atoms do not interact with $|\downarrow\rangle$, while $s$-wave
interactions dominate interaction with $|\uparrow\rangle$. This interaction
is characterized by the $s$-wave scattering length $a$ and can be
tuned via, e.g., Feshbach resonances \citep{Chin2010RMP}. The general
spirit of our scheme is similar to the original Ramsey interferometry,
where one uses radio-frequency (RF) pulses to manipulate the superposition
of pseudospin states. Throughout this work, we assume the RF pulses
to be infinitely fast rotations of pseudospin that do not perturb
the background Fermi gas. After some time of evolution, dynamical
phases accumulate for different pseudospin states, which reflects
the many-body responses to the different impurity-background interactions
and can be measured by the interference. However, there is one crucial
difference in our scheme: we use multiple pulses with different time
delays in between to drive the pseudo-spin through many different
quantum pathways that give nonlinear many-body responses in an MD
spectroscopy.

One example of a three-pulse scheme is shown in Fig. \ref{fig:Sketch}
(a), which is similar to one of the most common 2D NMR pulse sequences,
namely EXSY (EXchange SpectroscopY). In this scheme, we prepare the
impurity in the non-interacting state $|\downarrow\rangle$ initially,
and apply the first $\pi/2$ pulse that rotates the pseudospin state
to $(|\uparrow\rangle+|\downarrow\rangle)/\sqrt{2}$. After some time
$\tau$, we apply the second pulse. Subsequently, we wait for another
period of time, $T$, before applying the third pulse and carry out
a detection some time $t$ afterward. Following the same procedure
as EXSY in NMR, we take the Fourier transformation with respect to
both time variables $\tau$ and $t$ to generate a 2D spectrum as
a function of an absorption frequency $\omega_{\tau}$ and an emission
frequency $\omega_{t}$, respectively, whose physical interpretation
will become clear later (in the last paragraph of section II). The
mixing time $T$ allows many-body dynamical evolution between absorption
and emission. Figure \ref{fig:Sketch} (b) sketches a 2D spectrum
in the box, where diagonal peaks (red crosses) on the dashed line
mirror the singularities in the linear 1D spectrum (shown on the top
and to the right of the box). Coupled resonances give rise to the
off-diagonal cross-peaks (orange crosses) with the absorption frequency
of one resonantce and the emission frequency of the other, whereas
uncorrelated resonances produce no cross-peaks. The corss peaks are
thus the signature of correlations between resonances, which the 1D
spectrum cannot distinguish. Figure \ref{fig:Sketch} (c)-(g) shows
several different pulse schemes investigated in this work. Most of
the pulses in this work are $\pi/2$ pulses, except for the second
pulse in the scheme of Fig. \ref{fig:Sketch} (g), which is a $-\pi/2$
pulse. For convenience, we name (f) and (g) EXSY$+$ and EXSY$-$,
respectively.

Using the unit $\hbar=1$ hereafter, we write the many-body Hamiltonian
as,

\begin{equation}
\hat{\mathcal{H}}=\mathcal{\hat{H}}_{\uparrow}|\uparrow\rangle\langle\uparrow|+\mathcal{\hat{H}}_{\downarrow}|\downarrow\rangle\langle\downarrow|,
\end{equation}
where the non-interacting ($\mathcal{H}_{\downarrow}$) and interacting
($\mathcal{H}_{\uparrow}$) Hamiltonian are given by $\mathcal{\hat{H}}_{\downarrow}=\sum_{\mathbf{k}}\epsilon_{\mathbf{k}}c_{\mathbf{k}}^{\dagger}c_{\mathbf{k}}$
and $\hat{\mathcal{H}}_{\uparrow}=\mathcal{\hat{H}}_{\downarrow}+\sum_{\mathbf{k},\mathbf{q}}\tilde{V}(\mathbf{k}-\mathbf{q})c_{\mathbf{k}}^{\dagger}c_{\mathbf{q}}+\omega_{s}$.
Here, $\omega_{s}$ denotes the energy differences between the two
pseudospin levels. $c_{\mathbf{k}}^{\dagger}$ and $c_{\mathbf{k}}$
are creation and annihilation operators of the background fermions
with momentum $\mathbf{k}$, respectively. $\epsilon_{\mathbf{k}}=k^{2}/2m$
is the single-particle kinetic energy of the background fermions with
mass $m$. $\tilde{V}(\mathbf{k})$ is the Fourier transform of $V(\mathbf{r})$,
the interaction potential between $|\uparrow\rangle$ and the background
fermions. Initially, we prepare the impurity in $|\downarrow\rangle$
and the background fermions at some temperature $T^{\circ}$. The
background fermions can be described by a thermal density matrix $\rho_{{\rm FS}}=\exp[-(\mathcal{\hat{H}}_{\downarrow}-\mu\hat{N})/k_{B}T^{\circ}]/Z_{{\rm FS}}$,
where $\hat{N}=\sum_{\mathbf{k}}c_{\mathbf{k}}^{\dagger}c_{\mathbf{k}}$
is the number operator, $Z_{{\rm FS}}$ is a normalization constant,
and $k_{B}$ is the Boltzman constant. Here, $\mu\simeq E_{F}$ is
the chemical potential determined by number density $n$, where $E_{F}$
is the Fermi energy that also gives a typical many-body time-scale
$\tau_{F}=E_{F}^{-1}$ and momentum scale $k_{F}=\sqrt{2mE_{F}}$.

We aim to investigate the dynamics of the system under multiple RF
pulses with different time delays in between. As a concrete example,
we focus on a three-pulse EXSY$+$ scheme, as illustrated in Fig.
\ref{fig:Sketch} (f). The RF pulses can manipulate the spin-state
of the impurity within a much shorter time than the intrinsic time
scales of the background fermions $\tau_{F}$. As a result, one can
neglect the evolution of the Fermi sea during the pulse and describe
the pulse's effect as a rotation of the impurity's spin state. For
example, a pulse that achieves a $\pi/2$ rotation can be defined
as

\begin{equation}
R(\pi/2)\equiv\left(\begin{array}{cc}
R_{\uparrow\uparrow}^{(\pi/2)} & R_{\uparrow\downarrow}^{(\pi/2)}\\
R_{\downarrow\uparrow}^{(\pi/2)} & R_{\downarrow\downarrow}^{(\pi/2)}
\end{array}\right)=\frac{1}{\sqrt{2}}\left(\begin{array}{cc}
1 & 1\\
-1 & 1
\end{array}\right).
\end{equation}
The total time evolution in EXSY$+$ scheme is thus given by the unitary
transformation
\begin{equation}
\mathcal{U}(t,T,\tau)=U(t)R(\pi/2)U(T)R(\pi/2)U(\tau)R(\pi/2),
\end{equation}
where
\begin{equation}
U(t^{\prime})=\left(\begin{array}{cc}
e^{-i\hat{H}_{\uparrow}t^{\prime}} & 0\\
0 & e^{-i\hat{H}_{\downarrow}t^{\prime}}
\end{array}\right)
\end{equation}
gives the time evolution in between pulses. We denote the initial
state as $\rho_{i}=\rho_{{\rm FS}}\otimes|\downarrow\rangle\langle\downarrow|$
and arrive at the final density matrix as $\rho_{f}=\mathcal{U}\rho_{i}\mathcal{U}^{\dagger}$.
We can define a multidimensional response function in the time domain,
$S(\tau,T,t)$, by measuring

\begin{equation}
{\rm Re}[S(\tau,T,t)]=-\mathrm{Tr}\left(\sigma_{x}\rho_{f}\right),\ {\rm Im}[S(\tau,T,t)]=-\mathrm{Tr}\left(\sigma_{y}\rho_{f}\right),
\end{equation}
where $\sigma_{x}$ and $\sigma_{y}$ are the usual Pauli matrices
in the spin-basis. A tedious but straightforward manipulation of algebra
can give a close form

\begin{equation}
S(\tau,T,t)=\sum_{i=1}^{16}S_{i}(\tau,T,t)\equiv\frac{1}{4}\sum_{i=1}^{16}\mathrm{Tr}[I_{i}(\tau,T,t)\rho_{{\rm FS}}].\label{eq:StauTt}
\end{equation}
Here, $I_{i}(\tau,T,t)$, which we name as pathways, are a direct
product of six operators in the form of $e^{\pm i\mathcal{\hat{H}}t}$,
\begin{equation}
I_{i}(\tau,T,t)=c_{\vec{\sigma}_{i}}e^{i\mathcal{H}_{\sigma_{1i}^{\prime}}\tau}e^{i\mathcal{H}_{\sigma_{2i}^{\prime}}T}e^{i\mathcal{H}_{\uparrow}t}e^{-i\mathcal{H}_{\downarrow}t}e^{-i\mathcal{H}_{\sigma_{2i}}T}e^{-i\mathcal{H}_{\sigma_{1i}}\tau},
\end{equation}
where $\vec{\sigma}_{i}\equiv(\sigma_{1i},\sigma_{2i},\sigma_{1i}^{\prime},\sigma_{2i}^{\prime})$
is a collective index that takes sixteen different combinations, and
$c_{\vec{\sigma}_{i}}=-8R_{\uparrow\sigma_{2i}^{\prime}}^{(\pi/2)}R_{\sigma_{2i}^{\prime}\sigma_{1i}^{\prime}}^{(\pi/2)}R_{\sigma_{1}^{\prime}\downarrow}^{(\pi/2)}R_{\downarrow\sigma_{2i}}^{(\pi/2)}R_{\sigma_{2i}\sigma_{1i}}^{(\pi/2)}R_{\sigma_{1i}\downarrow}^{(\pi/2)}$
are coefficients that take values of $\pm1$. Here, we have applied
the relation $R(\pi/2)^{-1}=R(-\pi/2)=R(\pi/2)^{T}$ for the derivation,
where the superscript $T$ denotes the transpose of a matrix (and
should not be confused with mixing time $T$). The sorting of the
pathways can be arranged arbitrarily for convenience. Here, our first
four pathways are chosen as

\begin{equation}
I_{1}(\tau,T,t)=e^{i\hat{\mathcal{H}}_{\downarrow}\tau}e^{i\hat{\mathcal{H}}_{\uparrow}T}e^{i\hat{\mathcal{H}}_{\uparrow}t}e^{-i\hat{\mathcal{H}}_{\downarrow}t}e^{-i\hat{\mathcal{H}}_{\uparrow}T}e^{-i\hat{\mathcal{H}}_{\uparrow}\tau},
\end{equation}
\begin{equation}
I_{2}(\tau,T,t)=e^{i\hat{\mathcal{H}}_{\downarrow}\tau}e^{i\hat{\mathcal{H}}_{\downarrow}T}e^{i\hat{\mathcal{H}}_{\uparrow}t}e^{-i\hat{\mathcal{H}}_{\downarrow}t}e^{-i\hat{\mathcal{H}}_{\downarrow}T}e^{-i\hat{\mathcal{H}}_{\uparrow}\tau},
\end{equation}
\begin{equation}
I_{3}(\tau,T,t)=e^{i\hat{\mathcal{H}}_{\downarrow}\tau}e^{i\hat{\mathcal{H}}_{\downarrow}T}e^{i\hat{\mathcal{H}}_{\uparrow}t}e^{-i\hat{\mathcal{H}}_{\downarrow}t}e^{-i\hat{\mathcal{H}}_{\uparrow}T}e^{-i\hat{\mathcal{H}}_{\uparrow}\tau},
\end{equation}
and
\begin{equation}
I_{4}(\tau,T,t)=e^{i\hat{\mathcal{H}}_{\downarrow}\tau}e^{i\hat{\mathcal{H}}_{\uparrow}T}e^{i\hat{\mathcal{H}}_{\uparrow}t}e^{-i\hat{\mathcal{H}}_{\downarrow}t}e^{-i\hat{\mathcal{H}}_{\downarrow}T}e^{-i\hat{\mathcal{H}}_{\uparrow}\tau}.
\end{equation}
The expressions for other twelve pathways can be found in Appendix
\ref{sec:Appendix_Pathway}.

The contribution of each pathway, $S_{i}(\tau,T,t)$, can be calculated
exactly via FDA. To proceed, we define $\mathcal{H}_{\downarrow}\equiv\Gamma(h_{\downarrow})$
and $\mathcal{H}_{\uparrow}\equiv\Gamma(h_{\uparrow})+\omega_{s}$.
Here $\Gamma(h)\equiv\sum_{\mathbf{k},\mathbf{q}}h_{\mathbf{k}\mathbf{q}}c_{\mathbf{k}}^{\dagger}c_{\mathbf{q}}$
is a bilinear fermionic many-body Hamiltonian in the Fock space, and
$h_{\mathbf{k}\mathbf{q}}$ represents the matrix elements of the
corresponding operator in the single-particle Hilbert space. These
matrix elements are explicitly given by $(h_{\downarrow})_{\mathbf{k}\mathbf{q}}=\epsilon_{\mathbf{k}}\delta_{\mathbf{k}\mathbf{q}}$
and $(h_{\uparrow})_{\mathbf{k}\mathbf{q}}=\epsilon_{\mathbf{k}}\delta_{\mathbf{k}\mathbf{q}}+\tilde{V}(\mathbf{k}-\mathbf{q})$.
With these definitions, we can rewrite
\begin{equation}
S_{i}(\tau,T,t)=\frac{1}{4}\tilde{S}_{i}(\tau,T,t)e^{-i\omega_{s}f_{i}(t,T,\tau)},\label{eq:Si}
\end{equation}
where $e^{-i\omega_{s}f_{i}(t,T,\tau)}$ gives a simple phase and
$\tilde{S}_{i}(\tau,T,t)$ is a product of the exponentials of the
bilinear fermionic operator, both of which can be calculated exactly.
For example, we have $S_{1}(\tau,T,t)=\tilde{S}_{1}(\tau,T,t)e^{i\omega_{s}t}e^{-i\omega_{s}\tau}/4$,
where
\begin{equation}
\begin{aligned}\tilde{S}_{1}(\tau,T,t)= & {\rm Tr}[e^{i\Gamma(h_{\downarrow})\tau}e^{i\Gamma(h_{\uparrow})T}e^{i\Gamma(h_{\uparrow})t}\times\\
 & e^{-i\Gamma(h_{\downarrow})t}e^{-i\Gamma(h_{\uparrow})T}e^{-i\Gamma(h_{\uparrow})\tau}\rho_{{\rm FS}}]
\end{aligned}
.
\end{equation}
Applying Levitov\textquoteright s formula gives

\begin{equation}
\tilde{S}_{1}(\tau,T,t)={\rm det}[(1-\hat{n})+R_{1}(\tau,T,t)\hat{n}],
\end{equation}
with
\begin{equation}
R_{1}(\tau,T,t)=e^{ih_{\downarrow}\tau}e^{ih_{\uparrow}T}e^{ih_{\uparrow}t}e^{-ih_{\downarrow}t}e^{-ih_{\uparrow}T}e^{-ih_{\uparrow}\tau},
\end{equation}
and $\hat{n}=n_{\mathbf{k}}\delta_{\mathbf{k}\mathbf{k}'}$, where
$n_{\mathbf{k}}=1/(e^{\epsilon_{\mathbf{k}}/k_{B}T^{\circ}}+1)$ denotes
the single-particle occupation number operator. Calculations of other
pathway contributions are similar, which are presented in Appendix
\ref{sec:Appendix_Pathway}.

Numerical calculations are carried out in a finite system confined
in a sphere of radius $R$. Keeping the density constant, we increase
$R$ towards infinity until numerical results are converged. Typically,
we choose $k_{F}R=250\pi$ in a calculation. We focus on the $s$-wave
interaction channel between $|\downarrow\rangle$ and the background
fermions near a broad Feshbach resonance, which can be well mimicked
by a spherically symmetric and short-range van-der-Waals type potential
$V(r)=-C_{6}\exp(-r^{6}/r_{0}^{6})/r^{6}$. Here, $C_{6}$ determines
the van-der-Waals length $l_{{\rm vdW}}=(2mC_{6})^{1/4}/2$, and we
choose $l_{{\rm vdW}}k_{F}=0.01\ll1$, so the short-range details
are unimportant. The low-temperature many-body physics can be determined
by the $s$-wave energy-dependent scattering length $a(E_{F})=-\tan\eta(k_{F})/k_{F}$
at the Fermi energy $E_{F}$, with $\eta(E_{F})$ being an energy-dependent
$s$-wave scattering phase-shift tuned by adjusting $r_{0}$. For
the simplicity of notation, we denote $a\equiv a(E_{F})$ hereafter.
Consequently, $\tilde{S}_{i}(t,T,\tau)$ is a universal function of
$k_{B}T^{\circ}/E_{F}$, $k_{F}a$, $t/\tau_{F}$, and $\tau/\tau_{F}$
in the whole time domain.

A summation of the contributions of all pathways gives the total response
$S(t,T,\tau)$, and the spectrum in the frequency domain can be obtained
via a double Fourier transformation
\begin{equation}
A(\omega_{\tau},T,\omega_{t})=\frac{1}{\pi^{2}}\int_{0}^{\infty}\int_{0}^{\infty}dtd\tau e^{i\omega_{\tau}\tau}S(\tau,T,t)e^{-i\omega_{t}t},\label{eq:Aw2D}
\end{equation}
where $\omega_{t}$ and $\omega_{\tau}$ are interpreted as an absorption
and emission frequency, respectively. On the other hand, the $T$-dependence
of $A(\omega_{\tau},T,\omega_{t})$ can reveal the many-body coherent
and incoherent dynamics. We notice that $A(\omega_{\tau},T,\omega_{t})=\sum_{i=1}^{16}A_{i}(\omega_{\tau},T,\omega_{t})$
can also be expressed as a summation of sixteen pathways, where the
expression of each pathway is given by Eq. (\ref{eq:Aw2D}), with
$A$ and $S$ replaced by $A_{i}$ and $S_{i}$, respectively.

We emphasize that the MD spectroscopy contains all the information
of 1D spectroscopy. For example, one can examine the $T=t=0$ case,
where the pulse scheme becomes the same as the original 1D Ramsey
scheme shown in Fig. \ref{fig:Sketch} (f). In this case, $S(\tau,T=0,t=0)$
reduces to the 1D Ramsey response function $S_{a}(\tau)=\mathrm{Tr}(e^{i\mathcal{\hat{H}}_{\downarrow}\tau}e^{-i\mathcal{\hat{H}}_{\uparrow}\tau}\rho_{{\rm FS}})$,
which is also called the time-dependent overlap function. Similarly,
we have $S_{e}(t)\equiv S(\tau=0,T=0,t)=S_{a}^{*}(t)$, where the
superscript $^{*}$ denotes complex conjugate. Correspondingly, we
have $\int d\omega_{t}A(\omega_{\tau},T=0,\omega_{t})=A_{a}(\omega_{\tau})$,
where $A_{a}(\omega_{\tau})=\int d\tau S_{a}(\tau)e^{i\omega_{\tau}\tau}/\pi$
is the 1D absorption spectrum. Similarly, we have $A_{e}(\omega_{t})=\int d\omega_{\tau}A(\omega_{\tau},T=0,\omega_{t})=A_{a}^{*}(\omega_{t})$.
Since $A_{a}(\omega_{\tau})$ is the absorption spectrum, its complex
conjugate $A_{e}(\omega_{t})$ can thus be interpreted as an emission
spectrum. These interpretations are consistent with the fact that
the integration of $A(\omega_{\tau},T=0,\omega_{t})$ over the emission
frequency $\omega_{t}$ gives the 1D absorption spectrum $A_{a}(\omega_{\tau})$
and vice versa. The physical process underlying $A(\omega_{\tau},T,\omega_{t})$
can be interpreted as follows: the system first gets excited by absorbing
a photon with frequency $\omega_{\tau}$, after a period of mixing
time $T$, and then emits a photon with frequency $\omega_{t}$.

\section{Results}

\subsection{Two-dimensional spin-echo response}

\begin{figure}
\includegraphics[width=0.98\linewidth]{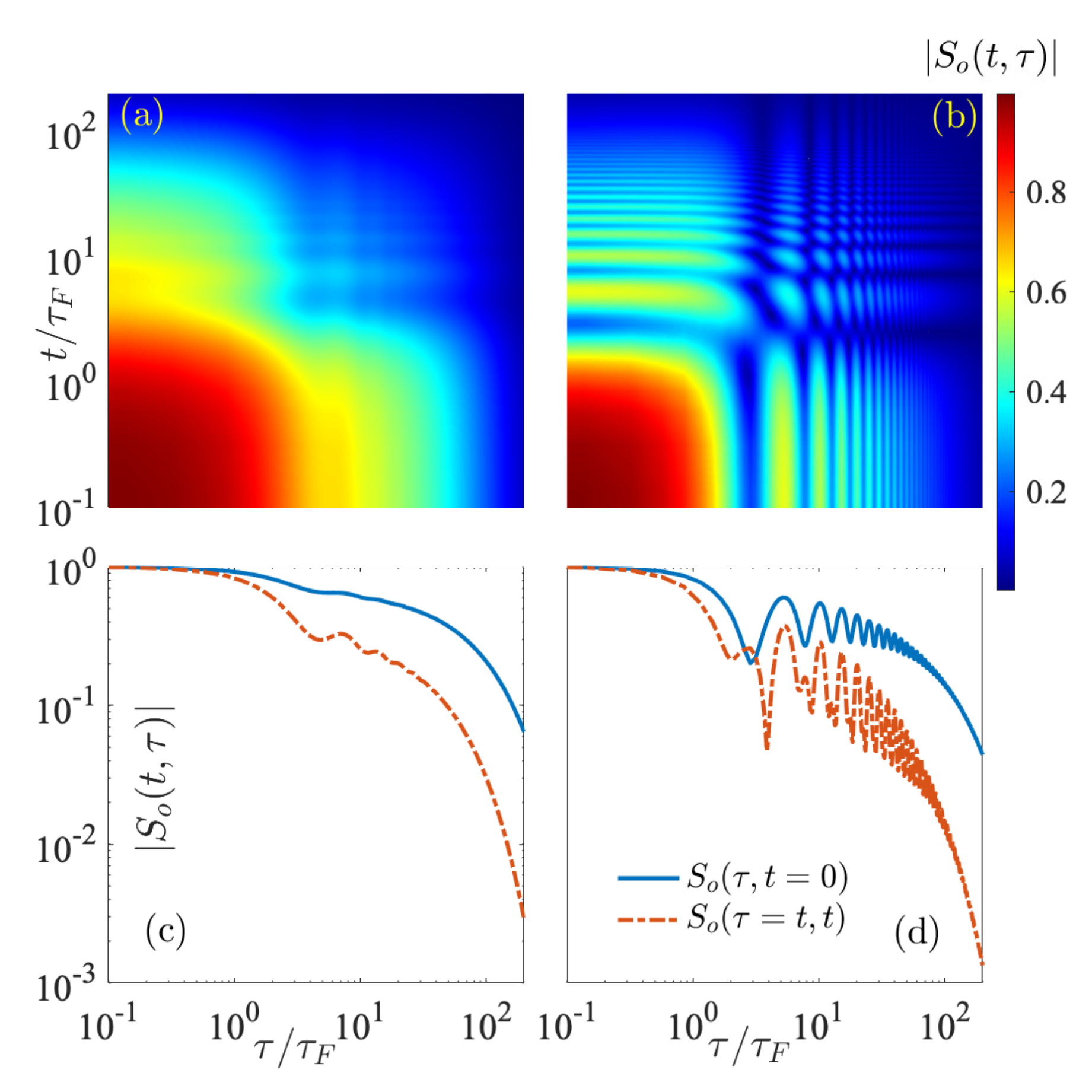}\caption{Universal 2D spin-echo response's amplitude $|S_{o}(\tau,t)|$, as
a function of $\tau$ and $t$, for (a) attractive interaction $k_{F}a=-0.5$
and (b) repulsive interaction $k_{F}a=0.5$. The temperature is set
as $k_{B}T^{\circ}=0.03E_{F}$. (c) shows $|S_{o}(\tau,t=0)|$ {[}the
slice of (a) along the $x$-axis{]} and $|S_{o}(\tau,t=\tau)|$ {[}the
slice of (a) along diagonal{]} as a function of $\tau$ in the blue
solid and red dash-dotted curves, respectively. (d) shows the same
slices for (b). \label{fig:Stplot}}
\end{figure}

Let us first investigate a relatively simple situation, $T=0$, which
is equivalent to the two-pulse scheme illustrated in Fig. \ref{fig:Sketch}
(e). The response function in the time domain is given by 
\begin{equation}
S_{o}(\tau,t)\equiv S(\tau,T=0,t)={\rm Tr}[I_{o}(\tau,t)\rho_{{\rm FS}}],\label{eq:Sotaut}
\end{equation}
where $I_{o}(\tau,t)=e^{i\hat{\mathcal{H}}_{\downarrow}\tau}e^{i\hat{\mathcal{H}}_{\uparrow}t}e^{-i\hat{\mathcal{H}}_{\downarrow}t}e^{-i\hat{\mathcal{H}}_{\uparrow}\tau}$.
We notice that when $t=\tau$, the scheme is equivalent to 1D spin-echo
scheme investigated in Refs. \citep{Demler2012PRX,Schmidt2018Review}
and illustrated in Fig. \ref{fig:Sketch} (d), hence naming our scheme
as a 2D spin echo scheme. We can examine that $S_{o}(t,t)$ reduce
to the 1D spin-echo response $S_{o}(t)=\mathrm{Tr}[e^{i\mathcal{H}_{\downarrow}t}e^{i\mathcal{H}_{\uparrow}t}e^{-i\mathcal{H}_{\downarrow}t}e^{-i\mathcal{H}_{\uparrow}t}\rho_{{\rm FS}}]$.

While we can also calculate $S_{o}(\tau,t)$ by using $S(\tau,T,t)$
in Eq. (\ref{eq:StauTt}) with $T=0$, a direct calculation of Eq.
(\ref{eq:Sotaut}) is more convenient. The expression of the 2D spin-echo
response can be written as $S_{o}(\tau,t)=e^{i\omega_{s}t}\tilde{S}_{o}(\tau,t)e^{-i\omega_{s}\tau}$,
where

\begin{equation}
\tilde{S}_{o}(\tau,t)={\rm Tr}[e^{i\Gamma(h_{\downarrow})\tau}e^{i\Gamma(h_{\uparrow})t}e^{-i\Gamma(h_{\downarrow})t}e^{-i\Gamma(h_{\uparrow})\tau}\rho_{{\rm FS}}]
\end{equation}
can be calculated exactly by applying Levitov\textquoteright s formula
in the FDA

\begin{equation}
\tilde{S}_{o}(\tau,t)={\rm det}[(1-\hat{n})+R_{o}(\tau,t)\hat{n}]
\end{equation}
with

\begin{equation}
R_{o}(\tau,t)=e^{ih_{\downarrow}\tau}e^{ih_{\uparrow}t}e^{-ih_{\downarrow}t}e^{-ih_{\uparrow}\tau}.
\end{equation}

Examples of this universal 2D response function $|S_{o}(t,\tau)|=|\tilde{S}_{o}(t,\tau)|$
with parameters $k_{F}a=-0.5$ and $k_{F}a=0.5$ at a finite temperature
$k_{B}T^{\circ}=0.03E_{F}$ are shown in Fig. \ref{fig:Stplot} (a)
and (b), respectively. The solid and dash-dotted curves in Fig. \ref{fig:Stplot}
(c) show $|S_{o}(\tau,t=0)|$ and $|S_{o}(\tau,t=\tau)|$ as a function
of $\tau$, i.e., the slice of (a) along the x-axis and diagonal,
respectively. Figure \ref{fig:Stplot} (d) shows the same slices for
(b). Fig. \ref{fig:Stplot} (c) indicates that $S_{o}(\tau,t=0)$
and $S_{o}(\tau,t=\tau)$ reduce to 1D Ramsey response $S_{a}(\tau)$
and 1D spin-echo signal $S_{o}(\tau)$, respectively.

\subsection{2D spin-echo spectrum}

\begin{figure*}
\includegraphics[width=0.98\columnwidth]{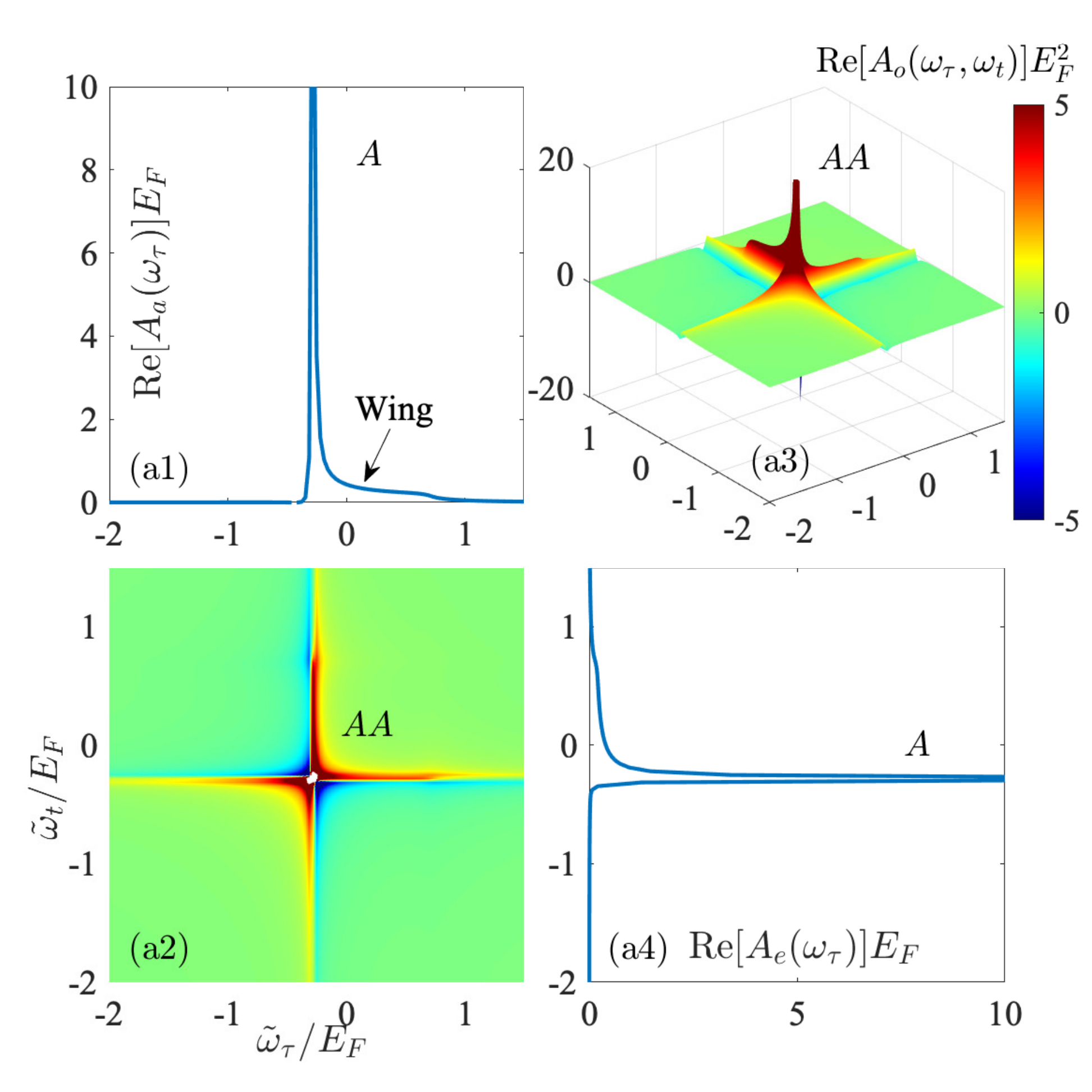}\includegraphics[width=0.98\columnwidth]{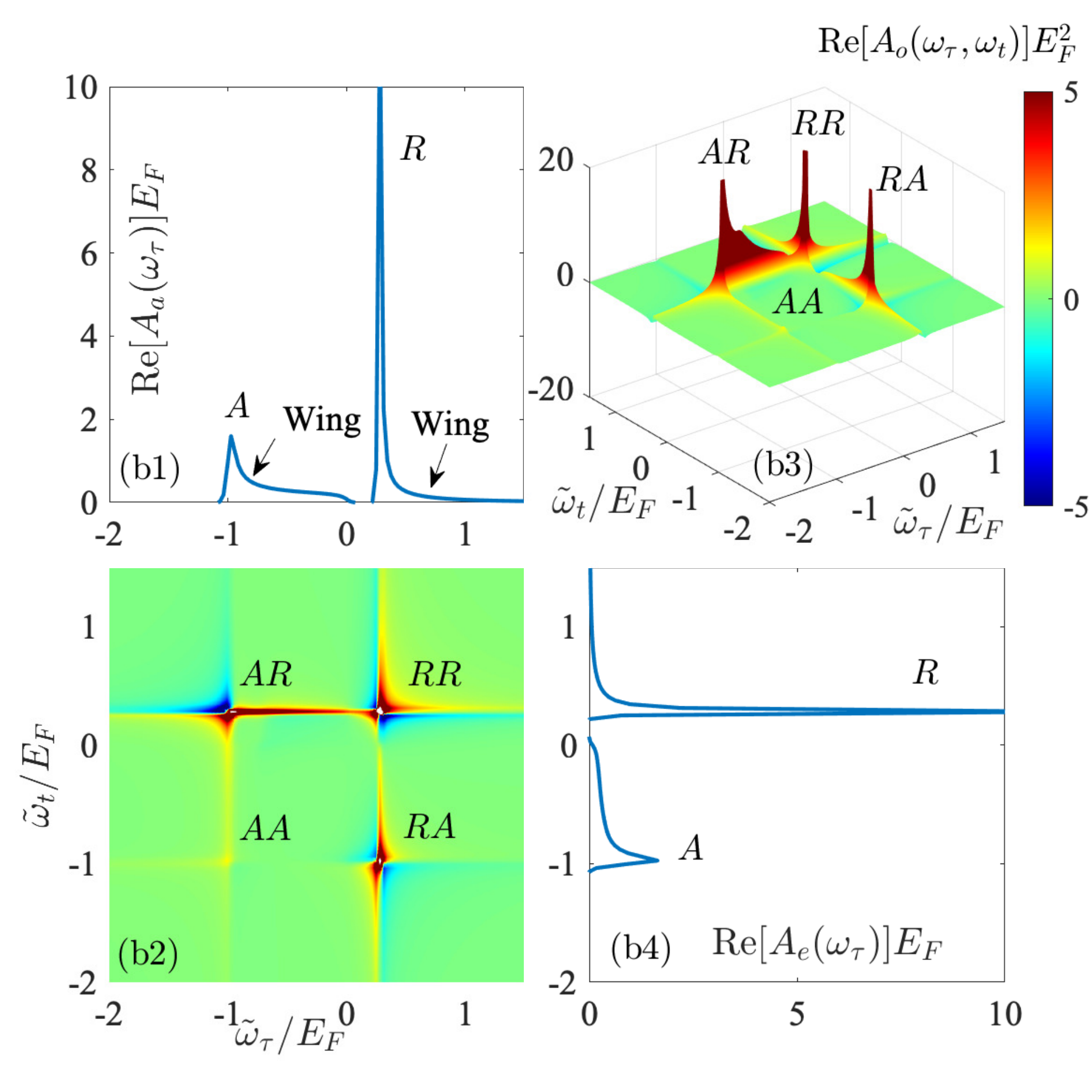}\caption{(a1) and (a4) shows the 1D absorption spectrum for attractive interaction
$k_{F}a=-0.05$ and finite temperature $k_{B}T^{\circ}=0.03E_{F}$.
The absorption singularity is denoted as $A$. (a2), and (a3) shows
the contour and 3D landscape of the 2D spin-echo spectrum ${\rm Re}[A_{o}(\omega_{\tau},\omega_{t})]$,
where the diagonal peak is denoted as $AA$. (b1)-(b4) are the same
as (a1)-(a4), correspondingly, but for repulsive interaction $k_{F}a=0.5$.
There are two singularities in (b1), the absorption spectrum, namely
repulsive and attractive singularities, which are denoted as $R$
and $A$. The corresponding diagonal peaks in (b2) and (b3) are denoted
as $AA$ and $RR$, while the off-diagonal cross-peaks are denoted
as $AR$ and $RA$. \label{fig:Aw}}
\end{figure*}

The 2D spin-echo spectrum in the frequency domain can be obtained
by applying a double Fourier transformation, Eq. (\ref{eq:Aw2D}),
with the relation $A_{o}(\omega_{\tau},\omega_{t})=A(\omega_{\tau},T=0,\omega_{t})$.
One can immediately observe that $A_{o}(\omega_{\tau},\omega_{t})=\int_{0}^{\infty}\int_{0}^{\infty}dtd\tau e^{i\tilde{\omega}_{\tau}\tau}\tilde{S}_{o}(\tau,t)e^{-i\tilde{\omega}_{t}t}/\pi^{2}$
with $\tilde{\omega}_{\tau}=\omega_{\tau}-\omega_{s}$ and $\tilde{\omega}_{t}=\omega_{t}-\omega_{s}$,
i.e., the energy differences between two spin-states only give simple
shifts of frequencies. Hereafter, unless specified otherwise, we denote
$\tilde{\omega}=\omega-\omega_{s}$ for any frequency variable $\omega$.
Figure \ref{fig:Aw} shows our finite temperature ($k_{B}T^{\circ}=0.03E_{F}$)
results for attractive $k_{F}a=-0.5$ and repulsive $k_{F}a=0.5$
interactions in (a1)-(a4) and (b1)-(b4), respectively. We present
the 2D contour of ${\rm Re}[A_{o}(\omega_{\tau},\omega_{t})]$ as
a function of $\tilde{\omega}_{\tau}$ and $\tilde{\omega}_{t}$ in
Figs. \ref{fig:Aw} (a2) and (b2) and the corresponding 3D landscape
in Figs. \ref{fig:Aw} (a3) and (b3). Here, we denote ${\rm Re}[\mathcal{C}]$
and ${\rm Im}[\mathcal{C}]$ as the real and imaginary parts of $\mathcal{C}$,
respectively. For comparison, we also show the 1D spectra ${\rm Re}[A_{a}(\omega_{\tau})]$
in Figs. \ref{fig:Aw} (a1) and (b1) and ${\rm Re}[A_{e}(\omega_{t})]={\rm Re}[A_{a}(\omega_{\tau})]$
in Figs. \ref{fig:Aw} (a4) and (b4). For completeness, we also show
the imaginary part and amplitude of the corresponding 2D spectroscopy
in Fig. \ref{fig:AwComplex} in Appendix \ref{sec:Appendix_IandA}.

The absorption spectrum has been well studied before \citep{Demler2012PRX,Schmidt2018Review}.
When the interaction is attractive $(k_{F}a<0)$, only one power-law
singularity ${\rm Re}[A_{a}(\omega_{\tau})]\sim\theta(\omega_{\tau}-\omega_{A-})|\omega_{\tau}-\omega_{A-}|^{-a_{A-}}$
with exponent coefficient $a_{A-}>0$ appears near $\omega_{A-}$
with a slight thermal broadening, as shown in Figs. \ref{fig:Aw}
(a1) and (a4). In contrast, Figs. \ref{fig:Aw} (b1) and (b4) show
two singularities for a repulsive interaction ($k_{F}a>0$). These
singularities are understood as manifestations of the well-known Anderson's
orthogonality catastrophe (OC). Due to the existence of multiple particle-hole
excitations of the background Fermi sea induced by the infinitely
massive impurity, the many-particle states with and without impurity
interactions are orthogonal, which leads to a vanishing quasiparticle
residue. Our recent studies further examined the scenario where a
mechanism, such as a superfluid gap or finite impurity mass, suppresses
those multiple particle-hole excitations \citep{JiaWang2022PRL,JiaWang2022PRA}.
In this case, OC can be prevented, and the singularities reduce back
to the so-called attractive or repulsive Fermi polarons. At the same
time, the ``wings'' attached to the singularity, indicated in Figs.
\ref{fig:Aw}(a1) and (b1), separate from the polaron signal and reduce
to the so-called molecule-hole continuums. Because of their close
relations to polaron resonances, we name these singularities as attractive
and repulsive singularities and denote them by $A$ and $R$, respectively,
in Figs. \ref{fig:Aw} (a1), (a4), (b1), and (b4).

The 2D spectrum in Figs. \ref{fig:Aw} (a2) and (a3) shows a double
dispersion lineshape commonly found in 2D NMR around $(\tilde{\omega}_{\tau},\tilde{\omega}_{t})\approx(\tilde{\omega}_{A-},\tilde{\omega}_{A-})$,
which is called a diagonal peak denoted as $AA$. For attractive interaction
$k_{F}a=-0.5$, the attractive singularity appears at $\tilde{\omega}_{A-}\approx-0.28E_{F}$
in the absorption spectrum. We have numerically verified that the
integration of 2D spectroscopy over emission frequency $\omega_{t}$
gives the 1D absorption spectrum $A_{a}(\omega_{\tau})$ (not shown
here). Interestingly, we can observe that there is no diagonal spectral
weight corresponding to the wing. Rather, the spectral weight on the
off-diagonal $A_{o}(\omega_{\tau},\omega_{t}\approx\omega_{A-})$
and $A_{o}(\omega_{\tau}\approx\omega_{A-},\omega_{t})$ is significant
and resembles the lineshape of the wing. This is a non-trivial manifestation
of OC in the 2D spectroscopy: the inhomogeneous and homogeneous lineshape
does not have the OC characteristic. Here, the inhomogeneous and homogeneous
lineshape refer to the lineshape near a singularity along the diagonal
or the direction perpendicular to the diagonal, which is better illustrated
in the amplitude of 2D spectroscopy shown in Fig. \ref{fig:AwComplex}
(c) in Appendix \ref{sec:Appendix_IandA}. As we can observe, the
widths of the singularity are much sharper along these two directions,
which might help experimental identification of the singularity, especially
at finite temperatures. The homogeneous and inhomogeneous broadenings
in MD spectroscopy also have their own experimental significance,
similar to their NMR or optical counterpart. In a realistic experiment,
the ensemble average of the impurity signal can give rise to a further
inhomogeneous broadening induced by the disorder of the local environment
(such as spatial magnetic field fluctuation). However, these disorders
are usually non-correlated and would not introduce homogeneous broadening
\citep{Cundiff2015JApplP,XiaoqinLi2016NL,XiaoqinLi2017NC}.

For repulsive interaction $k_{F}a=0.5$, there are two singularities,
the attractive and repulsive singularities, in the 1D absorption spectrum.
These singularities appear at $\tilde{\omega}_{A+}\approx-0.98E_{F}$
and $\tilde{\omega}_{R+}\approx0.28E_{F}$ in Figs. \ref{fig:Aw}
(b1) and (b4). As shown in Fig. \ref{fig:Aw} (b2) and (b3), there
are two diagonal peaks, $AA$ and $RR$, in the 2D spectroscopy that
mirror the attractive and repulsive singularities. In addition, there
are also two significant cross-peaks, $AR$ and $RA$, which indicate
a strong many-body quantum correlation between the attractive and
repulsive singularity. As far as we know, this is the first prediction
of many-body correlations between Fermi singularities in cold atom
systems. If the impurity has a finite mass or the background Fermi
gas is replaced by a superfluid with an excitation gap, we believe
these cross-peaks would remain and represent the correlations between
attractive and repulsive polarons.

\subsection{2D EXSY spectrum}

\begin{figure}
\includegraphics[width=0.98\linewidth]{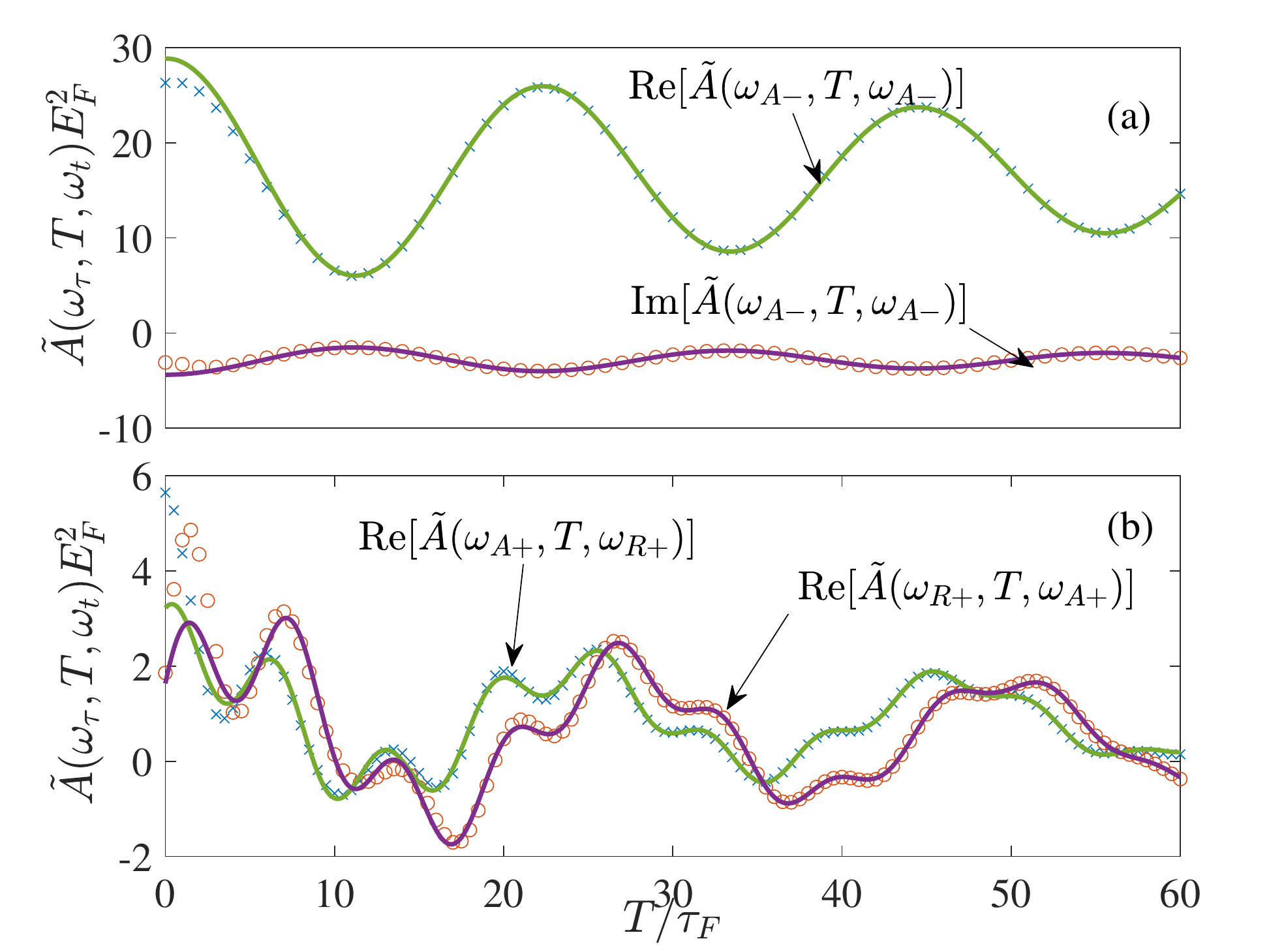}\caption{$T$-dependence of EXSY$+$ spectrum $\tilde{A}(\omega_{\tau},T,\omega_{t})$
for (a) attractive interaction $k_{F}a=-0.5$ and (b) repulsive interaction
$k_{F}a=0.5$, with temperature $k_{B}T^{\circ}=0.03E_{F}$. The cross
and circle symbols denote the real and imaginary parts for the diagonal
peak $\tilde{A}(\omega_{\tau}=\omega_{A-},T,\omega_{t}=\omega_{A-})$
in (a) and the real part of cross-peaks $\tilde{A}(\omega_{\tau}=\omega_{A+},T,\omega_{t}=\omega_{R+})$
and $\tilde{A}(\omega_{\tau}=\omega_{R+},T,\omega_{t}=\omega_{A+})$
in (b). The curves use the fitting formula in the main text to fit
numerical results at the late time $T\ge10\tau_{F}$. We, however,
also show the fitting curve at $T<10\tau_{F}$ to emphasize the differences
in the early time behaviors. \label{fig:Tdep}}
\end{figure}

In this section, we focus on the 2D spectrum, $A(\omega_{\tau},T,\omega_{t})$,
of the EXSY$+$ pulse scheme illustrated by Fig. \ref{fig:Sketch}
(f), which can be exactly calculated by Eqs. (\ref{eq:StauTt}), (\ref{eq:Si})
and (\ref{eq:Aw2D}) with the FDA. As mentioned above, in a 2D spin-echo
spectrum (and the 1D spectra), the trivial energy difference between
$|\downarrow\rangle$ and $|\uparrow\rangle$, $\omega_{s}$, only
introduces a frequency shift of the spectra as $(\omega_{\tau},\omega_{t})\rightarrow(\tilde{\omega}_{\tau},\tilde{\omega}_{t})$.
In contrast, the scenario is a bit more complicated for the EXSY$+$
spectrum, where the spectrum can be expressed as a summation of sixteen
pathway contributions, and each pathway is associated with a different
phase $e^{-i\omega_{s}f_{i}(t,T,\tau)}$ in Eq. (\ref{eq:Si}) (see
Appendix \ref{sec:Appendix_Pathway} for details). Consequently, each
pathway contribution $A_{i}(\omega_{\tau},T,\omega_{t})$ has shifted
to different centers in the frequency domain accordingly. The features
of Fermi singularities, in general, lie within a frequency range of
a few Fermi energy $E_{F}$ around $(\tilde{\omega}_{\tau},\tilde{\omega}_{t})=(0,0)$
in the 2D spectrum $A(\omega_{\tau},T,\omega_{t})$. In addition,
for a typical ultracold experiment, $\omega_{s}$ is usually much
larger than the Fermi energy $E_{F}$. As a result, only the first
four pathways associated with $e^{i\omega_{s}t}e^{-i\omega_{s}\tau}$
would give a non-negligible contribution, i.e., $A(\omega_{\tau},T,\omega_{t})\approx\tilde{A}(\omega_{\tau},T,\omega_{t})=\sum_{i=1}^{4}A_{i}(\omega_{\tau},T,\omega_{t})$,
within the frequency range in interest. A comparison between $A(\omega_{\tau},T,\omega_{t})$
and $\tilde{A}(\omega_{\tau},T,\omega_{t})$ for $\omega_{s}/E_{F}=2\pi$
is shown in Appendix \ref{sec:Appendix_Pathway}, where perfect agreement
is observed. Such reduction of pathways not only allows a faster calculation
but also helps us to identify the important pathways and further separate
them using a so-called ``phase cycling'' technique detailed in the
next section.

We find that the general landscape of the 2D spectrum $\tilde{A}(\omega_{\tau},T,\omega_{t})$
also shows strong off-diagonal contributions and cross-peaks (see
Fig. \ref{fig:AwCompare}, for example), similar to the 2D spin-echo
spectrum $A_{o}(\omega_{\tau},\omega_{t})$. However, the dependency
of $\tilde{A}(\omega_{\tau},T,\omega_{t})$ on the mixing time $T$
can give us further information on the many-body coherent and incoherent
dynamics. There is one additional complication, though: we observe
a fast oscillation with frequency $\omega_{s}$ in the $T$-dependency
of $\tilde{A}(\omega_{\tau},T,\omega_{t})$, which originates from
the interferences between the contribution of $I_{3}$ and $I_{4}$,
that is proportional to $e^{-i\omega_{s}T}$ and $e^{i\omega_{s}T}$
respectively. We are not interested in this trivial oscillation. Instead,
we would like to investigate the dynamic in the time scale of $\tau_{F}$
and choose to study the signals at $T_{M}\omega_{s}=2\pi M$, where
$M$ is an integer. Notice that since $\omega_{s}\gg E_{F}$, $T_{M}/\tau_{F}$
can be considered to be almost continuous. As we will see later, this
choice of $T_{M}$ is not necessary if we apply the phase cycling
to separate the pathways.

Figure \ref{fig:Tdep} (a) shows the real and imaginary part of $A(\omega_{\tau}\approx\omega_{A-},T,\omega_{t}\approx\omega_{A-})$
for $k_{F}a=-0.5$ as cross and circle symbols, showing a damping
oscillation behavior at a late time. This long-time behavior can be
fitted perfectly with a formula $F(T)=A_{a}\cos(\omega_{a}T+\varphi_{a})\exp(-T/T_{a})+B$
for the real and imaginary parts separately, both of which give $\omega_{a}\approx0.28E_{F}$
and $T_{a}\approx80\tau_{F}$. We also find this damping oscillation
behavior with the same $\omega_{a}$ and $T_{a}$ at other parts of
the spectrum. One can recognize $\omega_{a}\approx|\omega_{A-}|$
and the damping lifetime $T_{a}$ reflects a non-coherent many-body
dynamic, which might be related to the finite temperature $k_{B}T^{\circ}\approx0.03E_{F}$.
Figure \ref{fig:Tdep} (b) shows the $T$-dependence of ${\rm Re}[\tilde{A}(\omega_{\tau},T,\omega_{t})]$
with $k_{F}a=0.5$ for $AR$ and $RA$ cross-peaks as cross and circle
symbols, which can be fitted by a combination of two damping oscillations
$F(T)=A_{a}\cos(\omega_{a}T+\varphi_{a})\exp(-T/T_{a})+A_{r}\cos(\omega_{r}T+\varphi_{r})\exp(-T/T_{r})+B$
illustrated by the solid curves. The numerical fitting gives $\omega_{a}\approx|\omega_{A+}|$,
$\omega_{r}\approx|\omega_{R+}|$, $T_{a}\approx30\tau_{F}$ and $T_{r}\approx80\tau_{F}$.
These long-time damping oscillations indicate the non-trivial relaxation
process during the mixing time $T$, induced by multiple particle-hole
excitations. However, at a very early time, probably only a few particle-hole
pairs have been excited, and higher order correlation has not been
built up, which explains the deviation between the fitting results
and numerical calculation. It is also interesting to notice that $T_{a}$
and $T_{r}$ are different, which implies there might be an intrinsic
dynamical process between the attractive and repulsive singularities.

\subsection{Phase cycling}

\begin{figure}
\includegraphics[width=0.98\linewidth]{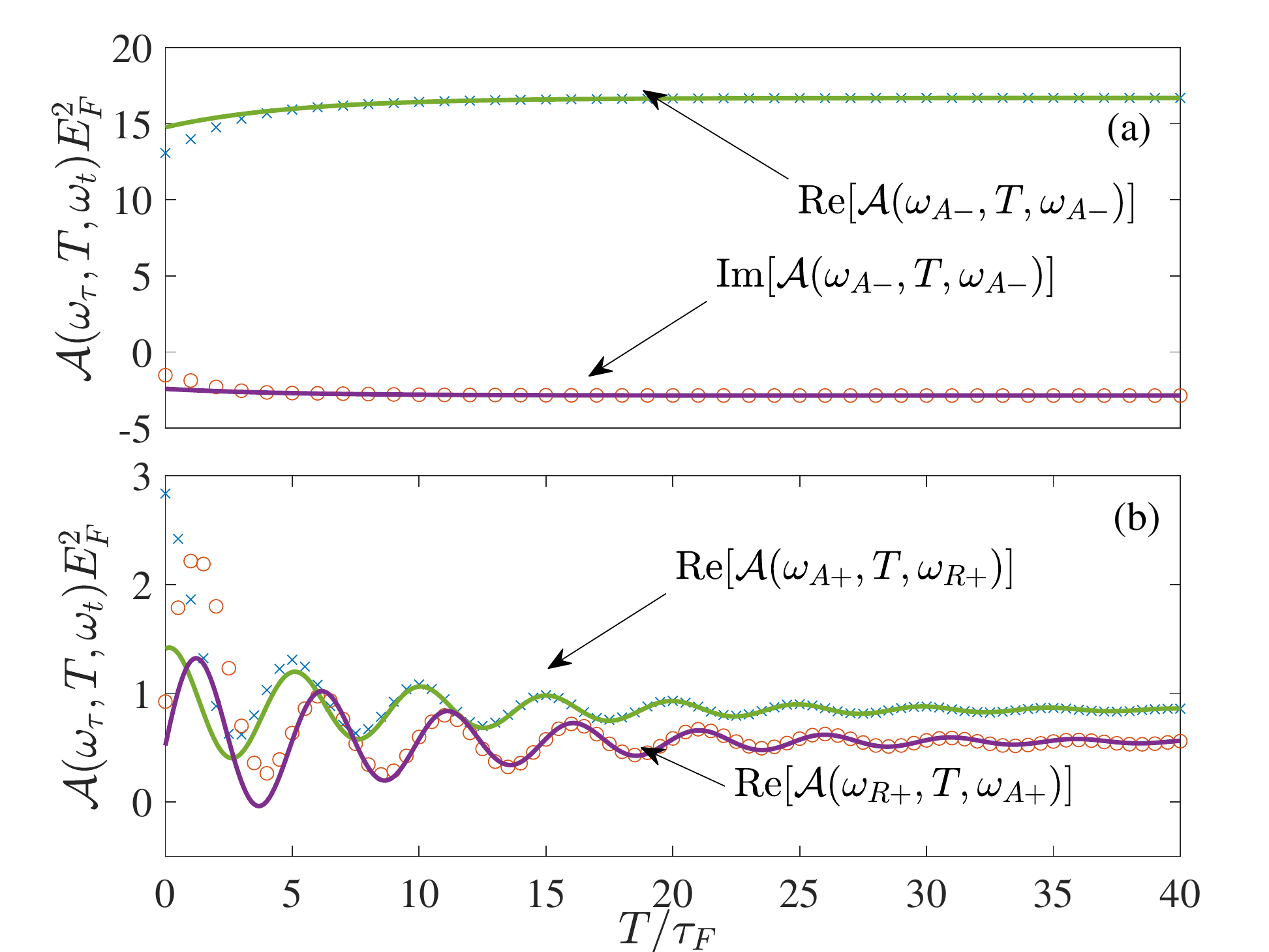}\caption{The same as Fig. \ref{fig:Tdep}, but for a phase cycling spectrum
$\mathcal{A}(\omega_{\tau},T,\omega_{t})$. \label{fig:sTdep}}
\end{figure}

Since the total spectrum is a summation of multiple pathway contributions,
it is sometimes important to be able to separate and measure one or
some of the pathway contributions. For example, we notice that the
third and fourth pathways, $I_{3}(\tau,T,t)=...e^{i\mathcal{H}_{\downarrow}T}...e^{-i\mathcal{H}_{\uparrow}T}...$
and $I_{4}(\tau,T,t)=...e^{i\mathcal{H}_{\uparrow}T}...e^{-i\mathcal{H}_{\downarrow}T}...$,
are the ones responsible for a fast oscillation in the mixing time
$T$ with the trivial frequency $\omega_{s}$. In addition, we have
$I_{3}(\tau,T,t)=I_{o}(\tau+T,t)$ and $I_{4}(\tau,T,t)=I_{o}(\tau,T+t)$,
implying these two pathways give the same information as the 2D spin-echo
sequence. Therefore, it would be interesting to be able to eliminate
the contributions of $I_{3}$ and $I_{4}$, which can be achieved
by following the same spirit as phase cycling, an important technique
in NMR.

Phase cycling is a technique that uses a linear combination of signals
(with possibly different weights) from different pulse schemes to
select the contribution of one or few coherent pathways. As a concrete
example, we define $\mathcal{A}(\omega_{\tau},T,\omega_{t})\equiv\tilde{A}(\omega_{\tau},T,\omega_{t})-\tilde{A}^{-}(\omega_{\tau},T,\omega_{t})$,
where $\tilde{A}^{-}(\omega_{\tau},T,\omega_{t})$ is the 2D Ramsey
response for the EXSY$-$ pulse scheme indicated in Fig. \ref{fig:Sketch}
(g). A manipulation of algebra gives (see Appendix \ref{sec:Appendix_Pathway})

\begin{equation}
\mathcal{A}(\omega_{\tau},T,\omega_{t})=2A_{1}(\omega_{\tau},T,\omega_{t})+2A_{2}(\omega_{\tau},T,\omega_{t}),
\end{equation}
which only includes the first two pathways. One can immediately notice
that both pathways, $I_{1}(\tau,T,t)$ and $I_{2}(\tau,T,t)$, are
independent of $\omega_{s}$, where we are no longer restricted to
measuring signals at $T=T_{M}$.

The corresponding double Fourier transformation $\mathcal{A}(\omega_{\tau},T,\omega_{t})$
is studied in Fig. \ref{fig:sTdep}. For the attractive ($k_{F}a=-0.5$)
interaction case shown in Fig. \ref{fig:sTdep} (a), a numerical fitting
with formula $F(T)=A\exp(-T/T_{a})+B$ indicates a pure exponential
relaxation with lifetime $T_{a}\approx5\tau_{F}$. On the contrary,
for the repulsive ($k_{F}a=+0.5$) interaction case shown in Fig.
\ref{fig:sTdep} (a), a damping oscillation $F(T)=A_{\Delta}\cos(\omega_{\Delta}T_{\Delta}+\varphi_{\Delta})\exp(-T/T_{\Delta})+B_{\Delta}$
of $\omega_{\Delta}\approx|\omega_{R+}-\omega_{A+}|$ and damping
lifetime $T_{\Delta}\approx10\tau_{F}$. This damping oscillation
indicates the intrinsic coherent and incoherent many-body dynamics
between the attractive and repulsive Fermi singularity. We notice
that a similar damping oscillation between exciton-polarons in TMDs
has previously been observed in the non-rephasing signal of an optical
2D spectroscopy \citep{XiaoqinLi2016NL} and explained by the nonlinear
Golden Rule \citep{Tempelaa2019NC}. More recently, this behavior
has also been observed in a FDA study \citep{Reichman2022arXiv}.
To our knowledge, our result is the first prediction of the coherent
and incoherent dynamic process between the two Fermi-edge singularities
in ultracold gases. We also believe the same procedure can be applied
to study the many-body dynamical process between attractive and repulsive
polarons.

\section{Conclusion}

In summary, we have investigated how to extend the 1D Ramsey spectrum
to multidimensional, which goes beyond the linear response regime
and can reveal correlations between many-body singularities and resonances.
Multidimensional spectroscopy also allows us to investigate the many-body
coherent and relaxation dynamics that are not accessible in 1D spectra.
Such a scheme is especially suitable and accessible in the clean and
controllable systems of ultracold gases.

As a concrete example, we investigate the Fermi singularity problem
and present a numerical exact many-body formalism for the simulation
of the multidimensional Ramsey spectrum of a heavy impurity in a Fermi
gas, both in the time domain and frequency domain. We believe this
is the first investigation of the nonlinear responses in such systems
and the first prediction of many-body correlations between attractive
and repulsive singularities, remnants of polaron resonances destroyed
by Anderson's orthogonal catastrophe. For the first time, we also
predict the many-body coherent dynamic and relaxation between the
two Fermi singularities.

We believe these many-body correlations and dynamics should also exist
between attractive and repulsive polarons, which can be calculated
exactly if the background gas is a Bardeen--Cooper--Schrieffer superfluid
\citep{JiaWang2022PRL,JiaWang2022PRA}. Another approach would be
to investigate mobile impurity with a Chevy ansatz. Although this
is an approximated approach, it might lead to intuitive understanding.
Finally, we argue that the application of multidimensional Ramsey
spectroscopy should not be limited to impurity systems, and the same
spirit can be generalized to other ultracold atom systems \citep{Bloch2004PRL_Ramsey,Boris2004PRA,Demler2010PRL_Ramsey,Demler2013PRL_Ramsey,Bloch2013NP,Demler2013PRL_Topo,Zwierlein2003PRL,Ketterle2003Science,Rey2009PRL,Yu2010PRL,Ye2013Science,Ye2013Nature}.

\section{Acknowledgments}

We are grateful to Hui Hu and Xia-Ji Liu for their insightful discussions
and critical reading of the manuscript. This research was supported
by the Australian Research Council's (ARC) Discovery Program, Grants
No. DE180100592 and No. DP190100815.

\appendix

\section{Pathway contributions\label{sec:Appendix_Pathway}}

In the main text, Eq. (\ref{eq:StauTt}) indicates that $S(\tau,T,t)$
can be written as a summation of sixteen different pathway contributions
$S(\tau,T,t)=\sum_{i=1}^{16}S_{i}(\tau,T,t)$ where $S_{i}(\tau,T,t)={\rm Tr}[I_{i}(\tau,T,t)\rho_{{\rm FS}}]/4$.
The sixteen pathways $I_{i}(\tau,T,t)$ can be written out explicitly:

\begin{subequations}
\begin{equation}
I_{1}(\tau,T,t)=e^{i\hat{\mathcal{H}}_{\downarrow}\tau}e^{i\hat{\mathcal{H}}_{\uparrow}T}e^{i\hat{\mathcal{H}}_{\uparrow}t}e^{-i\hat{\mathcal{H}}_{\downarrow}t}e^{-i\hat{\mathcal{H}}_{\uparrow}T}e^{-i\hat{\mathcal{H}}_{\uparrow}\tau},
\end{equation}
\begin{equation}
I_{2}(\tau,T,t)=e^{i\hat{\mathcal{H}}_{\downarrow}\tau}e^{i\hat{\mathcal{H}}_{\downarrow}T}e^{i\hat{\mathcal{H}}_{\uparrow}t}e^{-i\hat{\mathcal{H}}_{\downarrow}t}e^{-i\hat{\mathcal{H}}_{\downarrow}T}e^{-i\hat{\mathcal{H}}_{\uparrow}\tau},
\end{equation}
\begin{equation}
I_{3}(\tau,T,t)=e^{i\hat{\mathcal{H}}_{\downarrow}\tau}e^{i\hat{\mathcal{H}}_{\downarrow}T}e^{i\hat{\mathcal{H}}_{\uparrow}t}e^{-i\hat{\mathcal{H}}_{\downarrow}t}e^{-i\hat{\mathcal{H}}_{\uparrow}T}e^{-i\hat{\mathcal{H}}_{\uparrow}\tau},
\end{equation}
\begin{equation}
I_{4}(\tau,T,t)=e^{i\hat{\mathcal{H}}_{\downarrow}\tau}e^{i\hat{\mathcal{H}}_{\uparrow}T}e^{i\hat{\mathcal{H}}_{\uparrow}t}e^{-i\hat{\mathcal{H}}_{\downarrow}t}e^{-i\hat{\mathcal{H}}_{\downarrow}T}e^{-i\hat{\mathcal{H}}_{\uparrow}\tau}.
\end{equation}
\begin{equation}
I_{5}(\tau,T,t)=e^{i\hat{\mathcal{H}}_{\downarrow}\tau}e^{i\hat{\mathcal{H}}_{\uparrow}T}e^{i\hat{\mathcal{H}}_{\uparrow}t}e^{-i\hat{\mathcal{H}}_{\downarrow}t}e^{-i\hat{\mathcal{H}}_{\uparrow}T}e^{-i\hat{\mathcal{H}}_{\downarrow}\tau},
\end{equation}

\begin{equation}
I_{6}(\tau,T,t)=-e^{i\hat{\mathcal{H}}_{\downarrow}\tau}e^{i\hat{\mathcal{H}}_{\downarrow}T}e^{i\hat{\mathcal{H}}_{\uparrow}t}e^{-i\hat{\mathcal{H}}_{\downarrow}t}e^{-i\hat{\mathcal{H}}_{\downarrow}T}e^{-i\hat{\mathcal{H}}_{\downarrow}\tau},
\end{equation}
\begin{equation}
I_{7}(\tau,T,t)=e^{i\hat{\mathcal{H}}_{\downarrow}\tau}e^{i\hat{\mathcal{H}}_{\downarrow}T}e^{i\hat{\mathcal{H}}_{\uparrow}t}e^{-i\hat{\mathcal{H}}_{\downarrow}t}e^{-i\hat{\mathcal{H}}_{\uparrow}T}e^{-i\hat{\mathcal{H}}_{\downarrow}\tau},
\end{equation}
\begin{equation}
I_{8}(\tau,T,t)=-e^{i\hat{\mathcal{H}}_{\downarrow}\tau}e^{i\hat{\mathcal{H}}_{\uparrow}T}e^{i\hat{\mathcal{H}}_{\uparrow}t}e^{-i\hat{\mathcal{H}}_{\downarrow}t}e^{-i\hat{\mathcal{H}}_{\downarrow}T}e^{-i\hat{\mathcal{H}}_{\downarrow}\tau},
\end{equation}

\begin{equation}
I_{9}(\tau,T,t)=e^{i\hat{\mathcal{H}}_{\uparrow}\tau}e^{i\hat{\mathcal{H}}_{\uparrow}T}e^{i\hat{\mathcal{H}}_{\uparrow}t}e^{-i\hat{\mathcal{H}}_{\downarrow}t}e^{-i\hat{\mathcal{H}}_{\uparrow}T}e^{-i\hat{\mathcal{H}}_{\uparrow}\tau},
\end{equation}

\begin{equation}
I_{10}(\tau,T,t)=-e^{i\hat{\mathcal{H}}_{\uparrow}\tau}e^{i\hat{\mathcal{H}}_{\downarrow}T}e^{i\hat{\mathcal{H}}_{\uparrow}t}e^{-i\hat{\mathcal{H}}_{\downarrow}t}e^{-i\hat{\mathcal{H}}_{\downarrow}T}e^{-i\hat{\mathcal{H}}_{\uparrow}\tau},
\end{equation}
\begin{equation}
I_{11}(\tau,T,t)=-e^{i\hat{\mathcal{H}}_{\uparrow}\tau}e^{i\hat{\mathcal{H}}_{\downarrow}T}e^{i\hat{\mathcal{H}}_{\uparrow}t}e^{-i\hat{\mathcal{H}}_{\downarrow}t}e^{-i\hat{\mathcal{H}}_{\uparrow}T}e^{-i\hat{\mathcal{H}}_{\uparrow}\tau},
\end{equation}
\begin{equation}
I_{12}(\tau,T,t)=e^{i\hat{\mathcal{H}}_{\uparrow}\tau}e^{i\hat{\mathcal{H}}_{\uparrow}T}e^{i\hat{\mathcal{H}}_{\uparrow}t}e^{-i\hat{\mathcal{H}}_{\downarrow}t}e^{-i\hat{\mathcal{H}}_{\downarrow}T}e^{-i\hat{\mathcal{H}}_{\uparrow}\tau},
\end{equation}
\begin{equation}
I_{13}(\tau,T,t)=e^{i\hat{\mathcal{H}}_{\uparrow}\tau}e^{i\hat{\mathcal{H}}_{\uparrow}T}e^{i\hat{\mathcal{H}}_{\uparrow}t}e^{-i\hat{\mathcal{H}}_{\downarrow}t}e^{-i\hat{\mathcal{H}}_{\uparrow}T}e^{-i\hat{\mathcal{H}}_{\downarrow}\tau},
\end{equation}

\begin{equation}
I_{14}(\tau,T,t)=e^{i\hat{\mathcal{H}}_{\downarrow}\tau}e^{i\hat{\mathcal{H}}_{\downarrow}T}e^{i\hat{\mathcal{H}}_{\uparrow}t}e^{-i\hat{\mathcal{H}}_{\downarrow}t}e^{-i\hat{\mathcal{H}}_{\downarrow}T}e^{-i\hat{\mathcal{H}}_{\downarrow}\tau},
\end{equation}
\begin{equation}
I_{15}(\tau,T,t)=-e^{i\hat{\mathcal{H}}_{\uparrow}\tau}e^{i\hat{\mathcal{H}}_{\downarrow}T}e^{i\hat{\mathcal{H}}_{\uparrow}t}e^{-i\hat{\mathcal{H}}_{\downarrow}t}e^{-i\hat{\mathcal{H}}_{\uparrow}T}e^{-i\hat{\mathcal{H}}_{\downarrow}\tau},
\end{equation}

\begin{equation}
I_{16}(\tau,T,t)=-e^{i\hat{\mathcal{H}}_{\uparrow}\tau}e^{i\hat{\mathcal{H}}_{\uparrow}T}e^{i\hat{\mathcal{H}}_{\uparrow}t}e^{-i\hat{\mathcal{H}}_{\downarrow}t}e^{-i\hat{\mathcal{H}}_{\downarrow}T}e^{-i\hat{\mathcal{H}}_{\downarrow}\tau}.
\end{equation}
\end{subequations}

Since the dimensions of the many-body operator $\mathcal{H}_{\downarrow}$
and $\mathcal{H}_{\uparrow}$ grows exponentially with respect to
the number of particle $N$, a direct calculation of $S_{i}(\tau,T,t)$
is not accessible. However, by applying Levitov\textquoteright s formula
in FDA, we can show that $S_{i}(\tau,T,t)$ reduces to a determinant
in a single-particle Hilbert space that grows only linearly to $N$,
allowing an in-principle exact calculation. For this purpose, we rewrite
$\mathcal{H}_{\downarrow}\equiv\Gamma(h_{\downarrow})$ and $\mathcal{H}_{\uparrow}\equiv\Gamma(h_{\uparrow})+\omega_{s}$,
where $\Gamma(h)\equiv\sum_{\mathbf{k},\mathbf{q}}h_{\mathbf{k}\mathbf{q}}c_{\mathbf{k}}^{\dagger}c_{\mathbf{q}}$
is a bilinear fermionic many-body operator, and $h_{\mathbf{k}\mathbf{q}}$
represents the matrix elements corresponding to the single-particle
operator. As a result, each pathway's contribution can be written
as $S_{i}(\tau,T,t)=\tilde{S}_{i}(\tau,T,t)e^{-i\omega_{s}f_{i}(t,T,\tau)}/4,$
where $\tilde{S}_{i}(\tau,T,t)={\rm Tr}[\tilde{I}_{i}(\tau,T,t)\rho_{FS}]$.
Here, $\tilde{I}_{i}(\tau,T,t)$ has the same expression as $I_{i}(\tau,T,t)$
but with $\mathcal{H}_{\downarrow}$ and $\mathcal{H}_{\uparrow}$
replaced by $\Gamma(h_{\downarrow})$ and $\Gamma(h_{\uparrow})$,
respectively. Since both $\Gamma(h_{\downarrow})$ and $\Gamma(h_{\uparrow})$
are fermionic bilinear operators, applying Levitov\textquoteright s
formula gives
\begin{equation}
\tilde{S}_{i}(\tau,T,t)={\rm det}[(1-\hat{n})+R_{i}(\tau,T,t)\hat{n}],
\end{equation}
where $R_{i}(\tau,T,t)$ has the same expression as $I_{i}(\tau,T,t)$
but with $\mathcal{H}_{\downarrow}$ and $\mathcal{H}_{\uparrow}$
replaced by $h_{\downarrow}$ and $h_{\uparrow}$, respectively. The
phase factor $e^{-i\omega_{s}f_{i}(\tau,T,t)}$ can also be obtained
by replacing $e^{\pm\mathcal{H}_{\downarrow}t'}$and $e^{\pm i\mathcal{H}_{\uparrow}t'}$
with $1$ and $e^{\pm i\omega_{s}t'}$ in the expression of $I_{i}(\tau,T,t)$,
where $t'$ can be $\tau$, $T$, or $t$. This expression is now
ready for numerical calculation as mentioned in the main text.

\begin{figure}
\includegraphics[width=0.98\linewidth]{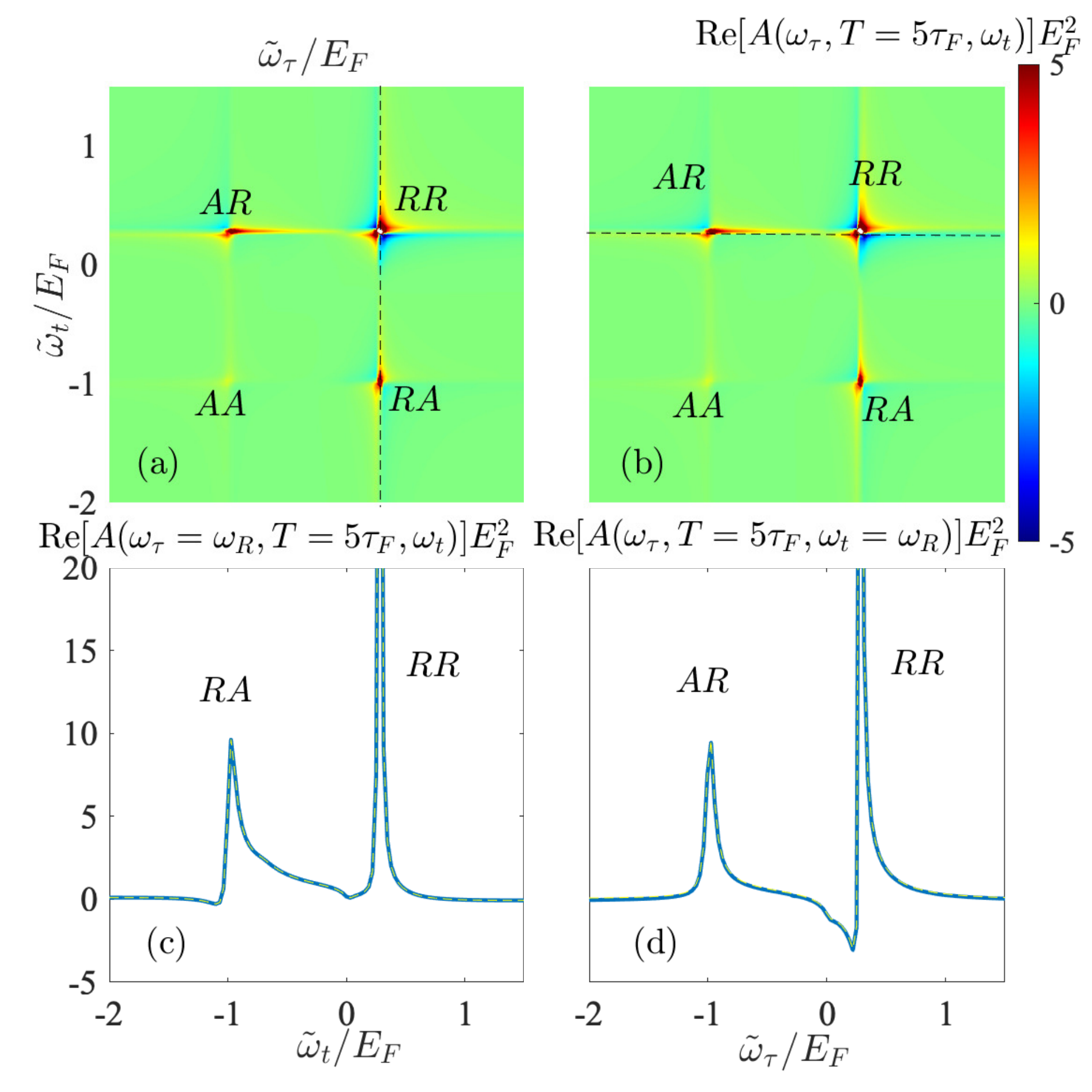}\caption{(a) shows the 2D spectroscopy $A(\omega_{\tau},T,\omega_{t})$ that
includes pathway contributions from all sixteen pathways, and (b)
$\tilde{A}(\omega_{\tau},T,\omega_{t})$ only includes the first four.
Solid and dashed curves in (c) compare $A(\omega_{\tau}=\omega_{R},T,\omega_{t})$
{[}the slice along the dashed line in (a){]} and $\tilde{A}(\omega_{\tau}=\omega_{R},T,\omega_{t})$,
respectively, and show excellent agreement by an essentially overlapping.
(d) shows the same comparison for $A(\omega_{\tau},T,\omega_{t}=\omega_{R})$
and $\tilde{A}(\omega_{\tau},T,\omega_{t}=\omega_{R})$ {[}the slice
along the dashed line in (b){]}. Other parameters are $k_{F}a=0.5$,
$k_{B}T^{\circ}=0.03E_{F}$, $\omega_{s}/E_{F}=2\pi$, and $T=5\tau_{F}$.
\label{fig:AwCompare}}
\end{figure}

\begin{figure}
\includegraphics[width=0.98\linewidth]{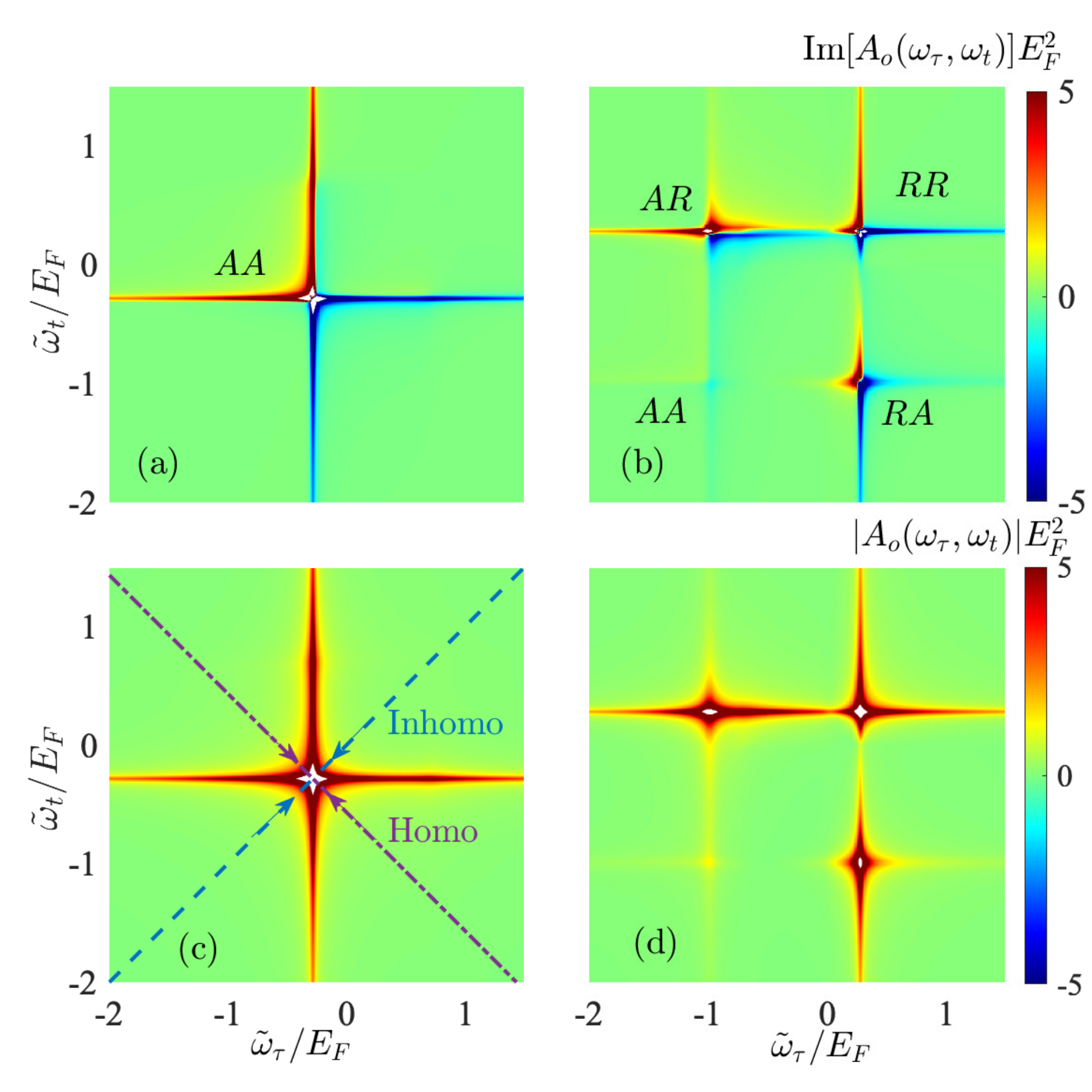}\caption{(a) and (b) shows contour plots of the imaginary part of the 2D spin-echo
spectrum $A_{o}(\omega_{\tau},\omega_{t})$ for attractive $k_{F}a=-0.5$
and repulsive $k_{F}a=0.5$ interactions, respectively. (c) and (d)
show the amplitude for $k_{F}a=-0.5$ and $k_{F}a=0.5$, respectively.
The temperature is set as $k_{B}T^{\circ}=0.03E_{F}$. \label{fig:AwComplex}}
\end{figure}

While $\tilde{S}_{i}(\tau,T,t)$ are universal functions of $k_{B}T^{\circ}/E_{F}$,
$k_{F}a$, $t/\tau_{F}$, and $\tau/\tau_{F}$, the total response
$S(\tau,T,t)$ involves the interference of phase factors $e^{-i\omega_{s}f_{i}(\tau,T,t)}$
between each contribution. The resulting oscillation in frequency
$\omega_{s}$ is not very interesting. Nevertheless, we find that
only a few pathways contribute to the singularities of $A(\omega_{\tau},T,\omega_{t})$
we are interested in. Our numerical results show that $\tilde{S}_{i}(\tau,T,t)$
oscillates at frequency $\sim E_{F}$, much slower than $\omega_{s}\gg E_{F}$
in usual ultracold experiments. Consequently, if we focus on the Fermi
singularity features that appear at $|\omega_{\tau}-\omega_{s}|,|\omega_{t}-\omega_{s}|\sim E_{F}$
in the 2D spectrum $A(\omega_{\tau},T,\omega_{t})$, only the pathways
$I_{1}$, $I_{2}$, $I_{3}$, and $I_{4}$ associated with the phase
factor $e^{i\omega_{s}t}e^{-i\omega_{s}\tau}$ contribute. The phase
$e^{i\omega_{s}t}e^{-i\omega_{s}\tau}$ only gives rise to a simple
frequency shift of $\omega_{t}\rightarrow\tilde{\omega}_{t}$ and
$\omega_{\tau}\rightarrow\tilde{\omega}_{\tau}$. Figure \ref{fig:AwCompare}
compares $A(\omega_{\tau},T,\omega_{t})$ and $\tilde{A}(\omega_{\tau},T,\omega_{t})=\sum_{i=1}^{4}A_{i}(\omega_{\tau},T,\omega_{t})$
that only includes contributions from the first four pathways, for
a set of chosen parameters: $k_{F}a=0.5$, $k_{B}T^{\circ}=0.03E_{F}$,
$\omega_{s}/E_{F}=2\pi$, and $T=5\tau_{F}$. Figure \ref{fig:AwCompare}
(a) and (b) shows the whole 2D spectrum for $A(\omega_{\tau},T,\omega_{t})$
and $\mathcal{A}(\omega_{\tau},T,\omega_{t})$, respectively. The
solid and dashed curves in Fig. \ref{fig:AwCompare} (c) show the
spectrum $A(\omega_{\tau},T,\omega_{t})$ and $\tilde{A}(\omega_{\tau},T,\omega_{t})$
along $\omega_{\tau}=\omega_{R}$, the slice indicated by the dashed
line in Fig. \ref{fig:AwCompare} (a). Fig. \ref{fig:AwCompare} (d)
shows the same comparison for $\omega_{t}=\omega_{R}$, the slice
indicated by the dashed line in Fig. \ref{fig:AwCompare} (b). All
comparison gives perfect agreement, e.g., the solid and dashed curves
essentially overlap in Fig. \ref{fig:AwCompare} (c) and (d).

Among the first four pathways, there is still a phase dependence on
$\omega_{s}T$. To be specific, while $A_{1}(\omega_{\tau},T,\omega_{t})$
and $A_{2}(\omega_{\tau},T,\omega_{t})$ are independent of $\omega_{s}$,
$A_{3}(\omega_{\tau},T,\omega_{t})$ and $A_{4}(\omega_{\tau},T,\omega_{t})$
has a phase dependence of $e^{-i\omega_{s}T}$ and $e^{i\omega_{s}T}$,
respectively. To investigate the dynamics in the time scale of $\tau_{F}$,
we can study the signals at $T_{M}\omega_{s}=2\pi M$, where $M$
is an integer. Notice that since $\omega_{s}\gg E_{F}$, $T_{M}/\tau_{F}$
can be considered to be almost continuous. Such $T$-dependence is
shown in Fig. \ref{fig:Tdep} in the main text.

Another way to observe the $T$-dependence that is independent of
$\omega_{s}$ is by using the so-call phase cycling, which uses a
linear combination of results from different pulse schemes to eliminate
some of the pathway contributions. For example, in the EXSY$-$ pulse
scheme illustrated in Fig. \ref{fig:Sketch} (g), we can carry out
the same calculation as EXSY$+$ with the middle pulse replaced by
a $-\pi/2$ rotation

\begin{equation}
R(-\pi/2)=R(\pi/2)^{T}=\frac{1}{\sqrt{2}}\left(\begin{array}{cc}
1 & -1\\
1 & 1
\end{array}\right).
\end{equation}
We also only need to focus on the first four pathways and can find
that $I_{1}^{-}=-I_{1}$, $I_{2}^{-}=-I_{2}$, $I_{3}^{-}=I_{3}$,
and $I_{4}^{-}=I_{4}$, where the superscript ``$-$'' indicates
the quantities for EXSY$-$ scheme. Consequently, we have $\tilde{A}^{-}=-A_{1}-A_{2}+A_{3}+A_{4}$.
As a result, in the differences between the EXSY$+$ and EXSY$-$,
$\mathcal{A}(\omega_{\tau},T,\omega_{t})\equiv\tilde{A}(\omega_{\tau},T,\omega_{t})-\tilde{A}^{-}(\omega_{\tau},T,\omega_{t})$,
only the contributions from the first two pathways remain, i.e.,
\begin{equation}
\mathcal{A}(\omega_{\tau},T,\omega_{t})=2A_{1}(\omega_{\tau},T,\omega_{t})+2A_{2}(\omega_{\tau},T,\omega_{t})
\end{equation}
which no longer depends on $\omega_{s}$.

\section{Imaginary part and Amplitude of the 2D spin-echo spectrum\label{sec:Appendix_IandA}}

For completeness, we show in Fig. \ref{fig:AwComplex} the imaginary
part and amplitude of the 2D spin-echo spectrum for the same parameters
in Fig. \ref{fig:Aw} in the main text. In particular, we indicate
the homogeneous and inhomogeneous lineshape in Fig. \ref{fig:AwComplex}
(c).

\bibliography{RefMDRamsey}

\begin{thebibliography}{93}%
\makeatletter
\providecommand \@ifxundefined [1]{%
 \@ifx{#1\undefined}
}%
\providecommand \@ifnum [1]{%
 \ifnum #1\expandafter \@firstoftwo
 \else \expandafter \@secondoftwo
 \fi
}%
\providecommand \@ifx [1]{%
 \ifx #1\expandafter \@firstoftwo
 \else \expandafter \@secondoftwo
 \fi
}%
\providecommand \natexlab [1]{#1}%
\providecommand \enquote  [1]{``#1''}%
\providecommand \bibnamefont  [1]{#1}%
\providecommand \bibfnamefont [1]{#1}%
\providecommand \citenamefont [1]{#1}%
\providecommand \href@noop [0]{\@secondoftwo}%
\providecommand \href [0]{\begingroup \@sanitize@url \@href}%
\providecommand \@href[1]{\@@startlink{#1}\@@href}%
\providecommand \@@href[1]{\endgroup#1\@@endlink}%
\providecommand \@sanitize@url [0]{\catcode `\\12\catcode `\$12\catcode
  `\&12\catcode `\#12\catcode `\^12\catcode `\_12\catcode `\%12\relax}%
\providecommand \@@startlink[1]{}%
\providecommand \@@endlink[0]{}%
\providecommand \url  [0]{\begingroup\@sanitize@url \@url }%
\providecommand \@url [1]{\endgroup\@href {#1}{\urlprefix }}%
\providecommand \urlprefix  [0]{URL }%
\providecommand \Eprint [0]{\href }%
\providecommand \doibase [0]{http://dx.doi.org/}%
\providecommand \selectlanguage [0]{\@gobble}%
\providecommand \bibinfo  [0]{\@secondoftwo}%
\providecommand \bibfield  [0]{\@secondoftwo}%
\providecommand \translation [1]{[#1]}%
\providecommand \BibitemOpen [0]{}%
\providecommand \bibitemStop [0]{}%
\providecommand \bibitemNoStop [0]{.\EOS\space}%
\providecommand \EOS [0]{\spacefactor3000\relax}%
\providecommand \BibitemShut  [1]{\csname bibitem#1\endcsname}%
\let\auto@bib@innerbib\@empty
\bibitem [{\citenamefont {W.~P.~Aue}\ and\ \citenamefont
  {Ernst}(1976)}]{ErnstJCP1976}%
  \BibitemOpen
  \bibfield  {author} {\bibinfo {author} {\bibfnamefont {E.~Bartholdi}\
  \bibnamefont {W.~P.~Aue}}\ and\ \bibinfo {author} {\bibfnamefont {R.~R.}\
  \bibnamefont {Ernst}},\ }\bibfield  {title} {\enquote {\bibinfo {title}
  {Two-dimensional spectroscopy. application to nuclear magnetic resonance},}\
  }\href@noop {} {\bibfield  {journal} {\bibinfo  {journal} {J. Chem. Phys.}\
  }\textbf {\bibinfo {volume} {64}},\ \bibinfo {pages} {2229} (\bibinfo {year}
  {1976})}\BibitemShut {NoStop}%
\bibitem [{\citenamefont {J.~Jeener}\ and\ \citenamefont
  {Ernst}(1979)}]{ErnstJCP1979}%
  \BibitemOpen
  \bibfield  {author} {\bibinfo {author} {\bibfnamefont {P.~Bachmann}\
  \bibnamefont {J.~Jeener}, \bibfnamefont {B.~H.~Meier}}\ and\ \bibinfo
  {author} {\bibfnamefont {R.~R.}\ \bibnamefont {Ernst}},\ }\bibfield  {title}
  {\enquote {\bibinfo {title} {Investigation of exchange processes by
  two-dimensional nmr spectroscopy},}\ }\href@noop {} {\bibfield  {journal}
  {\bibinfo  {journal} {J. Chem. Phys.}\ }\textbf {\bibinfo {volume} {71}},\
  \bibinfo {pages} {4546} (\bibinfo {year} {1979})}\BibitemShut {NoStop}%
\bibitem [{\citenamefont {Ernst}\ \emph {et~al.}(1987)\citenamefont {Ernst},
  \citenamefont {Bodenhausen},\ and\ \citenamefont {Wokaun}}]{Ernst1987Book}%
  \BibitemOpen
  \bibfield  {author} {\bibinfo {author} {\bibfnamefont {R.~R.}\ \bibnamefont
  {Ernst}}, \bibinfo {author} {\bibfnamefont {G.}~\bibnamefont {Bodenhausen}},
  \ and\ \bibinfo {author} {\bibfnamefont {A.}~\bibnamefont {Wokaun}},\
  }\href@noop {} {\emph {\bibinfo {title} {Principles of Nuclear Magnetic
  Resonance in One and Two Dimensions}}}\ (\bibinfo  {publisher} {Oxford U.
  Press},\ \bibinfo {address} {New York},\ \bibinfo {year} {1987})\BibitemShut
  {NoStop}%
\bibitem [{\citenamefont {Keeler}(2011)}]{Keeler2011Book}%
  \BibitemOpen
  \bibfield  {author} {\bibinfo {author} {\bibfnamefont {J.}~\bibnamefont
  {Keeler}},\ }\href@noop {} {\emph {\bibinfo {title} {Understanding NMR
  Spectroscopy}}},\ \bibinfo {edition} {2nd}\ ed.\ (\bibinfo  {publisher}
  {Wiley},\ \bibinfo {address} {United Kingdom},\ \bibinfo {year}
  {2011})\BibitemShut {NoStop}%
\bibitem [{\citenamefont {Tanimura}\ and\ \citenamefont
  {Mukamel}(1993)}]{Mukamel1993JCP}%
  \BibitemOpen
  \bibfield  {author} {\bibinfo {author} {\bibfnamefont {Yoshitaka}\
  \bibnamefont {Tanimura}}\ and\ \bibinfo {author} {\bibfnamefont {Shaul}\
  \bibnamefont {Mukamel}},\ }\bibfield  {title} {\enquote {\bibinfo {title}
  {Two-dimensional femtosecond vibrational spectroscopy of liquids},}\
  }\href@noop {} {\bibfield  {journal} {\bibinfo  {journal} {J. Chem. Phys.}\
  }\textbf {\bibinfo {volume} {99}},\ \bibinfo {pages} {9496} (\bibinfo {year}
  {1993})}\BibitemShut {NoStop}%
\bibitem [{\citenamefont {Nardin}\ \emph {et~al.}(2015)\citenamefont {Nardin},
  \citenamefont {Autry}, \citenamefont {Moody}, \citenamefont {Singh},
  \citenamefont {Li},\ and\ \citenamefont {Cundiff}}]{Cundiff2015JApplP}%
  \BibitemOpen
  \bibfield  {author} {\bibinfo {author} {\bibfnamefont {Ga{\"e}l}\
  \bibnamefont {Nardin}}, \bibinfo {author} {\bibfnamefont {Travis~M.}\
  \bibnamefont {Autry}}, \bibinfo {author} {\bibfnamefont {Galan}\ \bibnamefont
  {Moody}}, \bibinfo {author} {\bibfnamefont {Rohan}\ \bibnamefont {Singh}},
  \bibinfo {author} {\bibfnamefont {Hebin}\ \bibnamefont {Li}}, \ and\ \bibinfo
  {author} {\bibfnamefont {Steven~T.}\ \bibnamefont {Cundiff}},\ }\bibfield
  {title} {\enquote {\bibinfo {title} {Multi-dimensional coherent optical
  spectroscopy of semiconductor nanostructures: Collinear and non-collinear
  approaches},}\ }\href@noop {} {\bibfield  {journal} {\bibinfo  {journal} {J.
  Appl. Phys}\ }\textbf {\bibinfo {volume} {177}},\ \bibinfo {pages} {112804}
  (\bibinfo {year} {2015})}\BibitemShut {NoStop}%
\bibitem [{\citenamefont {Zanni}\ and\ \citenamefont
  {Hochstrasse}(2001)}]{Zanni2001COSB}%
  \BibitemOpen
  \bibfield  {author} {\bibinfo {author} {\bibfnamefont {Martin~T}\
  \bibnamefont {Zanni}}\ and\ \bibinfo {author} {\bibfnamefont {Robin~M}\
  \bibnamefont {Hochstrasse}},\ }\bibfield  {title} {\enquote {\bibinfo {title}
  {Two-dimensional infrared spectroscopy: a promising new method for the time
  resolution of structures},}\ }\href@noop {} {\bibfield  {journal} {\bibinfo
  {journal} {Curr. Opin. Struct. Biol.}\ }\textbf {\bibinfo {volume} {11}},\
  \bibinfo {pages} {516} (\bibinfo {year} {2001})}\BibitemShut {NoStop}%
\bibitem [{\citenamefont {Jonas}(2003)}]{Jonas2003AnnRevPhysChem}%
  \BibitemOpen
  \bibfield  {author} {\bibinfo {author} {\bibfnamefont {D.}~\bibnamefont
  {Jonas}},\ }\bibfield  {title} {\enquote {\bibinfo {title} {Two-dimensional
  fermtosecond spectroscopy},}\ }\href@noop {} {\bibfield  {journal} {\bibinfo
  {journal} {Ann. Rev. Phys. Chem.}\ }\textbf {\bibinfo {volume} {54}},\
  \bibinfo {pages} {425} (\bibinfo {year} {2003})}\BibitemShut {NoStop}%
\bibitem [{\citenamefont {hung Tseng}\ \emph {et~al.}(2009)\citenamefont {hung
  Tseng}, \citenamefont {Matsika},\ and\ \citenamefont
  {Weinacht}}]{Weinacht2009OE}%
  \BibitemOpen
  \bibfield  {author} {\bibinfo {author} {\bibfnamefont {Chien}\ \bibnamefont
  {hung Tseng}}, \bibinfo {author} {\bibfnamefont {Spiridoula}\ \bibnamefont
  {Matsika}}, \ and\ \bibinfo {author} {\bibfnamefont {Thomas~C}\ \bibnamefont
  {Weinacht}},\ }\bibfield  {title} {\enquote {\bibinfo {title}
  {Two-dimensional ultrafast fourier transform spectroscopy in the deep
  ultraviolet},}\ }\href@noop {} {\bibfield  {journal} {\bibinfo  {journal}
  {Opt. Express}\ }\textbf {\bibinfo {volume} {17}},\ \bibinfo {pages} {18788}
  (\bibinfo {year} {2009})}\BibitemShut {NoStop}%
\bibitem [{\citenamefont {West}\ and\ \citenamefont
  {Moran}(2012)}]{Moran2012JPCL}%
  \BibitemOpen
  \bibfield  {author} {\bibinfo {author} {\bibfnamefont {Brantley~A.}\
  \bibnamefont {West}}\ and\ \bibinfo {author} {\bibfnamefont {Andrew~M.}\
  \bibnamefont {Moran}},\ }\bibfield  {title} {\enquote {\bibinfo {title}
  {Two-dimensional electronic spectroscopy in the ultraviolet wavelength
  range},}\ }\href@noop {} {\bibfield  {journal} {\bibinfo  {journal} {J. Phys.
  Chem. Lett.}\ }\textbf {\bibinfo {volume} {3}},\ \bibinfo {pages} {2575}
  (\bibinfo {year} {2012})}\BibitemShut {NoStop}%
\bibitem [{\citenamefont {Dey}\ \emph {et~al.}(2015)\citenamefont {Dey},
  \citenamefont {Paul}, \citenamefont {Moody}, \citenamefont {Stevens},
  \citenamefont {Glikin}, \citenamefont {Kovalyuk}, \citenamefont {Kudrynskyi},
  \citenamefont {Romero}, \citenamefont {Cantarero}, \citenamefont {Hilton},\
  and\ \citenamefont {Karaiskaj}}]{Karaiskaj2015JCP}%
  \BibitemOpen
  \bibfield  {author} {\bibinfo {author} {\bibfnamefont {P.}~\bibnamefont
  {Dey}}, \bibinfo {author} {\bibfnamefont {J.}~\bibnamefont {Paul}}, \bibinfo
  {author} {\bibfnamefont {G.}~\bibnamefont {Moody}}, \bibinfo {author}
  {\bibfnamefont {C.~E.}\ \bibnamefont {Stevens}}, \bibinfo {author}
  {\bibfnamefont {N.}~\bibnamefont {Glikin}}, \bibinfo {author} {\bibfnamefont
  {Z.~D.}\ \bibnamefont {Kovalyuk}}, \bibinfo {author} {\bibfnamefont {Z.~R.}\
  \bibnamefont {Kudrynskyi}}, \bibinfo {author} {\bibfnamefont {A.~H.}\
  \bibnamefont {Romero}}, \bibinfo {author} {\bibfnamefont {A.}~\bibnamefont
  {Cantarero}}, \bibinfo {author} {\bibfnamefont {D.~J.}\ \bibnamefont
  {Hilton}}, \ and\ \bibinfo {author} {\bibfnamefont {D.}~\bibnamefont
  {Karaiskaj}},\ }\bibfield  {title} {\enquote {\bibinfo {title} {Biexciton
  formation and exciton coherent coupling in layered gase},}\ }\href@noop {}
  {\bibfield  {journal} {\bibinfo  {journal} {J. Chem. Phys.}\ }\textbf
  {\bibinfo {volume} {142}},\ \bibinfo {pages} {212422} (\bibinfo {year}
  {2015})}\BibitemShut {NoStop}%
\bibitem [{\citenamefont {Moody}\ \emph {et~al.}(2015)\citenamefont {Moody},
  \citenamefont {Dass}, \citenamefont {Hao}, \citenamefont {Chen},
  \citenamefont {Li}, \citenamefont {Singh}, \citenamefont {Tran},
  \citenamefont {Clark}, \citenamefont {Xu}, \citenamefont {Bergh{ \"a}user},
  \citenamefont {Malic}, \citenamefont {Knorr},\ and\ \citenamefont
  {Li}}]{XiaoqinLi2015NC}%
  \BibitemOpen
  \bibfield  {author} {\bibinfo {author} {\bibfnamefont {Galan}\ \bibnamefont
  {Moody}}, \bibinfo {author} {\bibfnamefont {Chandriker~Kavir}\ \bibnamefont
  {Dass}}, \bibinfo {author} {\bibfnamefont {Kai}\ \bibnamefont {Hao}},
  \bibinfo {author} {\bibfnamefont {Chang-Hsiao}\ \bibnamefont {Chen}},
  \bibinfo {author} {\bibfnamefont {Lain-Jong}\ \bibnamefont {Li}}, \bibinfo
  {author} {\bibfnamefont {Akshay}\ \bibnamefont {Singh}}, \bibinfo {author}
  {\bibfnamefont {Kha}\ \bibnamefont {Tran}}, \bibinfo {author} {\bibfnamefont
  {Genevieve}\ \bibnamefont {Clark}}, \bibinfo {author} {\bibfnamefont
  {Xiaodong}\ \bibnamefont {Xu}}, \bibinfo {author} {\bibfnamefont {Gunnar}\
  \bibnamefont {Bergh{ \"a}user}}, \bibinfo {author} {\bibfnamefont {Ermin}\
  \bibnamefont {Malic}}, \bibinfo {author} {\bibfnamefont {Andreas}\
  \bibnamefont {Knorr}}, \ and\ \bibinfo {author} {\bibfnamefont {Xiaoqin}\
  \bibnamefont {Li}},\ }\bibfield  {title} {\enquote {\bibinfo {title}
  {Intrinsic homogeneous linewidth and broadening mechanisms of excitons in
  monolayer transition metal dichalcogenides},}\ }\href@noop {} {\bibfield
  {journal} {\bibinfo  {journal} {Nat. Commun.}\ }\textbf {\bibinfo {volume}
  {6}},\ \bibinfo {pages} {8315} (\bibinfo {year} {2015})}\BibitemShut
  {NoStop}%
\bibitem [{\citenamefont {Dey}\ \emph {et~al.}(2016)\citenamefont {Dey},
  \citenamefont {Paul}, \citenamefont {Wang}, \citenamefont {Stevens},
  \citenamefont {Liu}, \citenamefont {Romero}, \citenamefont {Shan},
  \citenamefont {Hilton},\ and\ \citenamefont {Karaiskaj}}]{Karaiskaj2016PRL}%
  \BibitemOpen
  \bibfield  {author} {\bibinfo {author} {\bibfnamefont {P.}~\bibnamefont
  {Dey}}, \bibinfo {author} {\bibfnamefont {J.}~\bibnamefont {Paul}}, \bibinfo
  {author} {\bibfnamefont {Z.}~\bibnamefont {Wang}}, \bibinfo {author}
  {\bibfnamefont {C.~E.}\ \bibnamefont {Stevens}}, \bibinfo {author}
  {\bibfnamefont {C.}~\bibnamefont {Liu}}, \bibinfo {author} {\bibfnamefont
  {A.~H.}\ \bibnamefont {Romero}}, \bibinfo {author} {\bibfnamefont
  {J.}~\bibnamefont {Shan}}, \bibinfo {author} {\bibfnamefont {D.~J.}\
  \bibnamefont {Hilton}}, \ and\ \bibinfo {author} {\bibfnamefont
  {D.}~\bibnamefont {Karaiskaj}},\ }\bibfield  {title} {\enquote {\bibinfo
  {title} {Optical coherence in atomic-monolayer transition-metal
  dichalcogenides limited by electron-phonon interactions},}\ }\href@noop {}
  {\bibfield  {journal} {\bibinfo  {journal} {Phys. Rev. Lett.}\ }\textbf
  {\bibinfo {volume} {116}},\ \bibinfo {pages} {127402} (\bibinfo {year}
  {2016})}\BibitemShut {NoStop}%
\bibitem [{\citenamefont {Hao}\ \emph {et~al.}(2016{\natexlab{a}})\citenamefont
  {Hao}, \citenamefont {Moody}, \citenamefont {Wu}, \citenamefont {Dass},
  \citenamefont {Xu}, \citenamefont {Chen}, \citenamefont {Sun}, \citenamefont
  {Li}, \citenamefont {Li}, \citenamefont {MacDonald},\ and\ \citenamefont
  {Li}}]{XiaoqinLi2016NP}%
  \BibitemOpen
  \bibfield  {author} {\bibinfo {author} {\bibfnamefont {Kai}\ \bibnamefont
  {Hao}}, \bibinfo {author} {\bibfnamefont {Galan}\ \bibnamefont {Moody}},
  \bibinfo {author} {\bibfnamefont {Fengcheng}\ \bibnamefont {Wu}}, \bibinfo
  {author} {\bibfnamefont {Chandriker~Kavir}\ \bibnamefont {Dass}}, \bibinfo
  {author} {\bibfnamefont {Lixiang}\ \bibnamefont {Xu}}, \bibinfo {author}
  {\bibfnamefont {Chang-Hsiao}\ \bibnamefont {Chen}}, \bibinfo {author}
  {\bibfnamefont {Liuyang}\ \bibnamefont {Sun}}, \bibinfo {author}
  {\bibfnamefont {Ming-Yang}\ \bibnamefont {Li}}, \bibinfo {author}
  {\bibfnamefont {Lain-Jong}\ \bibnamefont {Li}}, \bibinfo {author}
  {\bibfnamefont {Allan~H.}\ \bibnamefont {MacDonald}}, \ and\ \bibinfo
  {author} {\bibfnamefont {Xiaoqin}\ \bibnamefont {Li}},\ }\bibfield  {title}
  {\enquote {\bibinfo {title} {Direct measurement of exciton valley coherence
  in monolayer wse$_2$},}\ }\href@noop {} {\bibfield  {journal} {\bibinfo
  {journal} {Nat. Phys.}\ }\textbf {\bibinfo {volume} {12}},\ \bibinfo {pages}
  {677} (\bibinfo {year} {2016}{\natexlab{a}})}\BibitemShut {NoStop}%
\bibitem [{\citenamefont {Hao}\ \emph {et~al.}(2017)\citenamefont {Hao},
  \citenamefont {Specht}, \citenamefont {Nagler}, \citenamefont {Xu},
  \citenamefont {Tran}, \citenamefont {Singh}, \citenamefont {Dass},
  \citenamefont {Sch{\"u}ller}, \citenamefont {Korn}, \citenamefont {Richter},
  \citenamefont {Knorr}, \citenamefont {Li},\ and\ \citenamefont
  {Moody}}]{XiaoqinLi2017NC}%
  \BibitemOpen
  \bibfield  {author} {\bibinfo {author} {\bibfnamefont {Kai}\ \bibnamefont
  {Hao}}, \bibinfo {author} {\bibfnamefont {Judith~F.}\ \bibnamefont {Specht}},
  \bibinfo {author} {\bibfnamefont {Philipp}\ \bibnamefont {Nagler}}, \bibinfo
  {author} {\bibfnamefont {Lixiang}\ \bibnamefont {Xu}}, \bibinfo {author}
  {\bibfnamefont {Kha}\ \bibnamefont {Tran}}, \bibinfo {author} {\bibfnamefont
  {Akshay}\ \bibnamefont {Singh}}, \bibinfo {author} {\bibfnamefont
  {Chandriker~Kavir}\ \bibnamefont {Dass}}, \bibinfo {author} {\bibfnamefont
  {Christian}\ \bibnamefont {Sch{\"u}ller}}, \bibinfo {author} {\bibfnamefont
  {Tobias}\ \bibnamefont {Korn}}, \bibinfo {author} {\bibfnamefont {Marten}\
  \bibnamefont {Richter}}, \bibinfo {author} {\bibfnamefont {Andreas}\
  \bibnamefont {Knorr}}, \bibinfo {author} {\bibfnamefont {Xiaoqin}\
  \bibnamefont {Li}}, \ and\ \bibinfo {author} {\bibfnamefont {Galan}\
  \bibnamefont {Moody}},\ }\bibfield  {title} {\enquote {\bibinfo {title}
  {Neutral and charged inter-valley biexcitons in monolayer mose{$_2$}},}\
  }\href@noop {} {\bibfield  {journal} {\bibinfo  {journal} {Nat. Commun.}\
  }\textbf {\bibinfo {volume} {8}},\ \bibinfo {pages} {15552} (\bibinfo {year}
  {2017})}\BibitemShut {NoStop}%
\bibitem [{\citenamefont {Hao}\ \emph {et~al.}(2016{\natexlab{b}})\citenamefont
  {Hao}, \citenamefont {Xu}, \citenamefont {Nagler}, \citenamefont {Singh},
  \citenamefont {Tran}, \citenamefont {Dass}, \citenamefont {Sch{\"u}ller},
  \citenamefont {Korn}, \citenamefont {Li},\ and\ \citenamefont
  {Moody}}]{XiaoqinLi2016NL}%
  \BibitemOpen
  \bibfield  {author} {\bibinfo {author} {\bibfnamefont {Kai}\ \bibnamefont
  {Hao}}, \bibinfo {author} {\bibfnamefont {Lixiang}\ \bibnamefont {Xu}},
  \bibinfo {author} {\bibfnamefont {Philipp}\ \bibnamefont {Nagler}}, \bibinfo
  {author} {\bibfnamefont {Akshay}\ \bibnamefont {Singh}}, \bibinfo {author}
  {\bibfnamefont {Kha}\ \bibnamefont {Tran}}, \bibinfo {author} {\bibfnamefont
  {Chandriker~Kavir}\ \bibnamefont {Dass}}, \bibinfo {author} {\bibfnamefont
  {Christian}\ \bibnamefont {Sch{\"u}ller}}, \bibinfo {author} {\bibfnamefont
  {Tobias}\ \bibnamefont {Korn}}, \bibinfo {author} {\bibfnamefont {Xiaoqin}\
  \bibnamefont {Li}}, \ and\ \bibinfo {author} {\bibfnamefont {Galan}\
  \bibnamefont {Moody}},\ }\bibfield  {title} {\enquote {\bibinfo {title}
  {Coherent and incoherent coupling dynamics between neutral and charged
  excitons in monolayer mose{$_2$}},}\ }\href@noop {} {\bibfield  {journal}
  {\bibinfo  {journal} {Nano Lett.}\ }\textbf {\bibinfo {volume} {16}},\
  \bibinfo {pages} {5109} (\bibinfo {year} {2016}{\natexlab{b}})}\BibitemShut
  {NoStop}%
\bibitem [{\citenamefont {Tempelaar}\ and\ \citenamefont
  {Berkelbach}(2019)}]{Tempelaa2019NC}%
  \BibitemOpen
  \bibfield  {author} {\bibinfo {author} {\bibfnamefont {Roel}\ \bibnamefont
  {Tempelaar}}\ and\ \bibinfo {author} {\bibfnamefont {Timothy~C.}\
  \bibnamefont {Berkelbach}},\ }\bibfield  {title} {\enquote {\bibinfo {title}
  {Many-body simulation of two-dimensional electronic spectroscopy of excitons
  and trions in monolayer transition metal dichalcogenides},}\ }\href@noop {}
  {\bibfield  {journal} {\bibinfo  {journal} {Nat. Commun.}\ }\textbf {\bibinfo
  {volume} {10}},\ \bibinfo {pages} {3419} (\bibinfo {year}
  {2019})}\BibitemShut {NoStop}%
\bibitem [{\citenamefont {Sidler}\ \emph {et~al.}(2016)\citenamefont {Sidler},
  \citenamefont {Back}, \citenamefont {Cotlet}, \citenamefont {Srivastava},
  \citenamefont {Fink}, \citenamefont {Kroner}, \citenamefont {Demler},\ and\
  \citenamefont {Imamoglu}}]{Imamoglu2016NP}%
  \BibitemOpen
  \bibfield  {author} {\bibinfo {author} {\bibfnamefont {Meinrad}\ \bibnamefont
  {Sidler}}, \bibinfo {author} {\bibfnamefont {Patrick}\ \bibnamefont {Back}},
  \bibinfo {author} {\bibfnamefont {Ovidiu}\ \bibnamefont {Cotlet}}, \bibinfo
  {author} {\bibfnamefont {Ajit}\ \bibnamefont {Srivastava}}, \bibinfo {author}
  {\bibfnamefont {Thomas}\ \bibnamefont {Fink}}, \bibinfo {author}
  {\bibfnamefont {Martin}\ \bibnamefont {Kroner}}, \bibinfo {author}
  {\bibfnamefont {Eugene}\ \bibnamefont {Demler}}, \ and\ \bibinfo {author}
  {\bibfnamefont {Atac}\ \bibnamefont {Imamoglu}},\ }\bibfield  {title}
  {\enquote {\bibinfo {title} {Fermi polaron-polaritons in charge-tunable
  atomically thin semiconductors},}\ }\href@noop {} {\bibfield  {journal}
  {\bibinfo  {journal} {Nat. Phys.}\ }\textbf {\bibinfo {volume} {10}},\
  \bibinfo {pages} {255} (\bibinfo {year} {2016})}\BibitemShut {NoStop}%
\bibitem [{\citenamefont {Tan}\ \emph {et~al.}(2020)\citenamefont {Tan},
  \citenamefont {Cotlet}, \citenamefont {Bergschneider}, \citenamefont
  {Schmidt}, \citenamefont {Back}, \citenamefont {Shimazaki}, \citenamefont
  {Kroner},\ and\ \citenamefont {\ifmmode \dot{I}\else
  \.{I}\fi{}mamo\ifmmode~\breve{g}\else \u{g}\fi{}lu}}]{Imamoglu2020PRX}%
  \BibitemOpen
  \bibfield  {author} {\bibinfo {author} {\bibfnamefont {Li~Bing}\ \bibnamefont
  {Tan}}, \bibinfo {author} {\bibfnamefont {Ovidiu}\ \bibnamefont {Cotlet}},
  \bibinfo {author} {\bibfnamefont {Andrea}\ \bibnamefont {Bergschneider}},
  \bibinfo {author} {\bibfnamefont {Richard}\ \bibnamefont {Schmidt}}, \bibinfo
  {author} {\bibfnamefont {Patrick}\ \bibnamefont {Back}}, \bibinfo {author}
  {\bibfnamefont {Yuya}\ \bibnamefont {Shimazaki}}, \bibinfo {author}
  {\bibfnamefont {Martin}\ \bibnamefont {Kroner}}, \ and\ \bibinfo {author}
  {\bibfnamefont {Ata\ifmmode\mbox{\c{c}}\else\c{c}\fi{}}\ \bibnamefont
  {\ifmmode \dot{I}\else \.{I}\fi{}mamo\ifmmode~\breve{g}\else \u{g}\fi{}lu}},\
  }\bibfield  {title} {\enquote {\bibinfo {title} {Interacting
  polaron-polaritons},}\ }\href@noop {} {\bibfield  {journal} {\bibinfo
  {journal} {Phys. Rev. X}\ }\textbf {\bibinfo {volume} {10}},\ \bibinfo
  {pages} {021011} (\bibinfo {year} {2020})}\BibitemShut {NoStop}%
\bibitem [{\citenamefont {Efimkin}\ and\ \citenamefont
  {MacDonald}(2017)}]{Dmitry2017PRB}%
  \BibitemOpen
  \bibfield  {author} {\bibinfo {author} {\bibfnamefont {Dmitry~K.}\
  \bibnamefont {Efimkin}}\ and\ \bibinfo {author} {\bibfnamefont {Allan~H.}\
  \bibnamefont {MacDonald}},\ }\bibfield  {title} {\enquote {\bibinfo {title}
  {Many-body theory of trion absorption features in two-dimensional
  semiconductors},}\ }\href@noop {} {\bibfield  {journal} {\bibinfo  {journal}
  {Phys. Rev. B}\ }\textbf {\bibinfo {volume} {95}},\ \bibinfo {pages} {035417}
  (\bibinfo {year} {2017})}\BibitemShut {NoStop}%
\bibitem [{\citenamefont {Efimkin}\ and\ \citenamefont
  {MacDonald}(2018)}]{Dmitry2018PRB}%
  \BibitemOpen
  \bibfield  {author} {\bibinfo {author} {\bibfnamefont {Dmitry~K.}\
  \bibnamefont {Efimkin}}\ and\ \bibinfo {author} {\bibfnamefont {Allan~H.}\
  \bibnamefont {MacDonald}},\ }\bibfield  {title} {\enquote {\bibinfo {title}
  {Exciton-polarons in doped semiconductors in a strong magnetic field},}\
  }\href@noop {} {\bibfield  {journal} {\bibinfo  {journal} {Phys. Rev. B}\
  }\textbf {\bibinfo {volume} {97}},\ \bibinfo {pages} {235432} (\bibinfo
  {year} {2018})}\BibitemShut {NoStop}%
\bibitem [{\citenamefont {Efimkin}\ \emph {et~al.}(2021)\citenamefont
  {Efimkin}, \citenamefont {Laird}, \citenamefont {Levinsen}, \citenamefont
  {Parish},\ and\ \citenamefont {MacDonald}}]{Dmitry2021PRB}%
  \BibitemOpen
  \bibfield  {author} {\bibinfo {author} {\bibfnamefont {Dmitry~K.}\
  \bibnamefont {Efimkin}}, \bibinfo {author} {\bibfnamefont {Emma~K.}\
  \bibnamefont {Laird}}, \bibinfo {author} {\bibfnamefont {Jesper}\
  \bibnamefont {Levinsen}}, \bibinfo {author} {\bibfnamefont {Meera~M.}\
  \bibnamefont {Parish}}, \ and\ \bibinfo {author} {\bibfnamefont {Allan~H.}\
  \bibnamefont {MacDonald}},\ }\bibfield  {title} {\enquote {\bibinfo {title}
  {Electron-exciton interactions in the exciton-polaron problem},}\ }\href@noop
  {} {\bibfield  {journal} {\bibinfo  {journal} {Phys. Rev. B}\ }\textbf
  {\bibinfo {volume} {103}},\ \bibinfo {pages} {075417} (\bibinfo {year}
  {2021})}\BibitemShut {NoStop}%
\bibitem [{\citenamefont {Muir}\ \emph {et~al.}()\citenamefont {Muir},
  \citenamefont {Levinsen}, \citenamefont {Earl}, \citenamefont {Conway},
  \citenamefont {Cole}, \citenamefont {Wurdack}, \citenamefont {Mishra},
  \citenamefont {Ing}, \citenamefont {Estrecho}, \citenamefont {Lu},
  \citenamefont {Efimkin}, \citenamefont {Tollerud}, \citenamefont
  {Ostrovskaya}, \citenamefont {Parish},\ and\ \citenamefont
  {Davis}}]{Muir2022arXiv}%
  \BibitemOpen
  \bibfield  {author} {\bibinfo {author} {\bibfnamefont {J.~B.}\ \bibnamefont
  {Muir}}, \bibinfo {author} {\bibfnamefont {J.}~\bibnamefont {Levinsen}},
  \bibinfo {author} {\bibfnamefont {S.~K.}\ \bibnamefont {Earl}}, \bibinfo
  {author} {\bibfnamefont {M.~A.}\ \bibnamefont {Conway}}, \bibinfo {author}
  {\bibfnamefont {J.~H.}\ \bibnamefont {Cole}}, \bibinfo {author}
  {\bibfnamefont {M.}~\bibnamefont {Wurdack}}, \bibinfo {author} {\bibfnamefont
  {R.}~\bibnamefont {Mishra}}, \bibinfo {author} {\bibfnamefont {D.~J.}\
  \bibnamefont {Ing}}, \bibinfo {author} {\bibfnamefont {E.}~\bibnamefont
  {Estrecho}}, \bibinfo {author} {\bibfnamefont {Y.}~\bibnamefont {Lu}},
  \bibinfo {author} {\bibfnamefont {D.~K.}\ \bibnamefont {Efimkin}}, \bibinfo
  {author} {\bibfnamefont {J.~O.}\ \bibnamefont {Tollerud}}, \bibinfo {author}
  {\bibfnamefont {E.~A.}\ \bibnamefont {Ostrovskaya}}, \bibinfo {author}
  {\bibfnamefont {M.~M.}\ \bibnamefont {Parish}}, \ and\ \bibinfo {author}
  {\bibfnamefont {J.~A.}\ \bibnamefont {Davis}},\ }\href@noop {} {\enquote
  {\bibinfo {title} {Exciton-polaron interactions in monolayer ws$_{2}$},}\
  }\bibinfo {note} {ArXiv:2206.12007 (2002)}\BibitemShut {NoStop}%
\bibitem [{\citenamefont {Landau}(1933)}]{Landau1933PhysZSoviet}%
  \BibitemOpen
  \bibfield  {author} {\bibinfo {author} {\bibfnamefont {L.}~\bibnamefont
  {Landau}},\ }\bibfield  {title} {\enquote {\bibinfo {title} {Uber die
  bewegung der elektronen im kristallgitter},}\ }\href@noop {} {\bibfield
  {journal} {\bibinfo  {journal} {Phys. Z. Soviet.}\ }\textbf {\bibinfo
  {volume} {3}},\ \bibinfo {pages} {664} (\bibinfo {year} {1933})}\BibitemShut
  {NoStop}%
\bibitem [{\citenamefont {Mahan}(2000)}]{Mahan2000Book}%
  \BibitemOpen
  \bibfield  {author} {\bibinfo {author} {\bibfnamefont {Gerald~D.}\
  \bibnamefont {Mahan}},\ }\href@noop {} {\emph {\bibinfo {title} {Many
  Particle Physics}}},\ \bibinfo {edition} {3rd}\ ed.\ (\bibinfo  {publisher}
  {Kluwer},\ \bibinfo {address} {New York},\ \bibinfo {year}
  {2000})\BibitemShut {NoStop}%
\bibitem [{\citenamefont {Schirotzek}\ \emph {et~al.}(2009)\citenamefont
  {Schirotzek}, \citenamefont {Wu}, \citenamefont {Sommer},\ and\ \citenamefont
  {Zwierlein}}]{Schirotzek2009PRL}%
  \BibitemOpen
  \bibfield  {author} {\bibinfo {author} {\bibfnamefont {Andr{\'e}}\
  \bibnamefont {Schirotzek}}, \bibinfo {author} {\bibfnamefont {Cheng-Hsun}\
  \bibnamefont {Wu}}, \bibinfo {author} {\bibfnamefont {Ariel}\ \bibnamefont
  {Sommer}}, \ and\ \bibinfo {author} {\bibfnamefont {Martin~W.}\ \bibnamefont
  {Zwierlein}},\ }\bibfield  {title} {\enquote {\bibinfo {title} {Observation
  of {Fermi} polarons in a tunable {Fermi} liquid of ultracold atoms},}\
  }\href@noop {} {\bibfield  {journal} {\bibinfo  {journal} {Phys. Rev. Lett.}\
  }\textbf {\bibinfo {volume} {102}},\ \bibinfo {pages} {230402} (\bibinfo
  {year} {2009})}\BibitemShut {NoStop}%
\bibitem [{\citenamefont {Zhang}\ \emph {et~al.}(2012)\citenamefont {Zhang},
  \citenamefont {Ong}, \citenamefont {Arakelyan},\ and\ \citenamefont
  {Thomas}}]{Zhang2012PRL}%
  \BibitemOpen
  \bibfield  {author} {\bibinfo {author} {\bibfnamefont {Y.}~\bibnamefont
  {Zhang}}, \bibinfo {author} {\bibfnamefont {W.}~\bibnamefont {Ong}}, \bibinfo
  {author} {\bibfnamefont {I.}~\bibnamefont {Arakelyan}}, \ and\ \bibinfo
  {author} {\bibfnamefont {J.~E.}\ \bibnamefont {Thomas}},\ }\bibfield  {title}
  {\enquote {\bibinfo {title} {Polaron-to-polaron transitions in the
  radio-frequency spectrum of a quasi-two-dimensional fermi gas},}\ }\href@noop
  {} {\bibfield  {journal} {\bibinfo  {journal} {Phys. Rev. Lett.}\ }\textbf
  {\bibinfo {volume} {108}},\ \bibinfo {pages} {235302} (\bibinfo {year}
  {2012})}\BibitemShut {NoStop}%
\bibitem [{\citenamefont {Kohstall}\ \emph {et~al.}(2012)\citenamefont
  {Kohstall}, \citenamefont {Zaccanti}, \citenamefont {Jag}, \citenamefont
  {Trenkwalder}, \citenamefont {Massignan}, \citenamefont {Bruun},
  \citenamefont {Schreck},\ and\ \citenamefont {Grimm}}]{Grimm2012Nature}%
  \BibitemOpen
  \bibfield  {author} {\bibinfo {author} {\bibfnamefont {C.}~\bibnamefont
  {Kohstall}}, \bibinfo {author} {\bibfnamefont {M.}~\bibnamefont {Zaccanti}},
  \bibinfo {author} {\bibfnamefont {M.}~\bibnamefont {Jag}}, \bibinfo {author}
  {\bibfnamefont {A.}~\bibnamefont {Trenkwalder}}, \bibinfo {author}
  {\bibfnamefont {P.}~\bibnamefont {Massignan}}, \bibinfo {author}
  {\bibfnamefont {G.~M.}\ \bibnamefont {Bruun}}, \bibinfo {author}
  {\bibfnamefont {F.}~\bibnamefont {Schreck}}, \ and\ \bibinfo {author}
  {\bibfnamefont {R.}~\bibnamefont {Grimm}},\ }\bibfield  {title} {\enquote
  {\bibinfo {title} {Metastability and coherence of repulsive polarons in a
  strongly interacting {Fermi} mixture},}\ }\href@noop {} {\bibfield  {journal}
  {\bibinfo  {journal} {Nature (London)}\ }\textbf {\bibinfo {volume} {485}},\
  \bibinfo {pages} {615} (\bibinfo {year} {2012})}\BibitemShut {NoStop}%
\bibitem [{\citenamefont {Koschorreck}\ \emph {et~al.}(2012)\citenamefont
  {Koschorreck}, \citenamefont {Pertot}, \citenamefont {Vogt}, \citenamefont
  {Fr{\"o}hlich}, \citenamefont {Feld},\ and\ \citenamefont
  {K{\"o}hl}}]{Kohl2012Nature}%
  \BibitemOpen
  \bibfield  {author} {\bibinfo {author} {\bibfnamefont {Marco}\ \bibnamefont
  {Koschorreck}}, \bibinfo {author} {\bibfnamefont {Daniel}\ \bibnamefont
  {Pertot}}, \bibinfo {author} {\bibfnamefont {Enrico}\ \bibnamefont {Vogt}},
  \bibinfo {author} {\bibfnamefont {Bernd}\ \bibnamefont {Fr{\"o}hlich}},
  \bibinfo {author} {\bibfnamefont {Michael}\ \bibnamefont {Feld}}, \ and\
  \bibinfo {author} {\bibfnamefont {Michael}\ \bibnamefont {K{\"o}hl}},\
  }\bibfield  {title} {\enquote {\bibinfo {title} {Attractive and repulsive
  {Fermi} polarons in two dimensions},}\ }\href@noop {} {\bibfield  {journal}
  {\bibinfo  {journal} {Nature (London)}\ }\textbf {\bibinfo {volume} {485}},\
  \bibinfo {pages} {619} (\bibinfo {year} {2012})}\BibitemShut {NoStop}%
\bibitem [{\citenamefont {Cetina}\ \emph {et~al.}(2016)\citenamefont {Cetina},
  \citenamefont {Jag}, \citenamefont {Lous}, \citenamefont {Fritsche},
  \citenamefont {Walraven}, \citenamefont {Grimm}, \citenamefont {Levinsen},
  \citenamefont {Parish}, \citenamefont {Schmidt}, \citenamefont {Knap},\ and\
  \citenamefont {Demler}}]{Demler2016Science}%
  \BibitemOpen
  \bibfield  {author} {\bibinfo {author} {\bibfnamefont {Marko}\ \bibnamefont
  {Cetina}}, \bibinfo {author} {\bibfnamefont {Michael}\ \bibnamefont {Jag}},
  \bibinfo {author} {\bibfnamefont {Rianne~S.}\ \bibnamefont {Lous}}, \bibinfo
  {author} {\bibfnamefont {Isabella}\ \bibnamefont {Fritsche}}, \bibinfo
  {author} {\bibfnamefont {Jook T.~M.}\ \bibnamefont {Walraven}}, \bibinfo
  {author} {\bibfnamefont {Rudolf}\ \bibnamefont {Grimm}}, \bibinfo {author}
  {\bibfnamefont {Jesper}\ \bibnamefont {Levinsen}}, \bibinfo {author}
  {\bibfnamefont {Meera~M.}\ \bibnamefont {Parish}}, \bibinfo {author}
  {\bibfnamefont {Richard}\ \bibnamefont {Schmidt}}, \bibinfo {author}
  {\bibfnamefont {Michael}\ \bibnamefont {Knap}}, \ and\ \bibinfo {author}
  {\bibfnamefont {Eugene}\ \bibnamefont {Demler}},\ }\bibfield  {title}
  {\enquote {\bibinfo {title} {Ultrafast many-body interferometry of impurities
  coupled to a fermi sea},}\ }\href@noop {} {\bibfield  {journal} {\bibinfo
  {journal} {Science}\ }\textbf {\bibinfo {volume} {354}},\ \bibinfo {pages}
  {6308} (\bibinfo {year} {2016})}\BibitemShut {NoStop}%
\bibitem [{\citenamefont {Hu}\ \emph {et~al.}(2016{\natexlab{a}})\citenamefont
  {Hu}, \citenamefont {Van~de Graaff}, \citenamefont {Kedar}, \citenamefont
  {Corson}, \citenamefont {Cornell},\ and\ \citenamefont {Jin}}]{Hu2016PRL}%
  \BibitemOpen
  \bibfield  {author} {\bibinfo {author} {\bibfnamefont {Ming-Guang}\
  \bibnamefont {Hu}}, \bibinfo {author} {\bibfnamefont {Michael~J.}\
  \bibnamefont {Van~de Graaff}}, \bibinfo {author} {\bibfnamefont {Dhruv}\
  \bibnamefont {Kedar}}, \bibinfo {author} {\bibfnamefont {John~P.}\
  \bibnamefont {Corson}}, \bibinfo {author} {\bibfnamefont {Eric~A.}\
  \bibnamefont {Cornell}}, \ and\ \bibinfo {author} {\bibfnamefont
  {Deborah~S.}\ \bibnamefont {Jin}},\ }\bibfield  {title} {\enquote {\bibinfo
  {title} {{Bose} polarons in the strongly interacting regime},}\ }\href@noop
  {} {\bibfield  {journal} {\bibinfo  {journal} {Phys. Rev. Lett.}\ }\textbf
  {\bibinfo {volume} {117}},\ \bibinfo {pages} {055301} (\bibinfo {year}
  {2016}{\natexlab{a}})}\BibitemShut {NoStop}%
\bibitem [{\citenamefont {J{\o{}}rgensen}\ \emph {et~al.}(2016)\citenamefont
  {J{\o{}}rgensen}, \citenamefont {Wacker}, \citenamefont {Skalmstang},
  \citenamefont {Parish}, \citenamefont {Levinsen}, \citenamefont
  {Christensen}, \citenamefont {Bruun},\ and\ \citenamefont
  {Arlt}}]{Jorgensen2016PRL}%
  \BibitemOpen
  \bibfield  {author} {\bibinfo {author} {\bibfnamefont {Nils~B.}\ \bibnamefont
  {J{\o{}}rgensen}}, \bibinfo {author} {\bibfnamefont {Lars}\ \bibnamefont
  {Wacker}}, \bibinfo {author} {\bibfnamefont {Kristoffer~T.}\ \bibnamefont
  {Skalmstang}}, \bibinfo {author} {\bibfnamefont {Meera~M.}\ \bibnamefont
  {Parish}}, \bibinfo {author} {\bibfnamefont {Jesper}\ \bibnamefont
  {Levinsen}}, \bibinfo {author} {\bibfnamefont {Rasmus~S.}\ \bibnamefont
  {Christensen}}, \bibinfo {author} {\bibfnamefont {Georg~M.}\ \bibnamefont
  {Bruun}}, \ and\ \bibinfo {author} {\bibfnamefont {Jan~J.}\ \bibnamefont
  {Arlt}},\ }\bibfield  {title} {\enquote {\bibinfo {title} {Observation of
  attractive and repulsive polarons in a {Bose-Einstein} condensate},}\
  }\href@noop {} {\bibfield  {journal} {\bibinfo  {journal} {Phys. Rev. Lett.}\
  }\textbf {\bibinfo {volume} {117}},\ \bibinfo {pages} {055302} (\bibinfo
  {year} {2016})}\BibitemShut {NoStop}%
\bibitem [{\citenamefont {Scazza}\ \emph {et~al.}(2017)\citenamefont {Scazza},
  \citenamefont {Valtolina}, \citenamefont {Massignan}, \citenamefont {Recati},
  \citenamefont {Amico}, \citenamefont {Burchianti}, \citenamefont {Fort},
  \citenamefont {Inguscio}, \citenamefont {Zaccanti},\ and\ \citenamefont
  {Roati}}]{Scazza2017PRL}%
  \BibitemOpen
  \bibfield  {author} {\bibinfo {author} {\bibfnamefont {F.}~\bibnamefont
  {Scazza}}, \bibinfo {author} {\bibfnamefont {G.}~\bibnamefont {Valtolina}},
  \bibinfo {author} {\bibfnamefont {P.}~\bibnamefont {Massignan}}, \bibinfo
  {author} {\bibfnamefont {A.}~\bibnamefont {Recati}}, \bibinfo {author}
  {\bibfnamefont {A.}~\bibnamefont {Amico}}, \bibinfo {author} {\bibfnamefont
  {A.}~\bibnamefont {Burchianti}}, \bibinfo {author} {\bibfnamefont
  {C.}~\bibnamefont {Fort}}, \bibinfo {author} {\bibfnamefont {M.}~\bibnamefont
  {Inguscio}}, \bibinfo {author} {\bibfnamefont {M.}~\bibnamefont {Zaccanti}},
  \ and\ \bibinfo {author} {\bibfnamefont {G.}~\bibnamefont {Roati}},\
  }\bibfield  {title} {\enquote {\bibinfo {title} {Repulsive fermi polarons in
  a resonant mixture of ultracold $^{6}\mathrm{Li}$ atoms},}\ }\href@noop {}
  {\bibfield  {journal} {\bibinfo  {journal} {Phys. Rev. Lett.}\ }\textbf
  {\bibinfo {volume} {118}},\ \bibinfo {pages} {083602} (\bibinfo {year}
  {2017})}\BibitemShut {NoStop}%
\bibitem [{\citenamefont {Yan}\ \emph {et~al.}(2019)\citenamefont {Yan},
  \citenamefont {Patel}, \citenamefont {Mukherjee}, \citenamefont {Fletcher},
  \citenamefont {Struck},\ and\ \citenamefont {Zwierlein}}]{Yan2019PRL}%
  \BibitemOpen
  \bibfield  {author} {\bibinfo {author} {\bibfnamefont {Zhenjie}\ \bibnamefont
  {Yan}}, \bibinfo {author} {\bibfnamefont {Parth~B.}\ \bibnamefont {Patel}},
  \bibinfo {author} {\bibfnamefont {Biswaroop}\ \bibnamefont {Mukherjee}},
  \bibinfo {author} {\bibfnamefont {Richard~J.}\ \bibnamefont {Fletcher}},
  \bibinfo {author} {\bibfnamefont {Julian}\ \bibnamefont {Struck}}, \ and\
  \bibinfo {author} {\bibfnamefont {Martin~W.}\ \bibnamefont {Zwierlein}},\
  }\bibfield  {title} {\enquote {\bibinfo {title} {Boiling a unitary fermi
  liquid},}\ }\href@noop {} {\bibfield  {journal} {\bibinfo  {journal} {Phys.
  Rev. Lett.}\ }\textbf {\bibinfo {volume} {122}},\ \bibinfo {pages} {093401}
  (\bibinfo {year} {2019})}\BibitemShut {NoStop}%
\bibitem [{\citenamefont {Yan}\ \emph {et~al.}(2020)\citenamefont {Yan},
  \citenamefont {Ni}, \citenamefont {Robens},\ and\ \citenamefont
  {Zwierlein}}]{Zwierlein2020Science}%
  \BibitemOpen
  \bibfield  {author} {\bibinfo {author} {\bibfnamefont {Z.~Z.}\ \bibnamefont
  {Yan}}, \bibinfo {author} {\bibfnamefont {Y.}~\bibnamefont {Ni}}, \bibinfo
  {author} {\bibfnamefont {C.}~\bibnamefont {Robens}}, \ and\ \bibinfo {author}
  {\bibfnamefont {M.W.}\ \bibnamefont {Zwierlein}},\ }\bibfield  {title}
  {\enquote {\bibinfo {title} {Bose polarons near quantum criticality},}\
  }\href@noop {} {\bibfield  {journal} {\bibinfo  {journal} {Science}\ }\textbf
  {\bibinfo {volume} {368}},\ \bibinfo {pages} {190} (\bibinfo {year}
  {2020})}\BibitemShut {NoStop}%
\bibitem [{\citenamefont {Ness}\ \emph {et~al.}(2020)\citenamefont {Ness},
  \citenamefont {Shkedrov}, \citenamefont {Florshaim}, \citenamefont {Diessel},
  \citenamefont {von Milczewski}, \citenamefont {Schmidt},\ and\ \citenamefont
  {Sagi}}]{Sagi2020PRX}%
  \BibitemOpen
  \bibfield  {author} {\bibinfo {author} {\bibfnamefont {Gal}\ \bibnamefont
  {Ness}}, \bibinfo {author} {\bibfnamefont {Constantine}\ \bibnamefont
  {Shkedrov}}, \bibinfo {author} {\bibfnamefont {Yanay}\ \bibnamefont
  {Florshaim}}, \bibinfo {author} {\bibfnamefont {Oriana~K.}\ \bibnamefont
  {Diessel}}, \bibinfo {author} {\bibfnamefont {Jonas}\ \bibnamefont {von
  Milczewski}}, \bibinfo {author} {\bibfnamefont {Richard}\ \bibnamefont
  {Schmidt}}, \ and\ \bibinfo {author} {\bibfnamefont {Yoav}\ \bibnamefont
  {Sagi}},\ }\bibfield  {title} {\enquote {\bibinfo {title} {Observation of a
  smooth polaron-molecule transition in a degenerate {Fermi} gas},}\
  }\href@noop {} {\bibfield  {journal} {\bibinfo  {journal} {Phys. Rev. X}\
  }\textbf {\bibinfo {volume} {10}},\ \bibinfo {pages} {041019} (\bibinfo
  {year} {2020})}\BibitemShut {NoStop}%
\bibitem [{\citenamefont {Massignan}\ \emph {et~al.}(2014)\citenamefont
  {Massignan}, \citenamefont {Zaccanti},\ and\ \citenamefont
  {Bruun}}]{Bruun2014Review}%
  \BibitemOpen
  \bibfield  {author} {\bibinfo {author} {\bibfnamefont {P.}~\bibnamefont
  {Massignan}}, \bibinfo {author} {\bibfnamefont {M.}~\bibnamefont {Zaccanti}},
  \ and\ \bibinfo {author} {\bibfnamefont {G.~M.}\ \bibnamefont {Bruun}},\
  }\bibfield  {title} {\enquote {\bibinfo {title} {Polarons, dressed molecules
  and itinerant ferromagnetism in ultra-cold {Fermi} gases},}\ }\href@noop {}
  {\bibfield  {journal} {\bibinfo  {journal} {Rep. Prog. Phys.}\ }\textbf
  {\bibinfo {volume} {77}},\ \bibinfo {pages} {034401} (\bibinfo {year}
  {2014})}\BibitemShut {NoStop}%
\bibitem [{\citenamefont {Chevy}(2006)}]{Chevy2006PRA}%
  \BibitemOpen
  \bibfield  {author} {\bibinfo {author} {\bibfnamefont {F.}~\bibnamefont
  {Chevy}},\ }\bibfield  {title} {\enquote {\bibinfo {title} {Universal phase
  diagram of a strongly interacting {Fermi} gas with unbalanced spin
  populations},}\ }\href@noop {} {\bibfield  {journal} {\bibinfo  {journal}
  {Phys. Rev. A}\ }\textbf {\bibinfo {volume} {74}},\ \bibinfo {pages} {063628}
  (\bibinfo {year} {2006})}\BibitemShut {NoStop}%
\bibitem [{\citenamefont {Lobo}\ \emph {et~al.}(2006)\citenamefont {Lobo},
  \citenamefont {Recati}, \citenamefont {Giorgini},\ and\ \citenamefont
  {Stringari}}]{Lobo2006PRL}%
  \BibitemOpen
  \bibfield  {author} {\bibinfo {author} {\bibfnamefont {C.}~\bibnamefont
  {Lobo}}, \bibinfo {author} {\bibfnamefont {A.}~\bibnamefont {Recati}},
  \bibinfo {author} {\bibfnamefont {S.}~\bibnamefont {Giorgini}}, \ and\
  \bibinfo {author} {\bibfnamefont {S.}~\bibnamefont {Stringari}},\ }\bibfield
  {title} {\enquote {\bibinfo {title} {Normal state of a polarized {Fermi} gas
  at unitarity},}\ }\href@noop {} {\bibfield  {journal} {\bibinfo  {journal}
  {Phys. Rev. Lett.}\ }\textbf {\bibinfo {volume} {97}},\ \bibinfo {pages}
  {200403} (\bibinfo {year} {2006})}\BibitemShut {NoStop}%
\bibitem [{\citenamefont {Combescot}\ \emph {et~al.}(2007)\citenamefont
  {Combescot}, \citenamefont {Recati}, \citenamefont {Lobo},\ and\
  \citenamefont {Chevy}}]{Combescot2007PRL}%
  \BibitemOpen
  \bibfield  {author} {\bibinfo {author} {\bibfnamefont {R.}~\bibnamefont
  {Combescot}}, \bibinfo {author} {\bibfnamefont {A.}~\bibnamefont {Recati}},
  \bibinfo {author} {\bibfnamefont {C.}~\bibnamefont {Lobo}}, \ and\ \bibinfo
  {author} {\bibfnamefont {F.}~\bibnamefont {Chevy}},\ }\bibfield  {title}
  {\enquote {\bibinfo {title} {Normal state of highly polarized {Fermi} gases:
  Simple many-body approaches},}\ }\href@noop {} {\bibfield  {journal}
  {\bibinfo  {journal} {Phys. Rev. Lett.}\ }\textbf {\bibinfo {volume} {98}},\
  \bibinfo {pages} {180402} (\bibinfo {year} {2007})}\BibitemShut {NoStop}%
\bibitem [{\citenamefont {Punk}\ \emph {et~al.}(2009)\citenamefont {Punk},
  \citenamefont {Dumitrescu},\ and\ \citenamefont {Zwerger}}]{Punk2009PRA}%
  \BibitemOpen
  \bibfield  {author} {\bibinfo {author} {\bibfnamefont {M.}~\bibnamefont
  {Punk}}, \bibinfo {author} {\bibfnamefont {P.~T.}\ \bibnamefont
  {Dumitrescu}}, \ and\ \bibinfo {author} {\bibfnamefont {W.}~\bibnamefont
  {Zwerger}},\ }\bibfield  {title} {\enquote {\bibinfo {title}
  {Polaron-to-molecule transition in a strongly imbalanced {Fermi} gas},}\
  }\href@noop {} {\bibfield  {journal} {\bibinfo  {journal} {Phys. Rev. A}\
  }\textbf {\bibinfo {volume} {80}},\ \bibinfo {pages} {053605} (\bibinfo
  {year} {2009})}\BibitemShut {NoStop}%
\bibitem [{\citenamefont {Cui}\ and\ \citenamefont {Zhai}(2010)}]{Cui2010PRA}%
  \BibitemOpen
  \bibfield  {author} {\bibinfo {author} {\bibfnamefont {Xiaoling}\
  \bibnamefont {Cui}}\ and\ \bibinfo {author} {\bibfnamefont {Hui}\
  \bibnamefont {Zhai}},\ }\bibfield  {title} {\enquote {\bibinfo {title}
  {Stability of a fully magnetized ferromagnetic state in repulsively
  interacting ultracold {Fermi} gases},}\ }\href@noop {} {\bibfield  {journal}
  {\bibinfo  {journal} {Phys. Rev. A}\ }\textbf {\bibinfo {volume} {81}},\
  \bibinfo {pages} {041602} (\bibinfo {year} {2010})}\BibitemShut {NoStop}%
\bibitem [{\citenamefont {Mathy}\ \emph {et~al.}(2011)\citenamefont {Mathy},
  \citenamefont {Parish},\ and\ \citenamefont {Huse}}]{Mathy2011PRL}%
  \BibitemOpen
  \bibfield  {author} {\bibinfo {author} {\bibfnamefont {Charles J.~M.}\
  \bibnamefont {Mathy}}, \bibinfo {author} {\bibfnamefont {Meera~M.}\
  \bibnamefont {Parish}}, \ and\ \bibinfo {author} {\bibfnamefont {David~A.}\
  \bibnamefont {Huse}},\ }\bibfield  {title} {\enquote {\bibinfo {title}
  {Trimers, molecules, and polarons in mass-imbalanced atomic {Fermi} gases},}\
  }\href@noop {} {\bibfield  {journal} {\bibinfo  {journal} {Phys. Rev. Lett.}\
  }\textbf {\bibinfo {volume} {106}},\ \bibinfo {pages} {166404} (\bibinfo
  {year} {2011})}\BibitemShut {NoStop}%
\bibitem [{\citenamefont {Schmidt}\ \emph {et~al.}(2012)\citenamefont
  {Schmidt}, \citenamefont {Enss}, \citenamefont {Pietil\"a},\ and\
  \citenamefont {Demler}}]{Schmidt2012PRA}%
  \BibitemOpen
  \bibfield  {author} {\bibinfo {author} {\bibfnamefont {Richard}\ \bibnamefont
  {Schmidt}}, \bibinfo {author} {\bibfnamefont {Tilman}\ \bibnamefont {Enss}},
  \bibinfo {author} {\bibfnamefont {Ville}\ \bibnamefont {Pietil\"a}}, \ and\
  \bibinfo {author} {\bibfnamefont {Eugene}\ \bibnamefont {Demler}},\
  }\bibfield  {title} {\enquote {\bibinfo {title} {Fermi polarons in two
  dimensions},}\ }\href@noop {} {\bibfield  {journal} {\bibinfo  {journal}
  {Phys. Rev. A}\ }\textbf {\bibinfo {volume} {85}},\ \bibinfo {pages} {021602}
  (\bibinfo {year} {2012})}\BibitemShut {NoStop}%
\bibitem [{\citenamefont {Rath}\ and\ \citenamefont
  {Schmidt}(2013)}]{Rath2013PRA}%
  \BibitemOpen
  \bibfield  {author} {\bibinfo {author} {\bibfnamefont {Steffen~Patrick}\
  \bibnamefont {Rath}}\ and\ \bibinfo {author} {\bibfnamefont {Richard}\
  \bibnamefont {Schmidt}},\ }\bibfield  {title} {\enquote {\bibinfo {title}
  {Field-theoretical study of the {Bose} polaron},}\ }\href@noop {} {\bibfield
  {journal} {\bibinfo  {journal} {Phys. Rev. A}\ }\textbf {\bibinfo {volume}
  {88}},\ \bibinfo {pages} {053632} (\bibinfo {year} {2013})}\BibitemShut
  {NoStop}%
\bibitem [{\citenamefont {Shashi}\ \emph {et~al.}(2014)\citenamefont {Shashi},
  \citenamefont {Grusdt}, \citenamefont {Abanin},\ and\ \citenamefont
  {Demler}}]{Shashi2014PRA}%
  \BibitemOpen
  \bibfield  {author} {\bibinfo {author} {\bibfnamefont {Aditya}\ \bibnamefont
  {Shashi}}, \bibinfo {author} {\bibfnamefont {Fabian}\ \bibnamefont {Grusdt}},
  \bibinfo {author} {\bibfnamefont {Dmitry~A.}\ \bibnamefont {Abanin}}, \ and\
  \bibinfo {author} {\bibfnamefont {Eugene}\ \bibnamefont {Demler}},\
  }\bibfield  {title} {\enquote {\bibinfo {title} {Radio-frequency spectroscopy
  of polarons in ultracold {Bose} gases},}\ }\href@noop {} {\bibfield
  {journal} {\bibinfo  {journal} {Phys. Rev. A}\ }\textbf {\bibinfo {volume}
  {89}},\ \bibinfo {pages} {053617} (\bibinfo {year} {2014})}\BibitemShut
  {NoStop}%
\bibitem [{\citenamefont {Li}\ and\ \citenamefont
  {Das~Sarma}(2014)}]{Li2014PRA}%
  \BibitemOpen
  \bibfield  {author} {\bibinfo {author} {\bibfnamefont {Weiran}\ \bibnamefont
  {Li}}\ and\ \bibinfo {author} {\bibfnamefont {S.}~\bibnamefont {Das~Sarma}},\
  }\bibfield  {title} {\enquote {\bibinfo {title} {Variational study of
  polarons in {Bose-Einstein} condensates},}\ }\href@noop {} {\bibfield
  {journal} {\bibinfo  {journal} {Phys. Rev. A}\ }\textbf {\bibinfo {volume}
  {90}},\ \bibinfo {pages} {013618} (\bibinfo {year} {2014})}\BibitemShut
  {NoStop}%
\bibitem [{\citenamefont {Kroiss}\ and\ \citenamefont
  {Pollet}(2015)}]{Kroiss2015PRL}%
  \BibitemOpen
  \bibfield  {author} {\bibinfo {author} {\bibfnamefont {Peter}\ \bibnamefont
  {Kroiss}}\ and\ \bibinfo {author} {\bibfnamefont {Lode}\ \bibnamefont
  {Pollet}},\ }\bibfield  {title} {\enquote {\bibinfo {title} {Diagrammatic
  monte carlo study of a mass-imbalanced fermi-polaron system},}\ }\href@noop
  {} {\bibfield  {journal} {\bibinfo  {journal} {Phys. Rev. B}\ }\textbf
  {\bibinfo {volume} {91}},\ \bibinfo {pages} {144507} (\bibinfo {year}
  {2015})}\BibitemShut {NoStop}%
\bibitem [{\citenamefont {Levinsen}\ \emph {et~al.}(2015)\citenamefont
  {Levinsen}, \citenamefont {Parish},\ and\ \citenamefont
  {Bruun}}]{Levinsen2015PRL}%
  \BibitemOpen
  \bibfield  {author} {\bibinfo {author} {\bibfnamefont {Jesper}\ \bibnamefont
  {Levinsen}}, \bibinfo {author} {\bibfnamefont {Meera~M.}\ \bibnamefont
  {Parish}}, \ and\ \bibinfo {author} {\bibfnamefont {Georg~M.}\ \bibnamefont
  {Bruun}},\ }\bibfield  {title} {\enquote {\bibinfo {title} {Impurity in a
  bose-einstein condensate and the efimov effect},}\ }\href@noop {} {\bibfield
  {journal} {\bibinfo  {journal} {Phys. Rev. Lett.}\ }\textbf {\bibinfo
  {volume} {115}},\ \bibinfo {pages} {125302} (\bibinfo {year}
  {2015})}\BibitemShut {NoStop}%
\bibitem [{\citenamefont {Wang}\ \emph {et~al.}(2015)\citenamefont {Wang},
  \citenamefont {Gacesa},\ and\ \citenamefont {C\^ot\'e}}]{JiaWang2015PRL}%
  \BibitemOpen
  \bibfield  {author} {\bibinfo {author} {\bibfnamefont {Jia}\ \bibnamefont
  {Wang}}, \bibinfo {author} {\bibfnamefont {Marko}\ \bibnamefont {Gacesa}}, \
  and\ \bibinfo {author} {\bibfnamefont {R.}~\bibnamefont {C\^ot\'e}},\
  }\bibfield  {title} {\enquote {\bibinfo {title} {Rydberg electrons in a
  bose-einstein condensate},}\ }\href@noop {} {\bibfield  {journal} {\bibinfo
  {journal} {Phys. Rev. Lett.}\ }\textbf {\bibinfo {volume} {114}},\ \bibinfo
  {pages} {243003} (\bibinfo {year} {2015})}\BibitemShut {NoStop}%
\bibitem [{\citenamefont {Hu}\ \emph {et~al.}(2016{\natexlab{b}})\citenamefont
  {Hu}, \citenamefont {Wang}, \citenamefont {Yi},\ and\ \citenamefont
  {Liu}}]{HuHui2016PRA}%
  \BibitemOpen
  \bibfield  {author} {\bibinfo {author} {\bibfnamefont {Hui}\ \bibnamefont
  {Hu}}, \bibinfo {author} {\bibfnamefont {An-Bang}\ \bibnamefont {Wang}},
  \bibinfo {author} {\bibfnamefont {Su}~\bibnamefont {Yi}}, \ and\ \bibinfo
  {author} {\bibfnamefont {Xia-Ji}\ \bibnamefont {Liu}},\ }\bibfield  {title}
  {\enquote {\bibinfo {title} {Fermi polaron in a one-dimensional quasiperiodic
  optical lattice: The simplest many-body localization challenge},}\
  }\href@noop {} {\bibfield  {journal} {\bibinfo  {journal} {Phys. Rev. A}\
  }\textbf {\bibinfo {volume} {93}},\ \bibinfo {pages} {053601} (\bibinfo
  {year} {2016}{\natexlab{b}})}\BibitemShut {NoStop}%
\bibitem [{\citenamefont {Goulko}\ \emph {et~al.}(2016)\citenamefont {Goulko},
  \citenamefont {Mishchenko}, \citenamefont {Prokof'ev},\ and\ \citenamefont
  {Svistunov}}]{Goulko2016PRA}%
  \BibitemOpen
  \bibfield  {author} {\bibinfo {author} {\bibfnamefont {Olga}\ \bibnamefont
  {Goulko}}, \bibinfo {author} {\bibfnamefont {Andrey~S.}\ \bibnamefont
  {Mishchenko}}, \bibinfo {author} {\bibfnamefont {Nikolay}\ \bibnamefont
  {Prokof'ev}}, \ and\ \bibinfo {author} {\bibfnamefont {Boris}\ \bibnamefont
  {Svistunov}},\ }\bibfield  {title} {\enquote {\bibinfo {title} {Dark
  continuum in the spectral function of the resonant fermi polaron},}\
  }\href@noop {} {\bibfield  {journal} {\bibinfo  {journal} {Phys. Rev. A}\
  }\textbf {\bibinfo {volume} {94}},\ \bibinfo {pages} {051605} (\bibinfo
  {year} {2016})}\BibitemShut {NoStop}%
\bibitem [{\citenamefont {Hu}\ \emph {et~al.}(2018)\citenamefont {Hu},
  \citenamefont {Mulkerin}, \citenamefont {Wang},\ and\ \citenamefont
  {Liu}}]{HuHui2018PRA}%
  \BibitemOpen
  \bibfield  {author} {\bibinfo {author} {\bibfnamefont {Hui}\ \bibnamefont
  {Hu}}, \bibinfo {author} {\bibfnamefont {Brendan~C.}\ \bibnamefont
  {Mulkerin}}, \bibinfo {author} {\bibfnamefont {Jia}\ \bibnamefont {Wang}}, \
  and\ \bibinfo {author} {\bibfnamefont {Xia-Ji}\ \bibnamefont {Liu}},\
  }\bibfield  {title} {\enquote {\bibinfo {title} {Attractive fermi polarons at
  nonzero temperatures with a finite impurity concentration},}\ }\href@noop {}
  {\bibfield  {journal} {\bibinfo  {journal} {Phys. Rev. A}\ }\textbf {\bibinfo
  {volume} {98}},\ \bibinfo {pages} {013626} (\bibinfo {year}
  {2018})}\BibitemShut {NoStop}%
\bibitem [{\citenamefont {Wang}\ \emph {et~al.}(2019)\citenamefont {Wang},
  \citenamefont {Liu},\ and\ \citenamefont {Hu}}]{JiaWang2019PRL}%
  \BibitemOpen
  \bibfield  {author} {\bibinfo {author} {\bibfnamefont {Jia}\ \bibnamefont
  {Wang}}, \bibinfo {author} {\bibfnamefont {Xia-Ji}\ \bibnamefont {Liu}}, \
  and\ \bibinfo {author} {\bibfnamefont {Hui}\ \bibnamefont {Hu}},\ }\bibfield
  {title} {\enquote {\bibinfo {title} {Roton-induced bose polaron in the
  presence of synthetic spin-orbit coupling},}\ }\href@noop {} {\bibfield
  {journal} {\bibinfo  {journal} {Phys. Rev. Lett.}\ }\textbf {\bibinfo
  {volume} {123}},\ \bibinfo {pages} {213401} (\bibinfo {year}
  {2019})}\BibitemShut {NoStop}%
\bibitem [{\citenamefont {Pe\~na Ardila}\ \emph {et~al.}(2019)\citenamefont
  {Pe\~na Ardila}, \citenamefont {J\o{}rgensen}, \citenamefont {Pohl},
  \citenamefont {Giorgini}, \citenamefont {Bruun},\ and\ \citenamefont
  {Arlt}}]{PenaArdila2019PRA}%
  \BibitemOpen
  \bibfield  {author} {\bibinfo {author} {\bibfnamefont {L.~A.}\ \bibnamefont
  {Pe\~na Ardila}}, \bibinfo {author} {\bibfnamefont {N.~B.}\ \bibnamefont
  {J\o{}rgensen}}, \bibinfo {author} {\bibfnamefont {T.}~\bibnamefont {Pohl}},
  \bibinfo {author} {\bibfnamefont {S.}~\bibnamefont {Giorgini}}, \bibinfo
  {author} {\bibfnamefont {G.~M.}\ \bibnamefont {Bruun}}, \ and\ \bibinfo
  {author} {\bibfnamefont {J.~J.}\ \bibnamefont {Arlt}},\ }\bibfield  {title}
  {\enquote {\bibinfo {title} {Analyzing a bose polaron across resonant
  interactions},}\ }\href {\doibase 10.1103/PhysRevA.99.063607} {\bibfield
  {journal} {\bibinfo  {journal} {Phys. Rev. A}\ }\textbf {\bibinfo {volume}
  {99}},\ \bibinfo {pages} {063607} (\bibinfo {year} {2019})}\BibitemShut
  {NoStop}%
\bibitem [{\citenamefont {Mulkerin}\ \emph {et~al.}(2019)\citenamefont
  {Mulkerin}, \citenamefont {Liu},\ and\ \citenamefont
  {Hu}}]{Mulkerin2019AnnPhys}%
  \BibitemOpen
  \bibfield  {author} {\bibinfo {author} {\bibfnamefont {B.~C.}\ \bibnamefont
  {Mulkerin}}, \bibinfo {author} {\bibfnamefont {X.-J.}\ \bibnamefont {Liu}}, \
  and\ \bibinfo {author} {\bibfnamefont {H.}~\bibnamefont {Hu}},\ }\bibfield
  {title} {\enquote {\bibinfo {title} {Breakdown of the fermi polaron
  description near fermi degeneracy at unitarity},}\ }\href@noop {} {\bibfield
  {journal} {\bibinfo  {journal} {Ann. Phys. (NY)}\ }\textbf {\bibinfo {volume}
  {407}},\ \bibinfo {pages} {29} (\bibinfo {year} {2019})}\BibitemShut
  {NoStop}%
\bibitem [{\citenamefont {Isaule}\ \emph {et~al.}(2021)\citenamefont {Isaule},
  \citenamefont {Morera}, \citenamefont {Massignan},\ and\ \citenamefont
  {Juli\'a-D\'{\i}az}}]{Isaule2021PRA}%
  \BibitemOpen
  \bibfield  {author} {\bibinfo {author} {\bibfnamefont {Felipe}\ \bibnamefont
  {Isaule}}, \bibinfo {author} {\bibfnamefont {Ivan}\ \bibnamefont {Morera}},
  \bibinfo {author} {\bibfnamefont {Pietro}\ \bibnamefont {Massignan}}, \ and\
  \bibinfo {author} {\bibfnamefont {Bruno}\ \bibnamefont {Juli\'a-D\'{\i}az}},\
  }\bibfield  {title} {\enquote {\bibinfo {title} {Renormalization-group study
  of bose polarons},}\ }\href@noop {} {\bibfield  {journal} {\bibinfo
  {journal} {Phys. Rev. A}\ }\textbf {\bibinfo {volume} {104}},\ \bibinfo
  {pages} {023317} (\bibinfo {year} {2021})}\BibitemShut {NoStop}%
\bibitem [{\citenamefont {Pessoa}\ \emph {et~al.}(2021)\citenamefont {Pessoa},
  \citenamefont {Vitiello},\ and\ \citenamefont {Ardila}}]{Pessoa2021PRA}%
  \BibitemOpen
  \bibfield  {author} {\bibinfo {author} {\bibfnamefont {Renato}\ \bibnamefont
  {Pessoa}}, \bibinfo {author} {\bibfnamefont {S.~A.}\ \bibnamefont
  {Vitiello}}, \ and\ \bibinfo {author} {\bibfnamefont {L.~A. Pe\~na}\
  \bibnamefont {Ardila}},\ }\bibfield  {title} {\enquote {\bibinfo {title}
  {Finite-range effects in the unitary fermi polaron},}\ }\href@noop {}
  {\bibfield  {journal} {\bibinfo  {journal} {Phys. Rev. A}\ }\textbf {\bibinfo
  {volume} {104}},\ \bibinfo {pages} {043313} (\bibinfo {year}
  {2021})}\BibitemShut {NoStop}%
\bibitem [{\citenamefont {Seetharam}\ \emph
  {et~al.}(2021{\natexlab{a}})\citenamefont {Seetharam}, \citenamefont
  {Shchadilova}, \citenamefont {Grusdt}, \citenamefont {Zvonarev},\ and\
  \citenamefont {Demler}}]{Seetharam2021PRL}%
  \BibitemOpen
  \bibfield  {author} {\bibinfo {author} {\bibfnamefont {Kushal}\ \bibnamefont
  {Seetharam}}, \bibinfo {author} {\bibfnamefont {Yulia}\ \bibnamefont
  {Shchadilova}}, \bibinfo {author} {\bibfnamefont {Fabian}\ \bibnamefont
  {Grusdt}}, \bibinfo {author} {\bibfnamefont {Mikhail~B.}\ \bibnamefont
  {Zvonarev}}, \ and\ \bibinfo {author} {\bibfnamefont {Eugene}\ \bibnamefont
  {Demler}},\ }\bibfield  {title} {\enquote {\bibinfo {title} {Dynamical
  quantum cherenkov transition of fast impurities in quantum liquids},}\
  }\href@noop {} {\bibfield  {journal} {\bibinfo  {journal} {Phys. Rev. Lett.}\
  }\textbf {\bibinfo {volume} {127}},\ \bibinfo {pages} {185302} (\bibinfo
  {year} {2021}{\natexlab{a}})}\BibitemShut {NoStop}%
\bibitem [{\citenamefont {Seetharam}\ \emph
  {et~al.}(2021{\natexlab{b}})\citenamefont {Seetharam}, \citenamefont
  {Shchadilova}, \citenamefont {Grusdt}, \citenamefont {Zvonarev},\ and\
  \citenamefont {Demler}}]{Kushal2021PRL}%
  \BibitemOpen
  \bibfield  {author} {\bibinfo {author} {\bibfnamefont {Kushal}\ \bibnamefont
  {Seetharam}}, \bibinfo {author} {\bibfnamefont {Yulia}\ \bibnamefont
  {Shchadilova}}, \bibinfo {author} {\bibfnamefont {Fabian}\ \bibnamefont
  {Grusdt}}, \bibinfo {author} {\bibfnamefont {Mikhail~B.}\ \bibnamefont
  {Zvonarev}}, \ and\ \bibinfo {author} {\bibfnamefont {Eugene}\ \bibnamefont
  {Demler}},\ }\bibfield  {title} {\enquote {\bibinfo {title} {Dynamical
  quantum cherenkov transition of fast impurities in quantum liquids},}\
  }\href@noop {} {\bibfield  {journal} {\bibinfo  {journal} {Phys. Rev. Lett.}\
  }\textbf {\bibinfo {volume} {127}},\ \bibinfo {pages} {185302} (\bibinfo
  {year} {2021}{\natexlab{b}})}\BibitemShut {NoStop}%
\bibitem [{\citenamefont {Hu}\ \emph {et~al.}(2022)\citenamefont {Hu},
  \citenamefont {Wang}, \citenamefont {Zhou},\ and\ \citenamefont
  {Liu}}]{HuHui2022PRA}%
  \BibitemOpen
  \bibfield  {author} {\bibinfo {author} {\bibfnamefont {Hui}\ \bibnamefont
  {Hu}}, \bibinfo {author} {\bibfnamefont {Jia}\ \bibnamefont {Wang}}, \bibinfo
  {author} {\bibfnamefont {Jing}\ \bibnamefont {Zhou}}, \ and\ \bibinfo
  {author} {\bibfnamefont {Xia-Ji}\ \bibnamefont {Liu}},\ }\bibfield  {title}
  {\enquote {\bibinfo {title} {Crossover polarons in a strongly interacting
  fermi superfluid},}\ }\href@noop {} {\bibfield  {journal} {\bibinfo
  {journal} {Phys. Rev. A}\ }\textbf {\bibinfo {volume} {105}},\ \bibinfo
  {pages} {023317} (\bibinfo {year} {2022})}\BibitemShut {NoStop}%
\bibitem [{\citenamefont {Bolger}\ \emph {et~al.}(1996)\citenamefont {Bolger},
  \citenamefont {Paul},\ and\ \citenamefont {Smirl}}]{Smirl1996PRB}%
  \BibitemOpen
  \bibfield  {author} {\bibinfo {author} {\bibfnamefont {J.~A.}\ \bibnamefont
  {Bolger}}, \bibinfo {author} {\bibfnamefont {A.~E.}\ \bibnamefont {Paul}}, \
  and\ \bibinfo {author} {\bibfnamefont {Arthur~L.}\ \bibnamefont {Smirl}},\
  }\bibfield  {title} {\enquote {\bibinfo {title} {Ultrafast ellipsometry of
  coherent processes and exciton-exciton interactions in quantum wells at
  negative delays},}\ }\href@noop {} {\bibfield  {journal} {\bibinfo  {journal}
  {Phys. Rev. B}\ }\textbf {\bibinfo {volume} {54}},\ \bibinfo {pages}
  {11666--11671} (\bibinfo {year} {1996})}\BibitemShut {NoStop}%
\bibitem [{\citenamefont {Smirl}\ \emph {et~al.}(1999)\citenamefont {Smirl},
  \citenamefont {Stevens}, \citenamefont {Chen},\ and\ \citenamefont
  {Buccafusca}}]{Buccafusca1999PRB}%
  \BibitemOpen
  \bibfield  {author} {\bibinfo {author} {\bibfnamefont {Arthur~L.}\
  \bibnamefont {Smirl}}, \bibinfo {author} {\bibfnamefont {Martin~J.}\
  \bibnamefont {Stevens}}, \bibinfo {author} {\bibfnamefont {X.}~\bibnamefont
  {Chen}}, \ and\ \bibinfo {author} {\bibfnamefont {O.}~\bibnamefont
  {Buccafusca}},\ }\bibfield  {title} {\enquote {\bibinfo {title} {Heavy-hole
  and light-hole oscillations in the coherent emission from quantum wells:
  Evidence for exciton-exciton correlations},}\ }\href@noop {} {\bibfield
  {journal} {\bibinfo  {journal} {Phys. Rev. B}\ }\textbf {\bibinfo {volume}
  {60}},\ \bibinfo {pages} {8267--8275} (\bibinfo {year} {1999})}\BibitemShut
  {NoStop}%
\bibitem [{\citenamefont {Shacklette}\ and\ \citenamefont
  {Cundiff}(2002)}]{Cundiff2002PRB}%
  \BibitemOpen
  \bibfield  {author} {\bibinfo {author} {\bibfnamefont {Justin~M.}\
  \bibnamefont {Shacklette}}\ and\ \bibinfo {author} {\bibfnamefont
  {Steven~T.}\ \bibnamefont {Cundiff}},\ }\bibfield  {title} {\enquote
  {\bibinfo {title} {Role of excitation-induced shift in the coherent optical
  response of semiconductors},}\ }\href@noop {} {\bibfield  {journal} {\bibinfo
   {journal} {Phys. Rev. B}\ }\textbf {\bibinfo {volume} {66}},\ \bibinfo
  {pages} {045309} (\bibinfo {year} {2002})}\BibitemShut {NoStop}%
\bibitem [{\citenamefont {Chang}\ \emph {et~al.}(2001)\citenamefont {Chang},
  \citenamefont {Shirley},\ and\ \citenamefont {Levine}}]{Zachary2001PRB}%
  \BibitemOpen
  \bibfield  {author} {\bibinfo {author} {\bibfnamefont {Eric~K.}\ \bibnamefont
  {Chang}}, \bibinfo {author} {\bibfnamefont {Eric~L.}\ \bibnamefont
  {Shirley}}, \ and\ \bibinfo {author} {\bibfnamefont {Zachary~H.}\
  \bibnamefont {Levine}},\ }\bibfield  {title} {\enquote {\bibinfo {title}
  {Excitonic effects on optical second-harmonic polarizabilities of
  semiconductors},}\ }\href@noop {} {\bibfield  {journal} {\bibinfo  {journal}
  {Phys. Rev. B}\ }\textbf {\bibinfo {volume} {65}},\ \bibinfo {pages} {035205}
  (\bibinfo {year} {2001})}\BibitemShut {NoStop}%
\bibitem [{\citenamefont {Attaccalite}\ \emph {et~al.}(2018)\citenamefont
  {Attaccalite}, \citenamefont {Gr\"uning}, \citenamefont {Amara},
  \citenamefont {Latil},\ and\ \citenamefont {Ducastelle}}]{Ducastelle2018PRB}%
  \BibitemOpen
  \bibfield  {author} {\bibinfo {author} {\bibfnamefont {Claudio}\ \bibnamefont
  {Attaccalite}}, \bibinfo {author} {\bibfnamefont {Myrta}\ \bibnamefont
  {Gr\"uning}}, \bibinfo {author} {\bibfnamefont {Hakim}\ \bibnamefont
  {Amara}}, \bibinfo {author} {\bibfnamefont {Sylvain}\ \bibnamefont {Latil}},
  \ and\ \bibinfo {author} {\bibfnamefont {Fran\ifmmode
  \mbox{\c{c}}\else~\c{c}\fi{}ois}\ \bibnamefont {Ducastelle}},\ }\bibfield
  {title} {\enquote {\bibinfo {title} {Two-photon absorption in two-dimensional
  materials: The case of hexagonal boron nitride},}\ }\href@noop {} {\bibfield
  {journal} {\bibinfo  {journal} {Phys. Rev. B}\ }\textbf {\bibinfo {volume}
  {98}},\ \bibinfo {pages} {165126} (\bibinfo {year} {2018})}\BibitemShut
  {NoStop}%
\bibitem [{\citenamefont {Prussel}\ and\ \citenamefont
  {V\'eniard}(2018)}]{Veniard2018PRB}%
  \BibitemOpen
  \bibfield  {author} {\bibinfo {author} {\bibfnamefont {Lucie}\ \bibnamefont
  {Prussel}}\ and\ \bibinfo {author} {\bibfnamefont {Val\'erie}\ \bibnamefont
  {V\'eniard}},\ }\bibfield  {title} {\enquote {\bibinfo {title} {Linear
  electro-optic effect in semiconductors: Ab initio description of the
  electronic contribution},}\ }\href@noop {} {\bibfield  {journal} {\bibinfo
  {journal} {Phys. Rev. B}\ }\textbf {\bibinfo {volume} {97}},\ \bibinfo
  {pages} {205201} (\bibinfo {year} {2018})}\BibitemShut {NoStop}%
\bibitem [{\citenamefont {Lindoy}\ \emph {et~al.}()\citenamefont {Lindoy},
  \citenamefont {Chang},\ and\ \citenamefont {Reichman}}]{Reichman2022arXiv}%
  \BibitemOpen
  \bibfield  {author} {\bibinfo {author} {\bibfnamefont {Lachlan~P}\
  \bibnamefont {Lindoy}}, \bibinfo {author} {\bibfnamefont {Yao-Wen}\
  \bibnamefont {Chang}}, \ and\ \bibinfo {author} {\bibfnamefont {David~R}\
  \bibnamefont {Reichman}},\ }\href@noop {} {\enquote {\bibinfo {title}
  {Two-dimensional spectroscopy of two-dimensional materials},}\ }\bibinfo
  {note} {ArXiv:2206.01799 (2022)}\BibitemShut {NoStop}%
\bibitem [{\citenamefont {Chin}\ \emph {et~al.}(2010)\citenamefont {Chin},
  \citenamefont {Grimm}, \citenamefont {Julienne},\ and\ \citenamefont
  {Tiesinga}}]{Chin2010RMP}%
  \BibitemOpen
  \bibfield  {author} {\bibinfo {author} {\bibfnamefont {Cheng}\ \bibnamefont
  {Chin}}, \bibinfo {author} {\bibfnamefont {Rudolf}\ \bibnamefont {Grimm}},
  \bibinfo {author} {\bibfnamefont {Paul}\ \bibnamefont {Julienne}}, \ and\
  \bibinfo {author} {\bibfnamefont {Eite}\ \bibnamefont {Tiesinga}},\
  }\bibfield  {title} {\enquote {\bibinfo {title} {Feshbach resonances in
  ultracold gases},}\ }\href@noop {} {\bibfield  {journal} {\bibinfo  {journal}
  {Rev. Mod. Phys.}\ }\textbf {\bibinfo {volume} {82}},\ \bibinfo {pages}
  {1225--1286} (\bibinfo {year} {2010})}\BibitemShut {NoStop}%
\bibitem [{\citenamefont {Knap}\ \emph {et~al.}(2012)\citenamefont {Knap},
  \citenamefont {Shashi}, \citenamefont {Nishida}, \citenamefont {Imambekov},
  \citenamefont {Abanin},\ and\ \citenamefont {Demler}}]{Demler2012PRX}%
  \BibitemOpen
  \bibfield  {author} {\bibinfo {author} {\bibfnamefont {Michael}\ \bibnamefont
  {Knap}}, \bibinfo {author} {\bibfnamefont {Aditya}\ \bibnamefont {Shashi}},
  \bibinfo {author} {\bibfnamefont {Yusuke}\ \bibnamefont {Nishida}}, \bibinfo
  {author} {\bibfnamefont {Adilet}\ \bibnamefont {Imambekov}}, \bibinfo
  {author} {\bibfnamefont {Dmitry~A.}\ \bibnamefont {Abanin}}, \ and\ \bibinfo
  {author} {\bibfnamefont {Eugene}\ \bibnamefont {Demler}},\ }\bibfield
  {title} {\enquote {\bibinfo {title} {Time-dependent impurity in ultracold
  fermions: Orthogonality catastrophe and beyond},}\ }\href@noop {} {\bibfield
  {journal} {\bibinfo  {journal} {Phys. Rev. X}\ }\textbf {\bibinfo {volume}
  {2}},\ \bibinfo {pages} {041020} (\bibinfo {year} {2012})}\BibitemShut
  {NoStop}%
\bibitem [{\citenamefont {Schmidt}\ \emph {et~al.}(2018)\citenamefont
  {Schmidt}, \citenamefont {Knap}, \citenamefont {Ivanov}, \citenamefont {You},
  \citenamefont {Cetina},\ and\ \citenamefont {Demler}}]{Schmidt2018Review}%
  \BibitemOpen
  \bibfield  {author} {\bibinfo {author} {\bibfnamefont {R.}~\bibnamefont
  {Schmidt}}, \bibinfo {author} {\bibfnamefont {M.}~\bibnamefont {Knap}},
  \bibinfo {author} {\bibfnamefont {D.~A.}\ \bibnamefont {Ivanov}}, \bibinfo
  {author} {\bibfnamefont {J.-S.}\ \bibnamefont {You}}, \bibinfo {author}
  {\bibfnamefont {M.}~\bibnamefont {Cetina}}, \ and\ \bibinfo {author}
  {\bibfnamefont {E.}~\bibnamefont {Demler}},\ }\bibfield  {title} {\enquote
  {\bibinfo {title} {Universal many-body response of heavy impurities coupled
  to a {Fermi} sea: a review of recent progress},}\ }\href@noop {} {\bibfield
  {journal} {\bibinfo  {journal} {Rep. Prog. Phys.}\ }\textbf {\bibinfo
  {volume} {81}},\ \bibinfo {pages} {024401} (\bibinfo {year}
  {2018})}\BibitemShut {NoStop}%
\bibitem [{\citenamefont {Anderson}(1967)}]{Anderson1967PRL}%
  \BibitemOpen
  \bibfield  {author} {\bibinfo {author} {\bibfnamefont {P.~W.}\ \bibnamefont
  {Anderson}},\ }\bibfield  {title} {\enquote {\bibinfo {title} {Infrared
  catastrophe in fermi gases with local scattering potentials},}\ }\href@noop
  {} {\bibfield  {journal} {\bibinfo  {journal} {Phys. Rev. Lett.}\ }\textbf
  {\bibinfo {volume} {18}},\ \bibinfo {pages} {1049--1051} (\bibinfo {year}
  {1967})}\BibitemShut {NoStop}%
\bibitem [{\citenamefont {Wang}\ \emph
  {et~al.}(2022{\natexlab{a}})\citenamefont {Wang}, \citenamefont {Liu},\ and\
  \citenamefont {Hu}}]{JiaWang2022PRL}%
  \BibitemOpen
  \bibfield  {author} {\bibinfo {author} {\bibfnamefont {Jia}\ \bibnamefont
  {Wang}}, \bibinfo {author} {\bibfnamefont {Xia-Ji}\ \bibnamefont {Liu}}, \
  and\ \bibinfo {author} {\bibfnamefont {Hui}\ \bibnamefont {Hu}},\ }\bibfield
  {title} {\enquote {\bibinfo {title} {Exact quasiparticle properties of a
  heavy polaron in bcs fermi superfluids},}\ }\href@noop {} {\bibfield
  {journal} {\bibinfo  {journal} {Phys. Rev. Lett.}\ }\textbf {\bibinfo
  {volume} {128}},\ \bibinfo {pages} {175301} (\bibinfo {year}
  {2022}{\natexlab{a}})}\BibitemShut {NoStop}%
\bibitem [{\citenamefont {Wang}\ \emph
  {et~al.}(2022{\natexlab{b}})\citenamefont {Wang}, \citenamefont {Liu},\ and\
  \citenamefont {Hu}}]{JiaWang2022PRA}%
  \BibitemOpen
  \bibfield  {author} {\bibinfo {author} {\bibfnamefont {Jia}\ \bibnamefont
  {Wang}}, \bibinfo {author} {\bibfnamefont {Xia-Ji}\ \bibnamefont {Liu}}, \
  and\ \bibinfo {author} {\bibfnamefont {Hui}\ \bibnamefont {Hu}},\ }\bibfield
  {title} {\enquote {\bibinfo {title} {Heavy polarons in ultracold atomic fermi
  superfluids at the bec-bcs crossover: Formalism and applications},}\
  }\href@noop {} {\bibfield  {journal} {\bibinfo  {journal} {Phys. Rev. A}\
  }\textbf {\bibinfo {volume} {105}},\ \bibinfo {pages} {043320} (\bibinfo
  {year} {2022}{\natexlab{b}})}\BibitemShut {NoStop}%
\bibitem [{\citenamefont {Levitov}\ and\ \citenamefont
  {Lee}(1996)}]{Leonid1996JMathPhys}%
  \BibitemOpen
  \bibfield  {author} {\bibinfo {author} {\bibfnamefont {Leonid~S.}\
  \bibnamefont {Levitov}}\ and\ \bibinfo {author} {\bibfnamefont {Hyunwoo}\
  \bibnamefont {Lee}},\ }\bibfield  {title} {\enquote {\bibinfo {title}
  {Electron counting statistics and coherent states of electric current},}\
  }\href@noop {} {\bibfield  {journal} {\bibinfo  {journal} {J. Math. Phys.}\
  }\textbf {\bibinfo {volume} {37}},\ \bibinfo {pages} {4845} (\bibinfo {year}
  {1996})}\BibitemShut {NoStop}%
\bibitem [{\citenamefont {Klich}(2003)}]{Klich2003Book}%
  \BibitemOpen
  \bibfield  {author} {\bibinfo {author} {\bibfnamefont {I.}~\bibnamefont
  {Klich}},\ }\href@noop {} {\emph {\bibinfo {title} {Full Counting Statistics:
  an Elementary Derivation of {Levitov's} Formula}}}\ (\bibinfo  {publisher}
  {Kluwer},\ \bibinfo {address} {Dordrecht},\ \bibinfo {year}
  {2003})\BibitemShut {NoStop}%
\bibitem [{\citenamefont {Sch{\"o}nhammer}(2007)}]{Schonhammer2007PRB}%
  \BibitemOpen
  \bibfield  {author} {\bibinfo {author} {\bibfnamefont {K.}~\bibnamefont
  {Sch{\"o}nhammer}},\ }\bibfield  {title} {\enquote {\bibinfo {title} {Full
  counting statistics for noninteracting fermions: Exact results and the
  {Levitov-Lesovik} formula},}\ }\href@noop {} {\bibfield  {journal} {\bibinfo
  {journal} {Phys. Rev. B}\ }\textbf {\bibinfo {volume} {75}},\ \bibinfo
  {pages} {205329} (\bibinfo {year} {2007})}\BibitemShut {NoStop}%
\bibitem [{\citenamefont {Ivanov}\ and\ \citenamefont
  {Abanov}(2013)}]{Ivanov2013JMathPhys}%
  \BibitemOpen
  \bibfield  {author} {\bibinfo {author} {\bibfnamefont {Dmitri~A}\
  \bibnamefont {Ivanov}}\ and\ \bibinfo {author} {\bibfnamefont {Alexander~G}\
  \bibnamefont {Abanov}},\ }\bibfield  {title} {\enquote {\bibinfo {title}
  {{Fisher-Hartwig} expansion for {Toeplitz} determinants and the spectrum of a
  single-particle reduced density matrix for one-dimensional free fermions},}\
  }\href@noop {} {\bibfield  {journal} {\bibinfo  {journal} {J. Phys. A: Math.
  Theor.}\ }\textbf {\bibinfo {volume} {46}},\ \bibinfo {pages} {375005}
  (\bibinfo {year} {2013})}\BibitemShut {NoStop}%
\bibitem [{\citenamefont {Widera}\ \emph {et~al.}(2004)\citenamefont {Widera},
  \citenamefont {Mandel}, \citenamefont {Greiner}, \citenamefont {Kreim},
  \citenamefont {H\"ansch},\ and\ \citenamefont {Bloch}}]{Bloch2004PRL_Ramsey}%
  \BibitemOpen
  \bibfield  {author} {\bibinfo {author} {\bibfnamefont {Artur}\ \bibnamefont
  {Widera}}, \bibinfo {author} {\bibfnamefont {Olaf}\ \bibnamefont {Mandel}},
  \bibinfo {author} {\bibfnamefont {Markus}\ \bibnamefont {Greiner}}, \bibinfo
  {author} {\bibfnamefont {Susanne}\ \bibnamefont {Kreim}}, \bibinfo {author}
  {\bibfnamefont {Theodor~W.}\ \bibnamefont {H\"ansch}}, \ and\ \bibinfo
  {author} {\bibfnamefont {Immanuel}\ \bibnamefont {Bloch}},\ }\bibfield
  {title} {\enquote {\bibinfo {title} {Entanglement interferometry for
  precision measurement of atomic scattering properties},}\ }\href@noop {}
  {\bibfield  {journal} {\bibinfo  {journal} {Phys. Rev. Lett.}\ }\textbf
  {\bibinfo {volume} {92}},\ \bibinfo {pages} {160406} (\bibinfo {year}
  {2004})}\BibitemShut {NoStop}%
\bibitem [{\citenamefont {Kuklov}\ \emph {et~al.}(2004)\citenamefont {Kuklov},
  \citenamefont {Prokof'ev},\ and\ \citenamefont {Svistunov}}]{Boris2004PRA}%
  \BibitemOpen
  \bibfield  {author} {\bibinfo {author} {\bibfnamefont {Anatoly}\ \bibnamefont
  {Kuklov}}, \bibinfo {author} {\bibfnamefont {Nikolay}\ \bibnamefont
  {Prokof'ev}}, \ and\ \bibinfo {author} {\bibfnamefont {Boris}\ \bibnamefont
  {Svistunov}},\ }\bibfield  {title} {\enquote {\bibinfo {title} {Detecting
  supercounterfluidity by ramsey spectroscopy},}\ }\href@noop {} {\bibfield
  {journal} {\bibinfo  {journal} {Phys. Rev. A}\ }\textbf {\bibinfo {volume}
  {69}},\ \bibinfo {pages} {025601} (\bibinfo {year} {2004})}\BibitemShut
  {NoStop}%
\bibitem [{\citenamefont {Kitagawa}\ \emph {et~al.}(2010)\citenamefont
  {Kitagawa}, \citenamefont {Pielawa}, \citenamefont {Imambekov}, \citenamefont
  {Schmiedmayer}, \citenamefont {Gritsev},\ and\ \citenamefont
  {Demler}}]{Demler2010PRL_Ramsey}%
  \BibitemOpen
  \bibfield  {author} {\bibinfo {author} {\bibfnamefont {Takuya}\ \bibnamefont
  {Kitagawa}}, \bibinfo {author} {\bibfnamefont {Susanne}\ \bibnamefont
  {Pielawa}}, \bibinfo {author} {\bibfnamefont {Adilet}\ \bibnamefont
  {Imambekov}}, \bibinfo {author} {\bibfnamefont {J\"org}\ \bibnamefont
  {Schmiedmayer}}, \bibinfo {author} {\bibfnamefont {Vladimir}\ \bibnamefont
  {Gritsev}}, \ and\ \bibinfo {author} {\bibfnamefont {Eugene}\ \bibnamefont
  {Demler}},\ }\bibfield  {title} {\enquote {\bibinfo {title} {Ramsey
  interference in one-dimensional systems: The full distribution function of
  fringe contrast as a probe of many-body dynamics},}\ }\href@noop {}
  {\bibfield  {journal} {\bibinfo  {journal} {Phys. Rev. Lett.}\ }\textbf
  {\bibinfo {volume} {104}},\ \bibinfo {pages} {255302} (\bibinfo {year}
  {2010})}\BibitemShut {NoStop}%
\bibitem [{\citenamefont {Knap}\ \emph {et~al.}(2013)\citenamefont {Knap},
  \citenamefont {Kantian}, \citenamefont {Giamarchi}, \citenamefont {Bloch},
  \citenamefont {Lukin},\ and\ \citenamefont {Demler}}]{Demler2013PRL_Ramsey}%
  \BibitemOpen
  \bibfield  {author} {\bibinfo {author} {\bibfnamefont {Michael}\ \bibnamefont
  {Knap}}, \bibinfo {author} {\bibfnamefont {Adrian}\ \bibnamefont {Kantian}},
  \bibinfo {author} {\bibfnamefont {Thierry}\ \bibnamefont {Giamarchi}},
  \bibinfo {author} {\bibfnamefont {Immanuel}\ \bibnamefont {Bloch}}, \bibinfo
  {author} {\bibfnamefont {Mikhail~D.}\ \bibnamefont {Lukin}}, \ and\ \bibinfo
  {author} {\bibfnamefont {Eugene}\ \bibnamefont {Demler}},\ }\bibfield
  {title} {\enquote {\bibinfo {title} {Probing real-space and time-resolved
  correlation functions with many-body ramsey interferometry},}\ }\href@noop {}
  {\bibfield  {journal} {\bibinfo  {journal} {Phys. Rev. Lett.}\ }\textbf
  {\bibinfo {volume} {111}},\ \bibinfo {pages} {147205} (\bibinfo {year}
  {2013})}\BibitemShut {NoStop}%
\bibitem [{\citenamefont {Atala}\ \emph {et~al.}(2013)\citenamefont {Atala},
  \citenamefont {Aidelsburger}, \citenamefont {Barreiro}, \citenamefont
  {Abanin}, \citenamefont {Kitagawa}, \citenamefont {Demler},\ and\
  \citenamefont {Bloch}}]{Bloch2013NP}%
  \BibitemOpen
  \bibfield  {author} {\bibinfo {author} {\bibfnamefont {Marcos}\ \bibnamefont
  {Atala}}, \bibinfo {author} {\bibfnamefont {Monika}\ \bibnamefont
  {Aidelsburger}}, \bibinfo {author} {\bibfnamefont {Julio~T.}\ \bibnamefont
  {Barreiro}}, \bibinfo {author} {\bibfnamefont {Dmitry}\ \bibnamefont
  {Abanin}}, \bibinfo {author} {\bibfnamefont {Takuya}\ \bibnamefont
  {Kitagawa}}, \bibinfo {author} {\bibfnamefont {Eugene}\ \bibnamefont
  {Demler}}, \ and\ \bibinfo {author} {\bibfnamefont {Immanuel}\ \bibnamefont
  {Bloch}},\ }\bibfield  {title} {\enquote {\bibinfo {title} {Direct
  measurement of the zak phase in topological bloch bands},}\ }\href@noop {}
  {\bibfield  {journal} {\bibinfo  {journal} {Nat. Phys.}\ }\textbf {\bibinfo
  {volume} {9}},\ \bibinfo {pages} {795} (\bibinfo {year} {2013})}\BibitemShut
  {NoStop}%
\bibitem [{\citenamefont {Abanin}\ \emph {et~al.}(2013)\citenamefont {Abanin},
  \citenamefont {Kitagawa}, \citenamefont {Bloch},\ and\ \citenamefont
  {Demler}}]{Demler2013PRL_Topo}%
  \BibitemOpen
  \bibfield  {author} {\bibinfo {author} {\bibfnamefont {Dmitry~A.}\
  \bibnamefont {Abanin}}, \bibinfo {author} {\bibfnamefont {Takuya}\
  \bibnamefont {Kitagawa}}, \bibinfo {author} {\bibfnamefont {Immanuel}\
  \bibnamefont {Bloch}}, \ and\ \bibinfo {author} {\bibfnamefont {Eugene}\
  \bibnamefont {Demler}},\ }\bibfield  {title} {\enquote {\bibinfo {title}
  {Interferometric approach to measuring band topology in 2d optical
  lattices},}\ }\href@noop {} {\bibfield  {journal} {\bibinfo  {journal} {Phys.
  Rev. Lett.}\ }\textbf {\bibinfo {volume} {110}},\ \bibinfo {pages} {165304}
  (\bibinfo {year} {2013})}\BibitemShut {NoStop}%
\bibitem [{\citenamefont {Zwierlein}\ \emph {et~al.}(2003)\citenamefont
  {Zwierlein}, \citenamefont {Hadzibabic}, \citenamefont {Gupta},\ and\
  \citenamefont {Ketterle}}]{Zwierlein2003PRL}%
  \BibitemOpen
  \bibfield  {author} {\bibinfo {author} {\bibfnamefont {Martin~W.}\
  \bibnamefont {Zwierlein}}, \bibinfo {author} {\bibfnamefont {Zoran}\
  \bibnamefont {Hadzibabic}}, \bibinfo {author} {\bibfnamefont {Subhadeep}\
  \bibnamefont {Gupta}}, \ and\ \bibinfo {author} {\bibfnamefont {Wolfgang}\
  \bibnamefont {Ketterle}},\ }\bibfield  {title} {\enquote {\bibinfo {title}
  {Spectroscopic insensitivity to cold collisions in a two-state mixture of
  fermions},}\ }\href@noop {} {\bibfield  {journal} {\bibinfo  {journal} {Phys.
  Rev. Lett.}\ }\textbf {\bibinfo {volume} {91}},\ \bibinfo {pages} {250404}
  (\bibinfo {year} {2003})}\BibitemShut {NoStop}%
\bibitem [{\citenamefont {Gupta}\ \emph {et~al.}(2003)\citenamefont {Gupta},
  \citenamefont {Hadzibabic}, \citenamefont {Zwierlein}, \citenamefont {Stan},
  \citenamefont {Dieckmann}, \citenamefont {Schunck}, \citenamefont {van
  Kempen}, \citenamefont {Verhaar},\ and\ \citenamefont
  {Ketterle}}]{Ketterle2003Science}%
  \BibitemOpen
  \bibfield  {author} {\bibinfo {author} {\bibfnamefont {S.}~\bibnamefont
  {Gupta}}, \bibinfo {author} {\bibfnamefont {Z.}~\bibnamefont {Hadzibabic}},
  \bibinfo {author} {\bibfnamefont {M.W.}\ \bibnamefont {Zwierlein}}, \bibinfo
  {author} {\bibfnamefont {C.A.}\ \bibnamefont {Stan}}, \bibinfo {author}
  {\bibfnamefont {K.}~\bibnamefont {Dieckmann}}, \bibinfo {author}
  {\bibfnamefont {C.H.}\ \bibnamefont {Schunck}}, \bibinfo {author}
  {\bibfnamefont {E.G.M.}\ \bibnamefont {van Kempen}}, \bibinfo {author}
  {\bibfnamefont {B.J.}\ \bibnamefont {Verhaar}}, \ and\ \bibinfo {author}
  {\bibfnamefont {W.}~\bibnamefont {Ketterle}},\ }\bibfield  {title} {\enquote
  {\bibinfo {title} {Radio-frequency spectroscopy of ultracold fermions},}\
  }\href@noop {} {\bibfield  {journal} {\bibinfo  {journal} {Science}\ }\textbf
  {\bibinfo {volume} {300}},\ \bibinfo {pages} {1723} (\bibinfo {year}
  {2003})}\BibitemShut {NoStop}%
\bibitem [{\citenamefont {Rey}\ \emph {et~al.}(2009)\citenamefont {Rey},
  \citenamefont {Gorshkov},\ and\ \citenamefont {Rubbo}}]{Rey2009PRL}%
  \BibitemOpen
  \bibfield  {author} {\bibinfo {author} {\bibfnamefont {A.~M.}\ \bibnamefont
  {Rey}}, \bibinfo {author} {\bibfnamefont {A.~V.}\ \bibnamefont {Gorshkov}}, \
  and\ \bibinfo {author} {\bibfnamefont {C.}~\bibnamefont {Rubbo}},\ }\bibfield
   {title} {\enquote {\bibinfo {title} {Many-body treatment of the collisional
  frequency shift in fermionic atoms},}\ }\href@noop {} {\bibfield  {journal}
  {\bibinfo  {journal} {Phys. Rev. Lett.}\ }\textbf {\bibinfo {volume} {103}},\
  \bibinfo {pages} {260402} (\bibinfo {year} {2009})}\BibitemShut {NoStop}%
\bibitem [{\citenamefont {Yu}\ and\ \citenamefont {Pethick}(2010)}]{Yu2010PRL}%
  \BibitemOpen
  \bibfield  {author} {\bibinfo {author} {\bibfnamefont {Zhenhua}\ \bibnamefont
  {Yu}}\ and\ \bibinfo {author} {\bibfnamefont {C.~J.}\ \bibnamefont
  {Pethick}},\ }\bibfield  {title} {\enquote {\bibinfo {title} {Clock shifts of
  optical transitions in ultracold atomic gases},}\ }\href@noop {} {\bibfield
  {journal} {\bibinfo  {journal} {Phys. Rev. Lett.}\ }\textbf {\bibinfo
  {volume} {104}},\ \bibinfo {pages} {010801} (\bibinfo {year}
  {2010})}\BibitemShut {NoStop}%
\bibitem [{\citenamefont {Martin}\ \emph {et~al.}(2013)\citenamefont {Martin},
  \citenamefont {Bishof}, \citenamefont {Swallows}, \citenamefont {Zhang},
  \citenamefont {Benko}, \citenamefont {von Stecher}, \citenamefont {Gorshkov},
  \citenamefont {Rey},\ and\ \citenamefont {Ye}}]{Ye2013Science}%
  \BibitemOpen
  \bibfield  {author} {\bibinfo {author} {\bibfnamefont {M.~J.}\ \bibnamefont
  {Martin}}, \bibinfo {author} {\bibfnamefont {M.}~\bibnamefont {Bishof}},
  \bibinfo {author} {\bibfnamefont {M.~D.}\ \bibnamefont {Swallows}}, \bibinfo
  {author} {\bibfnamefont {X.}~\bibnamefont {Zhang}}, \bibinfo {author}
  {\bibfnamefont {C.}~\bibnamefont {Benko}}, \bibinfo {author} {\bibfnamefont
  {J.}~\bibnamefont {von Stecher}}, \bibinfo {author} {\bibfnamefont {A.~V.}\
  \bibnamefont {Gorshkov}}, \bibinfo {author} {\bibfnamefont {A.~M.}\
  \bibnamefont {Rey}}, \ and\ \bibinfo {author} {\bibfnamefont {Jun}\
  \bibnamefont {Ye}},\ }\bibfield  {title} {\enquote {\bibinfo {title} {A
  quantum many-body spin system in an optical lattice clock},}\ }\href@noop {}
  {\bibfield  {journal} {\bibinfo  {journal} {Science}\ }\textbf {\bibinfo
  {volume} {341}},\ \bibinfo {pages} {632} (\bibinfo {year}
  {2013})}\BibitemShut {NoStop}%
\bibitem [{\citenamefont {Yan}\ \emph {et~al.}(2013)\citenamefont {Yan},
  \citenamefont {Moses}, \citenamefont {Gadway}, \citenamefont {Covey},
  \citenamefont {Hazzard}, \citenamefont {Rey}, \citenamefont {Jin},\ and\
  \citenamefont {Ye}}]{Ye2013Nature}%
  \BibitemOpen
  \bibfield  {author} {\bibinfo {author} {\bibfnamefont {Bo}~\bibnamefont
  {Yan}}, \bibinfo {author} {\bibfnamefont {Steven~A.}\ \bibnamefont {Moses}},
  \bibinfo {author} {\bibfnamefont {Bryce}\ \bibnamefont {Gadway}}, \bibinfo
  {author} {\bibfnamefont {Jacob~P.}\ \bibnamefont {Covey}}, \bibinfo {author}
  {\bibfnamefont {Kaden R.~A.}\ \bibnamefont {Hazzard}}, \bibinfo {author}
  {\bibfnamefont {Ana~Maria}\ \bibnamefont {Rey}}, \bibinfo {author}
  {\bibfnamefont {Deborah~S.}\ \bibnamefont {Jin}}, \ and\ \bibinfo {author}
  {\bibfnamefont {Jun}\ \bibnamefont {Ye}},\ }\bibfield  {title} {\enquote
  {\bibinfo {title} {Observation of dipolar spin-exchange interactions with
  lattice-confined polar molecules},}\ }\href@noop {} {\bibfield  {journal}
  {\bibinfo  {journal} {Nature}\ }\textbf {\bibinfo {volume} {501}},\ \bibinfo
  {pages} {521} (\bibinfo {year} {2013})}\BibitemShut {NoStop}%
\bibitem [{\citenamefont {Goold}\ \emph {et~al.}(2011)\citenamefont {Goold},
  \citenamefont {Fogarty}, \citenamefont {Lo~Gullo}, \citenamefont
  {Paternostro},\ and\ \citenamefont {Busch}}]{Goold2011PRA}%
  \BibitemOpen
  \bibfield  {author} {\bibinfo {author} {\bibfnamefont {J.}~\bibnamefont
  {Goold}}, \bibinfo {author} {\bibfnamefont {T.}~\bibnamefont {Fogarty}},
  \bibinfo {author} {\bibfnamefont {N.}~\bibnamefont {Lo~Gullo}}, \bibinfo
  {author} {\bibfnamefont {M.}~\bibnamefont {Paternostro}}, \ and\ \bibinfo
  {author} {\bibfnamefont {Th.}\ \bibnamefont {Busch}},\ }\bibfield  {title}
  {\enquote {\bibinfo {title} {Orthogonality catastrophe as a consequence of
  qubit embedding in an ultracold fermi gas},}\ }\href@noop {} {\bibfield
  {journal} {\bibinfo  {journal} {Phys. Rev. A}\ }\textbf {\bibinfo {volume}
  {84}},\ \bibinfo {pages} {063632} (\bibinfo {year} {2011})}\BibitemShut
  {NoStop}%
\bibitem [{\citenamefont {Mitchison}\ \emph {et~al.}(2020)\citenamefont
  {Mitchison}, \citenamefont {Fogarty}, \citenamefont {Guarnieri},
  \citenamefont {Campbell}, \citenamefont {Busch},\ and\ \citenamefont
  {Goold}}]{Goold2020PRL}%
  \BibitemOpen
  \bibfield  {author} {\bibinfo {author} {\bibfnamefont {Mark~T.}\ \bibnamefont
  {Mitchison}}, \bibinfo {author} {\bibfnamefont {Thom\'as}\ \bibnamefont
  {Fogarty}}, \bibinfo {author} {\bibfnamefont {Giacomo}\ \bibnamefont
  {Guarnieri}}, \bibinfo {author} {\bibfnamefont {Steve}\ \bibnamefont
  {Campbell}}, \bibinfo {author} {\bibfnamefont {Thomas}\ \bibnamefont
  {Busch}}, \ and\ \bibinfo {author} {\bibfnamefont {John}\ \bibnamefont
  {Goold}},\ }\bibfield  {title} {\enquote {\bibinfo {title} {In situ
  thermometry of a cold fermi gas via dephasing impurities},}\ }\href@noop {}
  {\bibfield  {journal} {\bibinfo  {journal} {Phys. Rev. Lett.}\ }\textbf
  {\bibinfo {volume} {125}},\ \bibinfo {pages} {080402} (\bibinfo {year}
  {2020})}\BibitemShut {NoStop}%
\bibitem [{\citenamefont {Bloch}\ \emph {et~al.}(2008)\citenamefont {Bloch},
  \citenamefont {Dalibard},\ and\ \citenamefont {Zwerger}}]{Bloch2008RMP}%
  \BibitemOpen
  \bibfield  {author} {\bibinfo {author} {\bibfnamefont {Immanuel}\
  \bibnamefont {Bloch}}, \bibinfo {author} {\bibfnamefont {Jean}\ \bibnamefont
  {Dalibard}}, \ and\ \bibinfo {author} {\bibfnamefont {Wilhelm}\ \bibnamefont
  {Zwerger}},\ }\bibfield  {title} {\enquote {\bibinfo {title} {Many-body
  physics with ultracold gases},}\ }\href@noop {} {\bibfield  {journal}
  {\bibinfo  {journal} {Rev. Mod. Phys.}\ }\textbf {\bibinfo {volume} {80}},\
  \bibinfo {pages} {885--964} (\bibinfo {year} {2008})}\BibitemShut {NoStop}%
\end{thebibliography}%

\end{document}